\documentclass[11pt,a4paper]{article}
\pdfoutput=1
\usepackage{jheppub}
\hypersetup{pdfencoding=unicode, bookmarksopen=true, bookmarksnumbered}
\usepackage[nomath]{lmodern}
\usepackage{microtype}
\usepackage{amsthm,amsbsy,amsfonts,mathrsfs,enumerate,float,wrapfig,amsmath}
\usepackage{multirow}
\usepackage{subfigure}
\usepackage{longtable}
\usepackage{adjustbox} 
\usepackage{physics}
\usepackage{dashrule}
\usepackage{tikz}
\DeclareMathOperator{\PE}{PE}
\DeclareMathOperator{\Li}{Li}

\title{Spectra of BPS Strings in 6d Supergravity and the Swampland}

\author[a]{Hirotaka Hayashi,}
\author[b,c]{Hee-Cheol Kim,}
\author[b]{Minsung Kim}

\affiliation[a]{Department of Physics, School of Science, Tokai University, 4-1-1 Kitakaname, Hiratsuka-shi, Kanagawa 259-1292, Japan}
\affiliation[b]{Department of Physics, POSTECH, Pohang 790-784, Korea}
\affiliation[c]{Asia Pacific Center for Theoretical Physics, Pohang 37673, Korea}

\emailAdd{h.hayashi@tokai.ac.jp}
\emailAdd{heecheol@postech.ac.kr}
\emailAdd{kms0323@postech.ac.kr}

\abstract{We explore BPS strings in supergravity theories in six-dimensions and related Swampland Conjectures. We first propose a general modular ansatz for bootstrapping elliptic genera of 2d worldvolume theories on strings in the 6d theories. By employing mirror symmetry on F-theory examples, we explicitly compute the elliptic genera and validate our ansatz. We extend this approach to investigate BPS strings and their spectrum in non-geometric 6d theories which have no known F-theory constructions, and confirm the Swampland conjectures, including the Weak Gravity Conjecture, Distance Conjecture, and Emergent String Conjecture. We also discuss tensionless little strings that emerge near infinite-distance limits of strong gauge coupling in the moduli space of certain special theories.}

\begin{document}

\maketitle

\section{Introduction}

Due to the absence of a complete formulation, quantum gravity remains the least-explored area of research in theoretical physics.  A prevailing strategy to address the challenges of quantum gravity involves a recent shift in focus toward low-energy physics where an effective field theory (EFT) offers a well-defined framework that allows us to systematically conduct quantum computations for gravitational phenomena. 
The effective field theory significantly deepens our grasp of quantum gravity, even though its applicability is confined to specific energy scale ranges set by a UV cut-off. 

The idea that any seemingly consistent EFTs can couple to quantum gravity is no longer universally accepted. In fact, only a highly special subset of EFTs can have consistent UV completions in the framework of quantum gravity. The Swampland program \cite{Vafa:2005ui} aims to pinpoint the boundaries of EFTs that can be consistently incorporated with quantum gravity. The term ``{\it swampland}'' refers to those EFTs that cannot be UV-completed in quantum gravity. Several criteria have been proposed to differentiate between EFTs in the swampland and those that can be completed in UV by quantum gravity.  Notably, the main Swampland Conjectures, which have significant implications for low-energy physics, include the Weak Gravity Conjecture (WGC), No Global Symmetry Conjecture, and the Distance Conjecture. 
The formulation for these conjectures lies in black hole physics, insights from string theory, and holographic considerations, all of which are supported by numerous examples and string theory arguments.

Recently, higher-dimensional objects, such as strings and branes, have been employed to shed light on new Swampland bounds in supersymmetric theories \cite{Kim:2019vuc,Lee:2019skh,Kim:2019ths,Katz:2020ewz,Angelantonj:2020pyr,Cheng:2021zjh,Hamada:2021bbz,Tarazi:2021duw,Martucci:2022krl}. They have also been used to confirm various Swampland conjectures in a broad category of EFTs that are formulated in string theory. In particular, the Weak Gravity Conjecture has been rigorously tested, and in some cases even proven, in the EFTs that are geometrically constructed through F-theory compactification on elliptically fibered Calabi-Yau (CY) manifolds \cite{Lee:2018spm,Lee:2019xtm,Lee:2019wij}. This is achieved by employing the critical strings that appears at infinite distance in the moduli space. In this context, the excitation spectrum of these critical strings plays a pivotal role in the proof of the Weak Gravity Conjecture.

In this work, we explore the properties of BPS strings in 6d $\mathcal{N}=(1,0)$ supergravity theories by explicitly calculating and utilizing their elliptic genera. We first suggest a general modular ansatz that serves as a foundation for bootstraping elliptic genera of BPS strings in 6d gravity. This is an extension of the modular bootstrap approach originally suggested in \cite{Huang:2015sta} for strings which come from wrapped D3-branes on 2-cycles within a compact elliptic Calabi-Yau threefolds $X$ in a 6d F-theory. This generalized ansatz proves to be remarkably effective, especially when considering D3-branes wrapping around curves of gauge singularities in the K\"ahler base $B$ of the elliptic 3-folds.  In the cases where the elliptic CY 3-fold admits a toric hypersurface construction, we are able to extract genus-zero Gopakumar-Vafa (GV) invariants \cite{Gopakumar:1998ki,Gopakumar:1998ii,Gopakumar:1998jq} in topological strings by employing the mirror symmetry technique \cite{Hosono:1993qy,Hosono:1994ax}. These mirror calculations, in turn, allow us to precisely determine the elliptic genus in certain cases, serving as a concrete means to verify our generalized modular ansatz. We discuss this process and provide a variety of intriguing examples in Section \ref{sec:bootstrap}.

In 6D supergravity, among the various BPS strings, those that become tensionless at infinite-distance points in the moduli space are especially intriguing.  The Emergent String Conjecture, as proposed in \cite{Lee:2018spm,Lee:2019xtm,Lee:2019wij}, anticipates  the presence of such strings in the asymptotic limits of the moduli space in the presence of gravity. In the context of six-dimensional compactifications of F-theory, these strings are identified as weakly coupled critical strings in a duality frame. These strings are of great importance for testing different versions of the Weak Gravity and Swampland Distance Conjectures. It is observed that there are exponentially light towers of particle excitations emerging from these tensionless critical strings. As a result, the effective field theory breaks down at the infinite-distance point in the moduli space, which is consistent with the predictions of the Swampland Distance Conjecture. Moreover, it can be rigorously proven that a tower of superextremal states, populating a sublattice of the charge lattice, exists. This proof utilizes the spectrum of the tensionless string and its behavior under modular transformations, as detailed in \cite{Lee:2018urn,Lee:2018spm}. This provides a solid validation of the Sublattice Weak Gravity Conjecture (sLWGC) proposed in \cite{Montero:2016tif,Heidenreich:2015nta,Heidenreich:2016aqi}, in a wide class of F-theory compactifications on elliptic Calabi-Yau 3-folds.

A detailed and systematic analysis of the weak coupling limits of gauge fields in 6D supergravity can be conducted within the framework of effective field theory descriptions, without necessarily referring to their geometric realizations in F-theory. Specifically, there exists a broad class of EFTs with gauge groups that can be Higgsed to yield another theory, which itself admits an F-theory construction. In these situations, the Higgsing procedure is related only to the vector and charged hypermultiplets, leaving the tensor branch moduli space in 6d supergravity unaffected before and after the Higgsing. This fact allows us to use the geometric analysis from the Higgsed theory to obtain a robust analysis of the tensor moduli space in the original theory. Especially, this implies that the charge lattice of BPS strings stays the same through the Higgsing process. So, if the Higgsed theory reveals infinite distance limits with tensionless strings in the moduli space, we can trace back and identify the same features, so the emergence of tensionless strings, in the original theory before the Higgsing.

Based on this argument, we will explore BPS strings in a specific 6d supergravity theory including  3-index symmetric representations of $SU(2)$ gauge group. This theory has been shown to be impossible to geometrically realize in F-theory \cite{Klevers:2017aku}. We will present  substantial evidence indicating that a tensionless string emerges at a weak gauge coupling limit, and furthermore, it becomes a critical heterotic string on a K3 surface in some duality frame.  Our discussion involves a thorough comparison of modular properties of its elliptic genus with those characteristic of heterotic strings in string theory. As we will illustrate, the elliptic genus of this tensionless string can be precisely determined using the modular bootstrap method, along with the properties of heterotic strings. This analysis allows us to identify charges and masses of particle states originating from the heterotic string and ultimately prove the Sublattice Weak Gravity Conjecture, even in a theory without geometric features. Our findings will highlight the impactful use of the BPS spectrum of strings in broader supergravity theories that lack geometric realizations.

In our analysis of the moduli space in 6d supergravity, we made an interesting observation. In certain theories, infinite distance limits, while keeping gravity dynamical, always correspond to strong coupling limits with asymptotically tensionless strings. This is still consistent with the Emergent String Conjecture. However, the strings that carry a charge under the gauge symmetry in these cases are not the critical strings we might anticipate. Instead, they turn out to be non-critical `little strings' found in 6d little string theories (LSTs) coupled to gravity. This means that, in these particular cases, the earlier proof of the Weak Gravity Conjecture in \cite{Lee:2018urn,Lee:2018spm}, which relies on the characteristics of charged heterotic strings, is not directly applicable here. We will illustrate this through a concrete example.

The rest of this paper is organized as follows. In Section \ref{sec:SUGRA}, we offer a review of 6d $(1,0)$ supergravity along with the characteristics of BPS strings. In Section \ref{sec:bootstrap}, we demonstrate how to calculate the elliptic genera of strings in 6d supergravity. To achieve this, we employ our generalized modular ansatz and mirror symmetry techniques, and we also present some intriguing examples. In Section \ref{sec:emergent-string}, we analyze the moduli space and spectrum of BPS strings in 6d theories that lack F-theory realizations. We provide a rigorous demonstration that these theories conform to various Swampland Conjectures.  Additionally, we provide illustrations of the cases where little strings emerge at the infinite-distance limit. In Section~\ref{sec:conclusion}, we summarize our results and discuss some future directions.

\section{\texorpdfstring{$\mathcal{N}=(1,0)$}{N=(1,0)} supergravity in six dimensions}\label{sec:SUGRA}
In this section, we will briefly review some basics of 6d $\mathcal{N}=(1,0)$ supergravity theories, focusing on the massless spectrum of the low energy effective theory, BPS strings, and their realizations in F-theory compactifications.

\subsection{Reviews of effective theory}
\label{sec:effective}
We will consider six-dimensional supergravity theories preserving 8 supersymmetries. The massless spectrum of each theory consists of a gravity multiplet, tensor multiplets, vector multiplets for gauge groups $G=\prod_iG_i$, and hypermultiplets transforming in a representation of $G$. The detailed spectrum is strongly constrained by anomaly cancellation conditions using Green-Schwarz-Sagnotti mechanism \cite{Sagnotti:1992qw}. The anomaly cancellation conditions for a generic 6d supergraivity theory can be summarized as follows:
\begin{align}\label{eq:6dconstraints}
    &H-V = 273 - 29T \, , \quad a\cdot a = 9-T \, , \nonumber \\
    &a\cdot b_i = \frac{\lambda_i}{6}\left(A^i_{\bf adj} - \sum_{\bf R} n^i_{\bf R} A^i_{\bf R}\right) \, , \quad B^i_{\bf adj} - \sum_{\bf R} n^i_{\bf R}B^i_{\bf R} = 0 \, ,\nonumber \\
    &b_i\cdot b_i = \frac{\lambda_i^2}{3}\left(\sum_{\bf R} n^i_{\bf R} C^i_R - C^i_{\bf adj}\right)  \, , \quad b_i\cdot b_j = 2\lambda_i\lambda_j\sum_{\bf R,S}n^{ij}_{\bf R,S}A^i_{\bf R} A^j_{\bf S} \quad (i\neq j) \, ,
\end{align}
where $T,V$ and $H$ are the number of tensor, vector, and hyper multiplets respectively. $a^\alpha$, $b_i^\alpha$ with $\alpha=0,1,\cdots T$ are the Green-Schwarz coefficients taking values in a tensor charge lattice of signature $(1,T)$, and the inner product of two vectors, i.e. $v\cdot w$, is defined with respect to the metric $\Omega_{\alpha\beta}$ in the tensor moduli space. $A_{\bf R}, B_{\bf R}, C_{\bf R}$ for the representation ${\bf R}$ are group theory coefficients defined as
\begin{align}
  {\rm tr}_{\bf R}F^2 = A_{\bf R}{\rm tr}F^2 \ , \quad {\rm tr}_{\bf R}F^4 = B_{\bf R}{\rm tr}F^4 + C_{\bf R}({\rm tr}F^2)^2 \ ,
\end{align}
and $n_{\bf R}$ is the number of hypermultiplets in the representation ${\bf R}$. Lastly, $\lambda_i$ denotes the group theory factor of $G_i$ which is summarized in \cite{Kumar:2009ae}.

When the anomaly cancellation conditions are satisfied, the 1-loop anomalies from the massless chiral fields factorize and we can cancel them by adding to the classical action the Green-Schwarz term of the form
\begin{align}
  S_{GS} = \int  \Omega_{\alpha\beta}B_2^{\alpha}\wedge X_4^\beta \ ,
\end{align}
with the (anti-)sef-dual 2-form fields $B_2^\alpha$ and
\begin{align}
X_4^\alpha &= \frac{1}{2}a^\alpha {\rm tr} R^2 + \frac{1}{4}\sum_i b_i^\alpha {\rm Tr} F_i^2 \ .
\end{align}

The moduli space of the 6d $\mathcal{N}=(1,0)$ supergravity theory is parametrized by the scalar fields in the tensor multiplets.  This space is a $T$-dimensional space, which locally takes the shape of $SO(1,T)/SO(T)$, described by a linear combination of the scalar fields $J_\alpha\in \mathbb{R}^{1,T}$ that satisfy the conditions
\begin{align}\label{condition_j}
  J\cdot J >0 \ , \quad J\cdot b_i >0 \ , \quad  - J \cdot a>0 \ .
\end{align}
The first condition is for ensuring the metric positivity on the tensor branch moduli space along $J$, while the second condition is responsible for the positivity of the gauge kinetic terms \cite{Sagnotti:1992qw}. The last condition is conjectured in \cite{Cheung:2016wjt,Hamada:2018dde} to secure the positivity of the Gauss-Bonnet term in gravity. Note that the 6d Planck mass $M_{\rm Pl}$ and the gauge coupling can be expressed in terms of the scalar fields as
\begin{align}
  M_{\rm Pl}^4 = \frac{1}{2}J\cdot J \ , \quad \frac{1}{g_{i,{\rm YM}}^2}= J\cdot b_i \ .
\end{align}

We are interested in the half-BPS strings on the tensor branch that are charged with respect to the 2-form symmetry of the gauge field $B_2^\alpha$. The worldvolume theories on these strings at low energy are described by 2d $\mathcal{N}=(0,4)$ superformal field theories (SCFTs). The central charges and the levels of flavor symmetries in the 2d worldvolume theories can be computed from the anomaly polynomial which can be uniquely identified by using anomaly inflows from the 6d bulk theory in the presence of the string sources \cite{Kim:2016foj,Shimizu:2016lbw,Kim:2019vuc}. For a given tensor charge $Q^\alpha$, the anomaly polynomial of the worldsheet CFT is
\begin{align}\label{eq:anomaly-polynomial}
	I_4 = -\frac{Q\cdot a}{4}p_1(T_2)+\frac{1}{4}\sum_i Q\cdot b_i {\rm Tr} F_i^2-\frac{Q\cdot Q -Q\cdot a}{2}c_2(R)+\frac{Q\cdot Q +Q\cdot a}{2}c_2(l) \ ,
\end{align}
where $p_1(T_2)$ is the first Pontryagin class of the two-manifold wrapped by the string, and $c_2(R)$ and $c_2(l)$ are the second Chern classes of the $SU(2)_R$ R-symmetry and the $SU(2)_l\subset SO(4)$ rotational symmetry, respectively, transverse to the worldsheet directions. We here defined `{\rm Tr}'  in such a way that integrating $\frac{1}{4}{\rm Tr}F^2$ over a 4-manifold yields an instanton number $n\in \mathbb{Z}$. 

From the anomaly polynomial, we can read the left-moving and the right-moving central charges, including the center of mass contributions, as
\begin{align}\label{eq:central-charge}
  c_L = 3Q\cdot Q-9Q\cdot a +6 \ , \quad c_R = 3Q\cdot Q - 3Q\cdot a +6\ ,
\end{align}
and the levels for the $SU(2)_l$ flavor symmetry and for the bulk gauge symmetries $G_i$ as
\begin{align}\label{eq:levels}
  k_l = \frac{1}{2}(Q\cdot Q+Q\cdot a) \ , \quad k_i = Q\cdot b_i \ ,
\end{align}
respectively.

It was recently argued based on the completeness hypothesis \cite{Polchinski:2003bq} that the string charge lattice $\Gamma$, which is unimodular and self-dual, spanned by a basis of BPS charges must be populated by BPS strings \cite{Kim:2019vuc,Tarazi:2021duw}. Furthermore, the condition that the left-moving central charge on the string probes should be large enough to allow unitary representations of the current algebras with $k_l,k_i$ provides upper bounds on the rank and the type of gauge groups, and the number of massless modes in 6d supergravity theories that can consistently couple to quantum gravity. This condition has served as a means to rule out a large number of supergravity theories that would otherwise appear to be consistent \cite{Kim:2019vuc,Lee:2019skh,Angelantonj:2020pyr,Cheng:2021zjh,Tarazi:2021duw}.

\subsection{F-theory realizations}

F-theory compactifications on genus-one fibered, compact Calabi-Yau threefolds may give rise to 6d $\mathcal{N}=(1,0)$ supergravity theories. Here we consider elliptically fibered Calabi-Yau threefolds which can be described by the Weierstrass form.\footnote{F-theory compactifications on genus-one fibered Calabi-Yau threefolds have been studied for example in \cite{Braun:2014oya}.}
Let $B$ be a base of an elliptically fibered Calabi-Yau threefold $X$ characterized by the Weierstrass form,
\begin{equation}\label{weierstrass}
y^2 = x^3 + f x+g\,,
\end{equation}
where $f, g$ are sections of $\mathcal{O}(-4K_B)$ and $\mathcal{O}(-6K_B)$ respectively where $K_B$ is the canonical divisor of $B$. We focus on the cases with the $\mathcal{N}=(1,0)$ supersymmetry and $h^{1,0}(B) = h^{2,0}(B) = 0$. 

The elliptic fiber of a Calabi-Yau threefold $X$ becomes singular when the discriminant of the Weierstrass form \eqref{weierstrass},
\begin{equation}\label{discriminant}
\Delta = 4f^3 + 27g^2\,,
\end{equation}
vanishes. The discriminant locus $\Delta = 0$ yields a divisor in $B$. Physically, the discriminant locus describes a location of 7-branes. We first consider elliptic fibers over generic points on the discriminant locus. Locally, one can describe the discriminant locus by $w=0$ with a local coordinate $w$ and the discriminant \eqref{discriminant} becomes $\Delta = w^N\Delta'$. $f, g$ can be expanded by $w$ on the local patch as
\begin{equation}
    f = f_0 + f_1 w + f_2 w^2 + \cdots, \qquad g = g_0 + g_1 w + g_2 w^2 + \cdots. 
\end{equation}
The structure of $f_k, g_k\; (k=1, \cdots)$ may determine the type of the singular elliptic fibers \cite{Vafa:1996xn, Morrison:1996na, Morrison:1996pp, Bershadsky:1996nh, Katz:2011qp}. When singularities of the elliptic fibers lead to singularities in the Calabi-Yau threefold $X$, then the resulting 6d supergravity theories have non-Abelian gauge groups. In other words, the singularities arise when 7-branes are on top of each other and the multiple 7-branes give the non-Abelian gauge groups. More generally the discriminant may have multiple components as
\begin{equation}
\Delta = \Delta_0\prod_{i=1}^n\Delta_i^{m_i},
\end{equation}
where $m_i$'s are positive integers. The elliptic fiber is of Kodaira type $I_1$ and $X$ is smooth on $\Delta_0=0$ while $X$ is singular on $\Delta_i=0$. Let $C_{\iota}$ be the divisor given by $\Delta_{\iota} = 0$ for $\iota=0,1, \cdots, n$. We also denote the non-Abelian gauge algebra associated to the elliptic fiber over $\Delta_i=0$ by $\mathfrak{g}_i$ for $i=1, \cdots, n$. Note that monodromies along $C_i$ may yield non-simply-laced gauge algebras. On the other hand, a $U(1)$ gauge group which is not related to the Cartan generators of non-Abelian gauge algebras arises when the elliptic fibration admits another section \cite{Morrison:1996pp, Aspinwall:1998xj, Aspinwall:2000kf}. The number of the $U(1)$ factors is given by the rank of the Mordell-Weil group of rational sections. The torsion part of the Mordell-Weil group is related to the global structure of non-Abelian gauge groups \cite{Aspinwall:1998xj, Mayrhofer:2014opa}.

Matters charged under non-Abelian gauge algebra arise in two ways, either from unlocalized matters or localized matters. As for the unlocalized matter, we have $g(C_i)$ hypermultiplets in the adjoint representation of $\mathfrak{g}_i$ where $g(C_i)$ is the geometric genus of $C_i$ \cite{Witten:1996qb}. Another unlocalized matter appears when $\mathfrak{g}_i$ is non-simply-laced. In cases where $\mathfrak{g}_i$ is non-simply-laced, the singular elliptic fiber is locally associated to a simply-laced gauge algebra $\widetilde{\mathfrak{g}}_i$. 
Then we have hypermultiplets in the representation ${\mathbf r}_i$ which appear in the decomposition \cite{Aspinwall:2000kf}
\begin{equation}
\text{adj}(\widetilde{\mathfrak{g}}_i) = \text{adj}(\mathfrak{g}_i) + (d_i-1){\mathbf r}_i,
\end{equation}
where $d_i$ is the order of the outer automorphism which reduces $\widetilde{\mathfrak{g}}_i$ to $\mathfrak{g}_i$. Their number is given by \cite{Aspinwall:2000kf, Grassi:2000we}
\begin{equation}
n_{{\mathbf r}_i} = (d_i-1)\left(g(C_i)-1\right) + \frac{1}{2}\text{deg}(R),
\end{equation}
where $R$ is the ramification divisor of $\widetilde{C}_i$ which is a degree $d_i$ cover of $C_i$.  

On the other hand, localized matters arise from codimension-two singularities on $B$. When $C_i$ intersects with $C_0$ or $C_j$, the singularity may be enhanced at the intersection points $Q_{i\iota}^{(r)}\; (\iota =0, j)$ where $r$ labels the intersection points. In a special case, $C_i$ may intersect with itself and then $C_i$ becomes singular. We assume that the vanishing order of $f$ is less than $4$ or that of $g$ is less than $6$. The number of the intersection points is related to the number of charged (half-)hypermultiplets localized at each point. To determine the representation of a hypermultiplet localized at each intersection point, we first read off the naive Kodaira type 
of the elliptic fiber on $Q_{i\iota}^{(r)}$ and determine the associated enhanced gauge algebra $\widetilde{\mathfrak{g}}_{Q_{i\iota}^{(r)}}$. Then we consider the decomposition of the adjoint representation of $\widetilde{\mathfrak{g}}_{Q_{i\iota}^{(r)}}$ under $\mathfrak{g}_i \oplus \mathfrak{g}_{\iota}$ where $\mathfrak{g}_{0}$ is the trivial representation. The representation of the hypermultiplet localized on $Q_{i\iota}^{(r)}$ is roughly given by the non-trivial irreducible representation in the decomposition except for the representations corresponding to the adjoint representations of $\widetilde{\mathfrak{g}}_i$ and $\widetilde{\mathfrak{g}}_{\iota}$ \cite{Katz:1996xe}. 
The precise statement requires extra care and the matter representations at codimension-two singularities under some genericity assumption has been determined in \cite{Grassi:2000we, Grassi:2011hq, Grassi:2018rva}. When $C_i$ intersects with itself, we consider decomposition under $\mathfrak{g}_i \subset \mathfrak{g}_i \oplus \mathfrak{g}_{i}$, which may, for example, give rise to the rank-2 symmetric representation of $\mathfrak{su}(N)$ \cite{Sadov:1996zm}. In the case of $\mathfrak{g}_i = \mathfrak{su}(2)$, a triple point singularity is possible, which lead to the rank-3 symmetric representation of $\mathfrak{su}(2)$ \cite{Klevers:2016jsz}.

There are also other types of codimension-two singularities. One type appears when the vanishing orders of $f$ and $g$ exceed the aforementioned bound at codimensin-two. In this case, a conformal matter \cite{DelZotto:2014hpa}, instead of ordinary hypermultiplets, may be localized at the singularity. Another type is $\mathbb{Q}$-factorial terminal singularities at codimension-two. Then the singularities may induce hypermultiplets uncharged under continuous gauge groups \cite{Arras:2016evy, Grassi:2018rva}.

The number of other uncharged hypermultiplets can be determined as follows. 
Let $X'$ be $X$ with all the singularities resolved, assuming that there are no $\mathbb{Q}$-factorial terminal singularities. 
We also denote the number of neutral hypermultiplets by $n_{{\mathbf 1}}$. Then $n_{{\mathbf 1}}$ is given by \cite{Vafa:1996xn, Morrison:1996na, Morrison:1996pp}
\begin{align}\label{num_singlet}
n_{{\mathbf 1}}&=h^{2,1}(X') + 1.
\end{align}
The number of tensor multiplets and the rank $r(V)$ of the gauge groups are also related to the Hodge numbers and they are given by
\begin{align}\label{tensor,rank}
    \begin{aligned}
        T &= h^{1,1}(B)-1, \\
        r(V)&=h^{1,1}(X') - h^{1,1}(B)-1.
    \end{aligned}
\end{align}

\paragraph{$SU(2)$ examples.} Let us consider an example with an $SU(2)$ gauge group. An $\mathfrak{su}(2)$ gauge algebra can be engineered by tuning the $f$ and $g$ in \eqref{weierstrass} as
\begin{align}
f &= -\frac{1}{3}\phi^2 + wf_1,\label{f_su2}\\
g &= \frac{2}{27}\phi^3 -\frac{1}{3}\phi w f_1 + w^2 g_2,\label{g_su2}
\end{align}
where $w$ is a local coordinate. Then the discriminant \eqref{discriminant} becomes
\begin{align}
\Delta = w^2\phi^2(4\phi g_2 - f_1^2) + \mathcal{O}(w^3).
\end{align}
The elliptic fiber is of Kodaira type $I_2$ on $w=0$ and it is associated to an $\mathfrak{su}(2)$. The type of the elliptic fiber over $w=0$ is enhanced at codimension-two loci $\phi=0$ or $4\phi g_2 - f_1^2=0$. The elliptic fiber becomes type $III$ and the associated gauge algebra is still $\mathfrak{su}(2)$ at $\{\phi =0\} \cap \{w=0\}$. Then there is no charged hypermultiplet localized at the codimension-two points. The elliptic fiber becomes type $I_3$ at $\{4\phi g_2 - f_1^2 = 0\}\cap \{w=0\}$ and the associated gauge algebra is $\mathfrak{su}(3)$. Then the decomposition 
\begin{align}
\text{adj}(\mathfrak{su}(3)) = \text{adj}(\mathfrak{su}(2)) + {\mathbf 1} + {\mathbf 2} + \overline{{\mathbf 2}}
\end{align}
implies that a hypermultiplet in the fundamental representation ${\mathbf 2}$ of $\mathfrak{su}(2)$ is localized at each point of the codimension-two singularities. As an example, suppose $B=\mathbb{P}^2$ and an $\mathfrak{su}(2)$ gauge algebra is realized on degree $d$ curve $d\ell$ which is assumed to be smooth. Here $\ell$ is a curve class in $\mathbb{P}^2$ with  self-intersection number $1$. The genus of the curve $d\ell$ can be computed by the adjunction formula and it gives rise to the number of the adjoint hypermultiplets,
\begin{align}\label{P2_3}
n_{{\mathbf 3}} = \frac{(d-1)(d-2)}{2}.
\end{align}
The curve $\{4\phi g_2 - f_1^2 = 0\}$ is a degree $(24 -2d)$ polynomial and the number of the intersection points $(24-2d) \ell \cap d\ell$ yields the number of the fundamental hypermultipelts,
\begin{align}\label{P2_2}
n_{{\mathbf 2}} = 2d(12-d).
\end{align}
Since $g_2$ is a degree $(18-2d)$ polynomial, $g_2$ becomes a constant when $d=9$. Then we can write $g_2 = g^2$ with a constant $g$ and the elliptic fibration described by the Weierstrass form \eqref{weierstrass} admits another section given by $(x, y)=\left(\frac{1}{3}\phi, wg\right)$ in the local patch where $f$ and $g$ are expressed as \eqref{f_su2} and \eqref{g_su2}. In this case there is an additional $U(1)$ factor. As for another example, consider $B=\mathbb{F}_1$ and an $\mathfrak{su}(2)$ gauge algebra is realized on a smooth curve $d(e+f)$ where $f$ is the fiber curve class and $e$ is the base curve class of $\mathbb{F}_1$. From the genus of the curve $d(e+f)$ the number of the adjoint hypermultiplets is given by
\begin{align}\label{F1_3}
    n_{{\mathbf 3}} = \frac{(d-1)(d-2)}{2}.
\end{align}
The curve class of $\{4\phi g_2 - f_1^2 = 0\}$ is $(16-2d)e+(24-2d)f$ and the number of the intersection points $(16-2d)e+(24-2d)f \cap d(e+f)$ corresponds to the number of the fundamental hypermultipelts,
\begin{align}\label{F1_2}
n_{{\mathbf 2}} = 2d(12-d).
\end{align}

Finally we comment on the realization of the BPS strings in F-theory. The BPS strings in 6d supergravity theories are realized by D3-branes wrapped on two-cycles in $B$ in F-theory compactifications. We can associate the quantities $a, b_i, J$ introduced in section \ref{sec:effective} with divisor classes in $B$. More concretely, we can associate them in the following way \cite{Sadov:1996zm, Kumar:2009ac, Kumar:2010ru},
\begin{align}\label{abj}
a \; \to \; K_B,\qquad b_i \; \to \; C_i, \qquad J \; \to \; j,
\end{align}
where $j$ is a K\"ahler class on $B$ and the inner product with the metric $\Omega^{\alpha\beta}$ is given by the intersection number of the divisor classes in $B$. We can see that the identification \eqref{abj} naturally realizes the conditions \eqref{condition_j}.

\section{Modular bootstrap for BPS strings}
\label{sec:bootstrap}
In this section, we review the modular bootstrap approach for calculating elliptic genera of BPS strings in 6d supersymmetric theories. We will demonstrate with concrete examples that this approach, when combined with the generalized modular ansatz for strings in supergravity theories which we propose below, becomes a very powerful tool for determining the topological string partition functions of elliptic threefolds that allow a toric hypersurface construction.

\subsection{General modular ansatz}
Consider a 6d $\mathcal{N}=(1,0)$ supergravity theory on a torus $T^2$ times $\Omega$-deformed $\mathbb{R}^4$. We can introduce BPS strings that wrap around the $T^2$ and are located at the origin of $\mathbb{R}^4$. The elliptic genera associated with these strings serve as a Witten index counting the BPS spectrum of the 2d worldsheet CFT. For a given tensor charge $Q$, the elliptic genus of the string is defined as
\begin{align}
  Z_Q(\tau,\phi,\epsilon) = {\rm Tr}_{RR}\left[(-1)^F q^{H_L}e^{2\pi i \epsilon J_l} e^{2\pi i \phi \cdot \Pi}\right] \ ,
\end{align}
 where $F$ is the Fermion number, $q=e^{2\pi i \tau}$ is the complex structure of the torus, $H_L$ is the left-moving Hamiltonian in the 2d worldsheet, $J_l$ is the Cartan generator of the $SU(2)_l\subset SO(4)$ Lorentz rotation on $\mathbb{R}^4$ with the $\Omega$-deformation parameter $\epsilon$, and $\phi, \Pi$ collectively denote the holonomies and charges, respectively, for the gauge symmetry $G$. The trace is taken in the RR sector.

Under a modular transformation, the elliptic genus transforms as a weight-zero Jacobi form of elliptic variables $z$ \cite{Benini:2013xpa},
\begin{align}
  Z_Q\left(\frac{a\tau+b}{c\tau+d},\frac{z}{c\tau+d}\right) = {\rm exp}\left(\frac{2\pi i c}{c\tau+d}f(z)\right)Z_Q(\tau,z) \ ,
\end{align}
with $ \smqty(a & b \\ c & d) \in \mathrm{SL}(2, \mathbb{Z}) $, up to a  constant phase factor. Here, the function $f(z)$ is the modular anomaly of the elliptic genus and it is determined by an equivariant integral of the anomaly polynomial $I_4$ of the 2d CFT \cite{DelZotto:2016pvm,Gu:2017ccq},
\begin{align}
	f(z) = \int_{\rm eq} I_4(z) \ .
\end{align}
Here the equivariant integral can be performed by substituting
the characteristic classes appearing in the $I_4$ with chemical potentials for the symmetries \cite{Bobev:2015kza}. Specifically, for the anomaly polynomial $I_4$ in \eqref{eq:anomaly-polynomial} of the worldsheet CFTs on 6d strings, the replacement rules for the characteristic classes are as follows:
\begin{align}\label{eq:replace-rule}
p_1(T_2) \mapsto 0 \, , \qquad c_2(l) \mapsto \epsilon^2 \,, \qquad c_2(R) \mapsto 0 \, , \qquad \frac{1}{2}\Tr F_a^2 \mapsto K_{a,ij} \phi_{a,i} \phi_{a,j} \, ,
\end{align}
where $K_{a,ij}$ denotes the Killing form of the symmetry group $G_a$. Thus, the modular anomaly of the elliptic genus for the BPS strings we are investigating is completely determined by the anomaly polynomial \eqref{eq:anomaly-polynomial}, along with the rule given in \eqref{eq:replace-rule}. 

The modular bootstrap for BPS strings in 6d supergravity theories was initially introduced in \cite{Huang:2015sta} and has been further explored in various references \cite{Haghighat:2015ega,Lee:2018urn,Cota:2020zse}. This idea has also been extended to cover 6d SCFTs and little string theories, and interested readers can find more details in \cite{DelZotto:2016pvm,Gu:2017ccq,Kim:2018gak,DelZotto:2018tcj}. Earlier works in this approach in 6d supergravity theories primarily focused on the strings which do not have an instantonic Higgs branch.  Here, we will give a brief review of these cases before moving on to discuss the generalization that incorporates instantonic strings.

First, the denominator of the elliptic genus can be fixed by the poles arising from the bosonic zero modes on the strings. In cases where the worldsheet theory lacks a Higgs branch associated to the moduli space of instantons of a bulk gauge symmetry, these poles can be fully characterized by the position zero modes along the transverse $\mathbb{R}^4$ directions.  By analyzing how the string can be broken down into its primitive components, we can figure out these poles and thus the factors in the denominator. Here, `primitive strings' refers to those that carry a basis charge in the tensor charge lattice.

Suppose that the string with tensor charge $Q$ can be expressed as a sum of multiples of primitive strings, each carrying a tensor charge $Q_\alpha$, such that $Q = \sum_{\alpha}n_\alpha Q_\alpha$. Then the elliptic genus for this string can be written as \cite{Huang:2015sta} 
\begin{align}\label{eq:ansatz-old}
	Z_Q(\tau,\epsilon,\phi) = \frac{\mathcal{N}_Q(\tau,\epsilon,\phi)}{\eta(\tau)^{2|c_L-c_R|}\prod_\alpha\prod_{s=1}^{n_\alpha}\varphi_{-2,1}(\tau,s\epsilon)} \ .
\end{align}
Here, the power of the Dedekind eta function $\eta(\tau)$ is determined by the vacuum Casimir energy of the worldsheet theory and $\varphi_{w,m}(\tau,z)$ is the weak Jacobi form of weight $w$ and index $m$. The detailed properties of these modular functions are summarized in Appendix~\ref{appendix:modular}.

We then need to determine the numerator $\mathcal{N}_Q$. The numerator is a weak Jacobi form whose weight and indices are determined as follows. Recall that the elliptic genus is a weight-zero Jacobi form and its indices for the global symmetries are fixed by the modular anomaly $f(z)$ or, equivalently, the anomaly polynomial $I_4$. Consequently, the weight and the indices for the symmetries of the numerator factor are fixed by the condition that the full elliptic genus in \eqref{eq:ansatz-old} has the appropriate weight and symmetry indices. A brief computation based on the ansatz \eqref{eq:ansatz-old} shows that the weight $w$ and the indices $m_i$ for $G_i$ of the numerator $\mathcal{N}_Q$ are given by
\begin{align}
	w = |c_L-c_R| -2\sum_\alpha n_\alpha \ , \quad
	m_l = k_l+\frac{1}{6}\sum_\alpha n_\alpha(2n_\alpha+1)(n_\alpha+1) \ , \quad m_i = k_i \ ,
\end{align}
where $k_l$ and $k_i$ are the levels for the global symmetries given in \eqref{eq:levels}.

It is then known that the ring of weak Jacobi forms is finitely generated \cite{Eichler:1995}. Therefore, once the weight and indices are fixed as mentioned earlier, it is possible to write the numerator $\mathcal{N}_Q$ as a finite sum of polynomials of suitable Jacobi forms with constant coefficients. For example, when the gauge group is $U(1)$ or $SU(2)$ with level $k$ and $m_l=0$, which corresponds to a rational primitive curve in geometry intersecting a 7-brane with the gauge group, the numerator takes the form:
\begin{align}\label{eq:numerator-nogauge}
  \mathcal{N}_Q(\tau,z) = \sum_{a_i\in \mathbb{Z}_{\ge0}} c_{a_1,a_2,a_3,a_4}E_4(\tau)^{a_1}E_6(\tau)^{a_2}\varphi_{-2,1}(\tau,z)^{a_3}\varphi_{0,1}(\tau,z)^{a_4} \ ,
\end{align}
with $w = 4a_1 + 6a_2 -2a_3$ and $k = a_3+a_4 $, where $E_4, E_6$ are the Eisenstein series defined in Appendix \ref{appendix:modular}.

The computation of the elliptic genus for the worldsheet CFT on BPS strings now reduces to the task of fixing the finite number of constant coefficients, such as $c_{a_1,a_2,a_3,a_4}$, in the polynomial ring. This task can be quite challenging in general, as it may require knowledge of the spectrum of certain states in the worldsheet theory. Nevertheless, for the theories from F-theory compactifications on Calabi-Yau hypersurfaces $X$ in toric ambient spaces, as discussed in \cite{Huang:2015sta}, the mirror symmetry technique developed in \cite{Hosono:1993qy,Hosono:1994ax} can be used to compute the  genus-zero Gopakumar-Vafa (GV) invariants on $X$, which, in turn, allows us to fix the coefficients. Various intriguing results utilizing this approach can be found, for example, in \cite{Haghighat:2015ega,Lee:2018urn,Cota:2020zse}. In addition, another constraints on the modular ansatz using the affine characters are studied in \cite{DelZotto:2018tcj, Duque:2022tub}.

We will now propose a generalized modular ansatz suitable for strings with Higgs branches describing the moduli space of instantons. As previously discussed, the presence of bosonic zero modes in the instanton moduli space gives rise to additional poles in the elliptic genus. These bosonic zero modes and their contributions to the elliptic genus have been extensively discussed for the elliptic genus of self-dual strings in local 6d SCFTs. See \cite{DelZotto:2016pvm,Gu:2017ccq,Kim:2018gak,DelZotto:2018tcj} for related discussions.
Instantonic strings in 6d supergravity theories have the same bosonic zero modes as the gauge theory instantons. Therefore, the poles in the elliptic genus arising from these zero modes exhibit the same structure. Based on this fact, we conjecture the modular ansatz as follows. 

Consider a string with a tensor charge represented as $Q = \sum_\alpha \kappa_i b_i+Q'$ with $b_i \subset Q$ but $b_i \not\subset Q'$. Here, the positive integer $\kappa_i$ denotes the instanton number for the gauge algebra $ \mathfrak{g}_i $ supported on the charge vector $b_i$, and $\kappa_i=0$ if $b_i\not\subset Q$. We propose the modular ansatz for the elliptic genus of this string as 
\begin{align}\label{eq:ansatz-new}
  Z_Q(\tau,\epsilon,\phi) = \frac{\mathcal{N}_Q(\tau,\epsilon,\phi)}{\eta(\tau)^{2|c_L-c_R|}\prod_\alpha\prod_{s=1}^{n_\alpha}\varphi_{-2,1}(\tau,s\epsilon)\prod_i \mathcal{D}_{\mathfrak{g}_i}(\tau,\epsilon,\phi)} \ .
\end{align}
Compared to the previous ansatz in \eqref{eq:ansatz-old}, this new ansatz involves additional factors in the denominator. Specifically, the factor $\mathcal{D}_{\mathfrak{g}_i}$ accounts for the contributions from bosonic zero modes arising from  $\kappa_i$  instantons of the $\mathfrak{g}_i$ symmetry. 
This factor has been studied for 6d SCFTs as well as little string theories in \cite{DelZotto:2016pvm, DelZotto:2017mee, Kim:2018gak, Kim:2023glm}. Similar to these cases, the factor $ \mathcal{D}_{\mathfrak{g}_i} $ is given by
\begin{align}
    \mathcal{D}_{\mathfrak{g}_i}(\tau,\epsilon,\phi) = \prod_{e \in \mathbf{R}^+_{\mathfrak{g}_i}} \prod_{s=1}^{\lfloor \kappa_i/\xi_e \rfloor} \prod_{l=0}^{s-1} \varphi_{-2,1}(\tau, (s-1-2l)\epsilon + e\cdot\phi) \, ,
\end{align}
where $ \mathbf{R}_{\mathfrak{g}_i}^+ $ is the set of positive roots of $ \mathfrak{g}_i $, $ \lfloor \cdot \rfloor $ is the floor function and
\begin{align}
    \xi_e = \left\{ \begin{array}{ll}
            1 & \quad \text{if $ e $ is a long root or $ \mathfrak{g}_i = A_n,D_n,E_n $ type} \\
            2 & \quad \text{if $ e $ is a short root and $ \mathfrak{g}_i = B_n, C_n, F_4 $ type} \\
            3 & \quad \text{if $ e $ is a short root and $ \mathfrak{g}_i=G_2 $ type.}
        \end{array}\right.
\end{align}
The numerator $ \mathcal{N}_Q $ of the modular ansatz is a Weyl invariant Jacobi form, with its weight and indices determined by the modular properties of the elliptic genus, and it can be expressed as a sum of polynomials of Jacobi forms. In the next subsection, we will test the validity of this ansatz by explicitly calculating the elliptic genera of instantonic strings in 6d supergravity theories that have F-theory realization on a Calabi-Yau hypersurface in a toric ambient space. 

We note that there is a connection between the elliptic genera of BPS strings and GV-invariants of the supergravity theory. One can express the full partition function $Z$ of a 6d supergravity as
\begin{align}\label{eq:GV-condition}
    Z = e^{\mathcal{E}} Z_{\mathrm{pert}} \sum_Q w^Q Z_Q = e^{\mathcal{E}} \PE\left[ \sum_{d,j_l,j_r} (-1)^{2(j_l+j_r)} (2j_r+1) N_{j_l,j_r}^d \frac{-y \chi_{j_l}(y)}{(1-y)^2} e^{2\pi i t \cdot d} \right] \, , 
\end{align}
where $ y\equiv e^{2\pi i \epsilon} $ and $ t $ collectively denotes the K\"ahler parameters. $ N_{j_l,j_r}^d $ is the number of BPS states with charge $ d $ and spin $ (j_l,j_r) $ for the $ SO(4)=SU(2)_l \times SU(2)_r $ rotation symmetry, and $ \chi_j $ is the $ SU(2) $ character of spin $ j $. The exponent $ \mathcal{E} $ denotes the polynomial terms in $t$, which come from the classical action and 1-loop regularization factor, $ Z_{\mathrm{pert}} $ is the BPS spectrum of perturbative states, and $ Z_Q $ is the elliptic genus of strings with tensor charge $Q$. The function $ \PE[f(x)] = \exp(\sum_n \frac{1}{n} f(x^n)) $ is the Plethystic exponential.

The number $ N_{j_l,j_r}^d $ is not really invariant for compact 3-folds. However, when we sum over the $ SU(2)_r $ representations, as in $ \sum_{j_r} (-1)^{2j_r} (2j_r+1) N_{j_l,j_r}^d $, it remains as an invariant in supergravity theories. The genus $ g $ GV-invariant $ \mathrm{GV}_d^{(g)} $ of the BPS state with charge $ d $ is defined as
\begin{align}
    \sum_{j_l,j_r} (-1)^{j_l} (2j_r+1) N_{j_l,j_r}^{d} [j_l] = \sum_g \mathrm{GV}_d^{(g)} \left( 2[0] + \left[\frac{1}{2}\right] \right)^{\otimes g} \, ,
\end{align}
where $ [j] $ is the spin-$ j $ representation. Then the partition function \eqref{eq:GV-condition} can be expressed using the GV-invariants as
\begin{align}\label{eq:GV-condition2}
    Z = e^{\mathcal{E}} \PE\left[ \sum_{d,g} (-1)^{g-1} \mathrm{GV}_d^{(g)} \left(\frac{y-1}{\sqrt{y}}\right)^{2g-2} e^{2\pi i t \cdot d} \right] \, .
\end{align}

The modular properties of elliptic genera on strings do not necessarily imply this expression in terms of GV-invariants. Actually, when attempting to transform the modular ansatz with generic coefficients into this expression, one usually finds higher-order poles at $y=1$ in the exponent of the Plethystic exponential. This is inconsistent with being the BPS spectrum of BPS strings in a 6d theory. As a result, the need for forms like  \eqref{eq:GV-condition} or \eqref{eq:GV-condition2} adds more constraints to the modular ansatz. We will refer to this particular constraint as the \emph{GV-invariant condition}. 

The genus $ g $ free energy $ F^{(g)} $ is defined by the expansion of the partition function in terms of $ \epsilon $ as
\begin{align}
    Z = \exp(\sum_{g=0}^\infty (i(2\pi i \epsilon))^{2g-2} F^{(g)}) \, .
\end{align}
For instance, genus zero and one free energies $ F^{(g=0,1)} $ are given by
\begin{align}
    F^{(g=0)}(t) &= \frac{1}{6} \kappa_{ijk} t_i t_j t_k +  \sum_{d} \mathrm{GV}_d^{(0)} \Li_3(e^{2\pi i t \cdot d}) \, , \label{eq:F0} \\
    F^{(g=1)}(t) &= \frac{t_i}{24} \int c_2 \wedge J_i + \sum_d \left(\frac{\mathrm{GV}_d^{(0)}}{12} + \mathrm{GV}_d^{(1)} \right) \Li_1(e^{2\pi i t \cdot d}) \, , \label{eq:F1}
\end{align}
where $ \Li_s(z) = \sum_k z^k/k^s $ is the polylogarithm. Here, the polynomial terms in $t_i$ come from the exponent $ \mathcal{E} $ of the partition function. Geometrically, $ \kappa_{ijk} $ represents the triple intersection numbers of divisors, $ c_2 $   stands for the second Chern class, and  $ J_i $ denotes the divisor classes of the Calabi-Yau threefold. We will have a closer look at these two free energies in the next subsection.

\subsection{Calabi-Yau hypersurfaces and mirror symmetry}\label{sec:hypersurface}

Typically, the modular properties and the modular ansatz alone are not sufficient to uniquely determine the elliptic genus. Additional information is required to completely fix the unknown coefficients in the modular ansatz.  In this subsection, we will explore the mirror symmetry technique to calculate genus zero and one GV-invariants for a group of Calabi-Yau manifolds known as Calabi-Yau hypersurfaces. As we will show through various examples, these GV-invariants offer the extra information needed to uniquely determine the elliptic genera of strings carrying small tensor charges.

Let us first review the mirror symmetry technique for Calabi-Yau hypersurfaces.
A \emph{lattice polytope} is a convex hull of a finite set of lattice points. Let $ N \cong \mathbb{Z}^n $ be a $ n $-dimensional integral lattice, $ M = N^* = \operatorname{Hom}(N,\mathbb{Z}) $ be its dual lattice. A lattice polytope $ \nabla \subset N $ is called \emph{reflexive} if its polar dual defined by
\begin{align}
    \nabla^* = \{ u \in M_{\mathbb{R}} \mid \langle u, v \rangle \geq -1 \text{ for all } v \in \nabla \}
\end{align}
is also a lattice polytope, where $ M_{\mathbb{R}} = M \otimes \mathbb{R} $ and $ \langle \cdot, \cdot \rangle $ is the Euclidean inner product. If $ \nabla $ is reflexive, its polar dual $ \nabla^* $ is also reflexive since $ (\nabla^*)^* = \nabla $. We will assume that $ \nabla $ contains the origin of $ N $ as an interior point. 

For a given reflexive polytope $ \nabla $, we first associate a \emph{toric fan} $ \Sigma $. A toric fan is a collection of strongly convex rational polyhedral cones\footnote{A strongly convex rational polyhedral cone $ \sigma \subset N_{\mathbb{R}} $ is a set $ \sigma = \{ \sum_{i=1}^k a_i v_k \mid a_i \geq 0\} $ for some vectors $ v_i $ in $ N $ such that $ \sigma \cap (-\sigma) = \{0\} $.} in $ N_{\mathbb{R}} = N \otimes \mathbb{R} $ satisfying (i) each face of a cone in $ \Sigma $ is also a cone in $ \Sigma $ and (ii) intersection of two cones in $ \Sigma $ is a face of each. We construct $ \Sigma $ by taking the 1-dimensional cones (which are called rays) to be the vertices of the reflexive polytope $ \nabla $. We can include more lattice points and subdivide faces of $ \nabla $ to refine the fan $ \Sigma $. The construction of the fan $ \Sigma $ depends on how we refine and triangulate the lattice polytope $ \nabla $. In our purpose, it sufficies to consider points of $ \nabla $ that are not interior to a codimension-1 face which is called a facet, and use specific triangulations called fine, regular, star triangulations \cite{Batyrev:1993oya,Altman:2014bfa}.

Let us denote $ \{v_1, \cdots, v_k\} $ as the set of rays of $ \Sigma $, and associate the homogeneous coordinate $ x_i $ for each ray $ v_i $. Let $ \mathcal{S} $ be a set of rays that do not span a cone of $ \Sigma $, $ V(\mathcal{S}) \subset \mathbb{C}^k $ be the zero set of polynomials $ x_i=0 $ for $ v_i \in \mathcal{S} $, and $ Z(\Sigma) $ be the union of all of the $ V(\mathcal{S}) $. Then $ n $-dimensional \emph{toric variety} from the toric fan is defined by a quotient space
\begin{align}
    X_\Sigma = \frac{\mathbb{C}^k \setminus Z(\Sigma)}{G} \, .
\end{align}
Here, the group $ G $ is the kernel of the map $ \phi : (\mathbb{C}^*)^k \to (\mathbb{C}^*)^n $ given by
\begin{align}
    \phi : (x_1, \cdots, x_k) \mapsto \left(\prod_{j=1}^k x_j^{v_{j,1}}, \cdots, \prod_{j=1}^k x_j^{v_{j,n}}  \right) \, ,
\end{align}
where $ v_{j,l} $ is the $ l $-th component of the ray $ v_j $. The anti-canonical class $ -K_\Sigma $ of the toric variety is given by
\begin{align}
    -K_\Sigma = \sum_{i=1}^k D_i \, ,
\end{align}
where $ D_i = \{x_i=0\} $ is the toric divisor associated to the ray $ v_i $. We refer \cite{Hori:2003ic} for the details of toric geometry.

Lastly, the $ (n-1) $-dimensional Calabi-Yau manifold $ X $, which is defined by the section of the anti-canonical class of the toric variety, is the zero locus of the polynomial
\begin{align}\label{hypersurface_eq}
    p = \sum_{i=1}^{\operatorname{pts}(\nabla^*)} c_i m_i \, , \quad
    m_i = \prod_j x_j^{\langle w_i, v_j \rangle + 1} \, , 
\end{align}
where $ \operatorname{pts}(\nabla^*) $ is the number of lattice points of $ \nabla^* $, $ w_i $ is the lattice point of $ \nabla^* $, $ c_i \in \mathbb{C} $ is a generic coefficient, and $ x_j $ is the homogeneous coordinate associated to the ray $ v_j $ of the toric fan $ \Sigma $ constructed from $ \nabla $. If we exchange the role of $ \nabla $ and $ \nabla^* $, we get a mirror dual $ \widetilde{X} $ of the Calabi-Yau manifold $ X $. This construction of Calabi-Yau manifolds is first done by Batyrev \cite{Batyrev:1993oya}. The family of Calabi-Yau manifolds of this type is called the Calabi-Yau hypersurfaces.

In this paper, we mainly consider 4-dimensional reflexive lattice polytopes which are fully classified in \cite{Kreuzer:2000xy} and related complex 3-dimensional Calabi-Yau hypersurfaces. The majority of these Calabi-Yau manifolds have an elliptic (or genus-1) fibration structure \cite{Huang:2018gpl, Huang:2018esr, Huang:2019pne}, which can be seen from the notion of the \emph{fibered polytopes}. A 4d reflexive polytope $ \nabla $ is called a fibered polytope if it has a 2-dimensional subpolytope $ \nabla_2 \subset \nabla $. If $ \nabla_2 $ itself is a 2d reflexive polytope, then the associated Calabi-Yau hypersurface is an elliptically fibered Calabi-Yau threefold \cite{Kreuzer:1997zg}. In this case, there is a projection map $ \pi : \nabla \to N_B $ such that $ N_B \cong \mathbb{Z}^2 $ and $ \pi^{-1}(0) = \nabla_2 $. This induces projection maps on fans $ \Sigma \to \Sigma_B $ and toric varieties $ X_\Sigma \to B $, where $ B $ is the 2d base toric variety defined by the fan $ \Sigma_B $. The hypersurface equation \eqref{hypersurface_eq} can be transformed to the Weierstrass form \eqref{weierstrass}. One can read gauge symmetry and charged matter contents from the singularities of the elliptic fiber \cite{Braun:2011ux,Huang:2018gpl}. We give some examples in the next subsection.

Many properties of Calabi-Yau manifolds can be extracted from the toric data. The hodge numbers of $ X $ and its mirror dual $ \widetilde{X} $ are given by
\begin{align}
    \begin{aligned}\label{eq:hodge}
    h^{1,1}(X) &= h^{n-2,1}(\widetilde{X}) = \operatorname{pts}(\nabla) - \sum_{\operatorname{codim}\theta=1} \operatorname{int}(\theta) + \sum_{\operatorname{codim}\theta=2} \operatorname{int}(\theta) \operatorname{int}(\theta^*) - 5 \, , \\
    h^{1,1}(\widetilde{X}) &= h^{n-2,1}(X) = \operatorname{pts}(\nabla^*) - \sum_{\operatorname{codim}\tilde{\theta}=1} \operatorname{int}(\tilde{\theta}) + \sum_{\operatorname{codim}\tilde{\theta}=2} \operatorname{int}(\tilde{\theta}) \operatorname{int}(\tilde{\theta}^*) - 5 \, ,
    \end{aligned}
\end{align}
where $ \theta $ and $ \tilde{\theta} $ are faces of $ \nabla $ and $ \nabla^* $, respectively, and $ \operatorname{int}(\cdot) $ denotes the number of interior points of a face. For a given $ m $-dimensional face $ \theta $ of $ \nabla $, its dual face $ \theta^* $ is a $ (n-m-1) $-dimensional face of $ \nabla^* $ defined by $ \{ v \in \nabla^* \mid \langle v, w \rangle = -1,\ \forall\, \text{vertices $ w $ of $ \theta $} \} $. The last sum in \eqref{eq:hodge} for codimension-2 faces is the summation over such dual pairs of faces. Another useful information is triple intersection numbers of divisors. For a given basis $ J_i \in H^{1,1}(X) $ which are Poincar\'e\ dual to the divisors, the triple intersection numbers $ \kappa_{ijk} $ are
\begin{align}
    \kappa_{ijk} = J_i \cdot J_j \cdot J_k \equiv \int_X J_i \wedge J_j \wedge J_k = -\int_{X_\Sigma} J_i \wedge J_j \wedge J_k \wedge K_\Sigma \, ,
\end{align}
where $ -K_\Sigma $ is the anti-canonical class of the ambient toric variety.

In addition, we can find the Mori cone $ \mathcal{M} $ of the variety. For the rays $ v_i $ of the toric fan $ \Sigma $, there are linear relations $ \sum_i \tilde{l}_i v_i = 0 $ for some vectors $ \tilde{l} \in \mathbb{Z}^k $. The cone defined by the vectors $ \tilde{l} $ is called the Mori cone. We denote $ l^\alpha $ as the minimal set of generators of the cone. The systematic way to find $ l^\alpha $ is given in \cite{Oda:1991,Berglund:1995gd,Grimm:2011fx,Kashani-Poor:2019jyo}. Suppose we fix a particular star triangulation of $ \nabla $ which means that we subdivide $ \nabla $ into $ n $-dimensional simplices with apex at the origin. Then, for each pair of the $ n $-dimensional simplicies which have a common $ (n-1) $-dimensional simplex, find a linear relation $ \sum_i m_i v_i = 0 $ among vertices $ v_i $ of the pair of simplices, such that $ m_i $ are minimal integers and $ m_i \geq 0 $ if $ v_i $ is not a common point of the two simplices. The generators $ l^\alpha $ of the Mori cone are the minimal integer vectors that can represent every coefficient vector $ (m_i) $ as a positive linear combination thereof.
In general, $ l^\alpha $ generated by this procedure form the Mori cone of the ambient variety which may be larger than the Mori cone of the Calabi-Yau hypersurface. To fine a better approximation of the Mori cone for the hypersurface, we need to consider other triangulations of the reflexive polytope $ \nabla $. When different triangulations correspond to flop transition of a subvariety of the toric variety which is unrelated to the Calabi-Yau hypersurface, we take the intersection of the Mori cones associated with the ambient varieties from the distinct triangulations. This procedure provides a refined approximation $ \mathcal{M}_{\cap} $ for the Mori cone $ \mathcal{M} $ of the Calabi-Yau hypersurface. If all generators of $ \mathcal{M}_\cap $ has non-zero GV-invariants, then $ \mathcal{M}_\cap = \mathcal{M} $. 

In a computational side, there are several useful tools for working with reflexive polytopes, triangulations and toric geometries such as \textsf{PALP}, \textsf{TOPCOM} and \textsf{SINGULAR} \cite{Kreuzer:2002uu,TOPCOM,singular} which are currently implemented in \textsf{sage} \cite{sagemath}. Additionally, a recently developed package \textsf{cytools} \cite{Demirtas:2022hqf} has more efficient algorithm to handle reflexive polytopes for large $ h^{1,1} $. Furthermore, it is beneficial to utilize pre-computed database available in \cite{CYdata, Altman:2014bfa}.

The toric construction of the Calabi-Yau hypersurfaces allows us to compute their genus zero GV-invariants \cite{Hosono:1993qy,Hosono:1994ax}. Let us focus on the mirror pair $ (X, \widetilde{X}) $ of Calabi-Yau threefold, and introduce an integral basis $ (\alpha_i, \beta^i) $ of the cohomology group $ H^3(\widetilde{X}, \mathbb{Z}) $ and its dual basis $ (A^i, B_i) $ of $ H_3(\widetilde{X}, \mathbb{Z}) $, where $ i=0,1,\cdots,h^{2,1}(\tilde{X}) $ such that intersection numbers are given by $ A^i \cdot B_j = \delta^i_j $, $ A^i \cdot A_j = B^i \cdot B_j = 0 $ and
\begin{align}
    \int_{\tilde{X}} \alpha_j \wedge \beta^i = \int_{A^i} \alpha_j = -\int_{B_j} \beta^i = \delta^{i}_j \, ,
\end{align}
with the other integrals vanishing. Let $ \Omega(z) $ be the unique nowhere vanishing $ (3,0) $ form which depends on the complex structure moduli $ z = (z_1, \cdots, z_{h^{2,1}(\tilde{X})}) $. Using the above integral basis, $ \Omega $ can be written as $ \Omega = X^i \alpha_i - \mathcal{F}_i \beta^i $, where each entry of
\begin{align}
    \Pi(z) = \mqty( \int_{A^i} \Omega \\ \int_{B_i} \Omega) = \mqty( X^i(z) \\ \mathcal{F}_i(z) )
\end{align}
is called the period of $ \Omega $. The periods $ X^i $ are regarded as the homogeneous coordinates of the complex structure moduli space and $ \mathcal{F}_i $ are the derivatives of the function as $ \mathcal{F}_i = \partial \mathcal{F} / \partial X^i $, where $ \mathcal{F} = \frac{1}{2} X^i \mathcal{F}_i $ is homogeneous of degree 2 in $ X^i $ \cite{Strominger:1990pd}. The function $ \mathcal{F} $ is called the \emph{prepotential}, and it encodes information about genus zero GV-invariants.

The periods are solutions to a set of differential equations called Picard-Fuchs equations given by\footnote{More precisely, $ \mathcal{D}_\alpha $ defines the Gel'fand-Kapranov-Zelevinsky (GKZ) hypergeometric system which may have more than $ 2h^{2,1}(\widetilde{X})+2 $ solutions. One can obtain a Picard-Fuchs system which has $ 2h^{2,1}(\widetilde{X})+2 $ independent solutions by factoring the operators and providing additional data as \cite{Hosono:1993qy}.} \cite{Gelfand:1989:hypergeometric,Gelfand:1990:Generalized}
\begin{align}
    \mathcal{D}_\alpha \qty(\frac{\Pi(z)}{a_0}) = 0 \, , \quad
    \mathcal{D}_\alpha = \prod_{l^\alpha_i > 0} \qty(\frac{\partial}{\partial a_i})^{l^\alpha_i} - \prod_{l^\alpha_i < 0} \qty(\frac{\partial}{\partial a_i})^{-l^\alpha_i} \, .
\end{align}
These equations only depend on $ z_\alpha = \prod_{i\geq 0} a_i^{l_i^\alpha} $ for the Mori cone vectors $ l^\alpha $. Here, the 0-th component of the Mori cone vector is defined as $ l^\alpha_0 = -\sum_{i>0} l^\alpha_i $. The solutions to the differential equations near $ z_\alpha=0 $ is well-studied using the theory of Frobenius \cite{Hosono:1993qy, Hosono:1994ax}. Let $ n $ be the lattice points of the Mori cone $ \mathcal{M} $.\footnote{In practice, we can use the approximation $ \mathcal{M}_\cap $ instead of the true Mori cone $ \mathcal{M} $. The Mori cones written in section \ref{sec:examples} are the approximation $\mathcal{M}_{\cap}$.} If $ \mathcal{M} $ is simplicial, meaning that it is generated by linearly independent vectors $ l^\alpha $, then $ n \in \mathbb{Z}_{\geq 0}^{\dim(\mathcal{M})} $. Otherwise, $ \mathcal{M} $ is called non-simplicial and some components of $ n $ can be negative. For a function $ \varpi_0 $ defined by
\begin{align}\label{eq:fundamental-period}
    \varpi_0(z, \rho) = \sum_n c(n+\rho) \prod_\alpha z_\alpha^{n_\alpha+\rho_\alpha} \, , \quad
    c(n) = \frac{(-\sum_\alpha l^\alpha_0 n_\alpha)!}{\prod_{i>0} (\sum_\alpha l^\alpha_i n_\alpha)!} \, ,
\end{align}
the periods which are solutions to Picard-Fuchs equations are
\begin{align}\label{eq:Period-series}
    \begin{aligned}
    X^0(z) &= \varpi_0(z, 0) \, , \quad
    X^\alpha(z) = \frac{1}{2\pi i} \left. \frac{\partial}{\partial \rho_\alpha} \varpi_0(z, \rho) \right|_{\rho=0} \, , \\
    \mathcal{F}_0(z) &= -\frac{1}{3!(2\pi i)^3} \sum_{\alpha,\beta,\gamma} \kappa_{\alpha\beta\gamma} \left. \frac{\partial^3 \varpi_0(z,\rho)}{\partial \rho_\alpha \partial \rho_\beta \partial \rho_\gamma} \right|_{\rho=0} \, , \\
    \mathcal{F}_\alpha(z) &= \frac{1}{2(2\pi i)^2} \sum_{\beta,\gamma} \kappa_{\alpha\beta\gamma} \left. \frac{\partial^2 \varpi_0(z,\rho)}{\partial \rho_\beta \partial \rho_\gamma} \right|_{\rho=0} \, ,
    \end{aligned}
\end{align}
where $ \kappa_{\alpha\beta\gamma} $ is the triple intersection numbers of divisors dual to Mori cone generators.

The mirror symmetry identifies the prepotential of two manifolds $ X $ and its mirror $ \widetilde{X} $ up to exchanging the role of the K\"ahler moduli and complex structure moduli. The map from complex structure moduli to K\"ahler moduli is given by
\begin{align}\label{eq:mirror}
    t_j = \frac{X^j(z)}{X^0(z)} = \frac{1}{2\pi i} \log z_j + \mathcal{O}(z) \, ,
\end{align}
which is called the \emph{mirror map}. After parametrizing $ \mathcal{F} = (X^0)^2 F(t) $ using the homogeneity of $ \mathcal{F} $, the vector $ \Pi $ can be written as
\begin{align}
    \Pi = \mqty( X^0 \\ X^i \\ \mathcal{F}_0 \\ \mathcal{F}_i) = X^0 \mqty( 1 \\ t_i \\ 2F - t_i \partial F / \partial t_i \\ \partial F/\partial t_i ) \, ,
\end{align}
where $ F(t) $ is identified as the genus zero free energy $ F^{(0)} $ in \eqref{eq:F0} that encodes genus zero GV-invariants. In the mirror Calabi-Yau side, the Yukawa coupling defined by
\begin{align}\label{eq:Yukawa}
    \widetilde{C}_{ijk}(z) = \int \Omega \wedge \partial_{z_i} \partial_{z_j} \partial_{z_k} \Omega = \sum_{l=0}^{h^{2,1}(\widetilde{X})+1}  \left( \mathcal{F}_l \partial_{z_i} \partial_{z_j} \partial_{z_k} X^l - X^l \partial_{z_i} \partial_{z_j} \partial_{z_k} \mathcal{F}_l \right)
\end{align}
can be written as the ratio of polynomials on $ z $:
\begin{align}\label{eq:Yukawa-rational}
    \widetilde{C}_{ijk} = \frac{\kappa_{ijk} + \mathcal{O}(z)}{z_i z_j z_k \prod_a \Delta_a(z)} \, .
\end{align}
Here, the polynomials $ \Delta_a(z) $ are called the discriminants of the Picard-Fuchs operators, and the numerator of $ \widetilde{C}_{ijk} $ is also a finite polynomial on $ z $. The Yukawa coupling is related with the genus zero GV-invariants by
\begin{align}
    C_{ijk}(t) = \frac{\partial^3 F^{(0)}(t)}{\partial t_i \partial t_j \partial t_k} = \frac{1}{(X^0)^2} \frac{\partial z_a}{\partial t_i} \frac{\partial z_b}{\partial t_j} \frac{\partial z_c}{\partial t_k} \widetilde{C}_{abc}(z(t)) = \kappa_{ijk} + \sum_{d} \frac{\mathrm{GV}_d^{(0)} d_i d_j d_k}{1- e^{2\pi i t \cdot d}} \, .
\end{align}
This computation method of genus zero GV-invariants is extensively studied in \cite{Hosono:1993qy} and generalized into non-simplicial Mori cones and large hodge numbers in \cite{Demirtas:2023als}.

The higher genus free energies $ F^{(g)} $ satisfy the holomorphic anomaly equations \cite{Bershadsky:1993ta, Bershadsky:1993cx} which are the recursive differential equations connecting $ F^{(g)} $ and $ F^{(g'<g)} $. However, it is in general hard to extract higher genus GV-invariants by solving it. The major problem of the solving holomorphic anomaly equation is fixing the integration constants called holomorphic ambiguities. For a large $ g $, we do not know enough constraints for $ F^{(g)} $ that fix all holomorphic ambiguities. But when $ g=1 $, the general solution is known and given by
\begin{align}\label{eq:genus1}
    F^{(g=1)} = -\frac{1}{2}\left(3 + h^{1,1} - \frac{\chi}{12}\right) K - \frac{1}{2} \log\det G_{i\bar{j}}^{-1} + \frac{1}{24} \sum_i s_i \log z_i + \frac{1}{12} \sum_a \log\Delta_a \, .
\end{align}
Here, $ \chi $ is the Euler characteristic, $ s_i $ are some constants, $ \Delta_a $ are the discriminants of the Picard-Fuchs operators, $ K $ is the K\"ahler potential of the manifold given by
\begin{align}
    e^{-K} = i \int_{\tilde{X}} \Omega \wedge \Omega = i \Pi^\dagger \eta \Pi \, , \quad \eta = \mqty( 0 & \mathbf{I} \\ -\mathbf{I} & 0 ) \, ,
\end{align}
and $ G_{i\bar{j}} = \partial_i \partial_{\bar{j}} K $ is the K\"ahler metric. To extract the genus one GV-invariants, we use the mirror map to express $ z_i $ in terms of $ t_i $, and take the holomorphic limit $ \bar{t}_j \to i\infty $ while keeping $ t_j $ fixed. In this limit, the K\"ahler potential and K\"ahler metric become 
\begin{align}
    K \to -\log X^0 \, , \quad
    \det G_{i\bar{j}}^{-1} \to \det(\frac{\partial t_i}{\partial z_j}) \, ,
\end{align}
and we can fix $ s_i $ by using the polynomial part of the genus one free energy \eqref{eq:F1} determined by the second Chern class.

\subsection{Examples}
\label{sec:examples}

We now provide examples of GV-invariant calculations and the modular bootstrap for elliptic genera of strings in various 6d supergravity theories arising from F-theory compactifications on elliptically fibered Calabi-Yau hypersurfaces.

\subsubsection{\texorpdfstring{$ B=\mathbb{P}^2 $}{B=P2} without gauge symmetry} \label{subsubsec:ellP2}

The first example is the 6d $(1,0) $ supergravity theory obtained by the F-theory compactified on elliptic 3-folds over the complex projective space $ \mathbb{P}^2 $ serving as its base. The anomaly cancellation condition \eqref{eq:6dconstraints} is satisfied with
\begin{align}
    a = -3 \, , \quad
    b = 0 \, , \quad
    \Omega = 1 \, .
\end{align}
This theory has no tensor multiplet, $ T=0 $, and there are 273 neutral hypermultiplets $ H=n_{\mathbf{1}} = 273 $. The geometry and mirror symmetry of this theory are studied in \cite{Candelas:1994hw, Hosono:1993qy} and the modular bootstrap for elliptic genera of strings is considered in \cite{Huang:2015sta}, which we briefly review from now on.

Let us consider the Calabi-Yau hypersurface whose defining 4d reflexive polytope $ \nabla $ and two Mori cone generators $ l^{\alpha=1,2} $ are given as follows:
\begin{align}\label{eq:ellP2-toric}
    \begin{array}{c|rrrr|rr}
        & \multicolumn{4}{c|}{\nabla} & l^1 & l^2 \\ \hline
        y & 0 & 0 & 0 & -1 & 3 & 0 \\ 
        x & 0 & 0 & -1 & 0 & 2 & 0 \\
        b_1 & 1 & 0 & 2 & 3 & 0 & 1 \\
        b_2 & 0 & 1 & 2 & 3 & 0 & 1 \\
        b_3 & -1 & -1 & 2 & 3 & 0 & 1 \\
        z & 0 & 0 & 2 & 3 & 1 & -3
    \end{array}
\end{align}
The rays of $ \nabla $ in \eqref{eq:ellP2-toric} correspond to the toric divisors and we assign the homogeneous coordinates $ \{y,x,b_{1,2,3},z\} $ to the rays. The vectors $ l^\alpha $ are associated with the Mori cone curves $ C_\alpha $, and each entry of $ l^\alpha $ represents the intersection number between $ C_\alpha $ and each toric divisor. We consider the Calabi-Yau hypersurface defined by \eqref{eq:ellP2-toric} in following reasons. First, $ \nabla $ is a fibered polytope that contains 2d reflexive subpolytope $ \nabla_2 $ with vertices $ \{(-1,0),(0,-1),(2,3)\} $, and thus \eqref{eq:ellP2-toric} defines an elliptically fibered Calabi-Yau threefold. Second, there is a projection map $ \pi $ of $ \nabla $ onto its first two coordinates satisfying $ \nabla_2=\pi^{-1}(0) $, which defines a base fan $ \Sigma_B $ consisting of three rays $ \{(1,0),(0,1),(-1,-1)\} $. This suggests that the base surface $ B $ is the complex projective space $ \mathbb{P}^2 $. Third, its hodge numbers and Euler characteristics are $ (h^{1,1}, h^{2,1}, \chi) = (2,272,-540) $, which are expected from \eqref{num_singlet} and \eqref{tensor,rank}. Lastly, the hypersurface equation \eqref{hypersurface_eq} is factorized into the Tate form
\begin{align}\label{Tate}
    y^2 + a_1 xyz + a_3 yz^3 = x^3 + a_2 x^2z^2 + a_4 xz^4  + a_6 z^6 \, ,
\end{align}
where $ a_i $ are functions of the base coordinates $ b_{1,2,3} $. The discriminant and vanishing orders of $ a_i $ in the Tate form suggests that there is no non-trivial gauge symmetry in this model \cite{Katz:2011qp}. Thus, the Calabi-Yau hypersurface constructed from the toric data \eqref{eq:ellP2-toric} is the model we are after.

We can compute genus zero and one GV-invariants from the toric data \eqref{eq:ellP2-toric}. The triple intersection numbers $ \kappa_{\alpha\beta\gamma} = J_\alpha \cdot J_\beta \cdot J_\gamma $ of the divisors $ J_\alpha $ dual to $ C_\beta $ (i.e., $ J_\alpha $ intersecting with $ C_\alpha $ at one point and not intersecting with $ C_{\beta \neq \alpha} $) are
\begin{align}
    \kappa_{111} = 9 \, , \quad
    \kappa_{112} = 3 \, , \quad
    \kappa_{122} = 1 \, , \quad
    \kappa_{222} = 0 \, .
\end{align}
Using the solutions \eqref{eq:Period-series} to the Picard-Fuchs equations and the mirror map \eqref{eq:mirror} from the Mori cone vectors $ l^\alpha $ and triple intersection numbers $ \kappa_{\alpha\beta\gamma} $, we can compute the genus zero free energy and the genus zero GV-invariants. We summarize the results in Table~\ref{table:ellP2}, where $ (n_1,n_2) $ in the table labels lattice point $ l^1 n_1 + l^2 n_2 $ of the Mori cone. The Yukawa couplings \eqref{eq:Yukawa} is given by
\begin{alignat}{2}
    \begin{aligned}
        &\widetilde{C}_{111} = \frac{9}{z_1^3 \Delta_1} \, , \quad
        &&\widetilde{C}_{112} = \frac{3(1-432z_1)}{z_1^2 z_2 \Delta_1} \, , \\
        &\widetilde{C}_{122} = \frac{(1-432z_1)^2}{z_1 z_2^2 \Delta_1} \, , \quad
        &&\widetilde{C}_{222} = \frac{9(1-3(432z_1)+3(432z_1)^2)}{z_2^2 \Delta_1 \Delta_2} \, ,
    \end{aligned}
\end{alignat}
where
\begin{align}
    \Delta_1 = (1-432z_1)^3 - 27z_2(432z_1)^3 \, , \quad
    \Delta_2 = 1+27z_2
\end{align}
are discriminants of the Picard-Fuchs operators. Using the discriminants $ \Delta_1 $ and $ \Delta_2 $ together with the divisor integrals of the second Chern class
\begin{align}
    \int c_2 \wedge J_1 = 102 \, , \quad
    \int c_2 \wedge J_2 = 36 \, ,
\end{align}
we can extract genus one GV-invariants via \eqref{eq:genus1} and \eqref{eq:F1}. We summarize the result in Table~\ref{table:ellP2g1}.
\begin{table}
    \centering
    \footnotesize
    \begin{tabular}{c|cccc}
        $ n_1 \setminus n_2 $ & $ 0 $ & $ 1 $ & $ 2 $ & $ 3 $ \\ \hline
        $ 0 $ &  & $ 3 $ & $ -6 $ & $ 27 $ \\
        $ 1 $ & $ 540 $ & $ -1080 $ & $ 2700 $ & $ -17280 $  \\
        $ 2 $ & $ 540 $ & $ 143370 $ & $ -574560 $ & $ 5051970 $ \\
        $ 3 $ & $ 540 $ & $ 204071184 $ & $ 74810520 $ & $ -913383000 $ \\
        $ 4 $ & $ 540 $ & $ 21772947555 $ & $ -49933059660 $ & $ 224108858700 $
    \end{tabular}
    \caption{Genus zero GV invariants of $ B=\mathbb{P}^2 $ with elliptic fibration} \label{table:ellP2}
\end{table}

\begin{table}
    \centering
    \footnotesize
    \begin{tabular}{c|cccc}
        $ n_1 \setminus n_2 $ & $ 0 $ & $ 1 $ & $ 2 $ & $ 3 $ \\ \hline
        $ 0 $ & & & & $ -10 $ \\
        $ 1 $ & $ 3 $ & $ -6 $ & $ 15 $ & $ 4764 $ \\
        $ 2 $ & $ 3 $ & $ 2142 $ & $ -8574 $ & $ -1079298 $ \\
        $ 3 $ & $ 3 $ & $ -280284 $ & $ 2126358 $ & $ 152278986 $ \\
        $ 4 $ & $ 3 $ & $ -408993990 $ & $ 521856996 $ & $ -16704086880 $
    \end{tabular}
    \caption{Genus one GV invariants of $ B=\mathbb{P}^2 $ with elliptic fibration} \label{table:ellP2g1}
\end{table}

We now move on to the partition function which contains higher genus GV-invariants. The 1-loop perturbative contribution from a gravity multiplet and 273 neutral hypermultiplets is given by
\begin{align}
    Z_{\mathrm{pert}} = \PE\left[\frac{3y(y+y^{-1})}{(1-y)^2} \frac{1}{1-q} - \frac{273y}{(1-y)^2} \frac{1+q}{1-q}\right] \, .
\end{align}
By comparing $ Z_{\mathrm{pert}} $ and GV-invariants given in Table~\ref{table:ellP2} and \ref{table:ellP2g1}, we can identify the Mori cone vectors: $ l^1 $ is associated with the K\"ahler parameter $ \tau $, and $ l^2 $ is related with the curve class inside the base $ \mathbb{P}^2 $. 

Next, the anomaly polynomial \eqref{eq:anomaly-polynomial} of the 2d CFT on a string with charge $n$ is
\begin{align}
    I_4 = \frac{3n}{4} p_1(T_2) - \frac{n(n+3)}{2} c_2(R) + \frac{n(n-3)}{2} c_2(l) \, .
\end{align} 
The central charges of this string are
\begin{align}
    c_L = 3n^2 + 27n + 6 \, , \quad
    c_R = 3n^2 + 9n + 6 \, , \quad
    k_l = \frac{n(n-3)}{2}
\end{align}
from \eqref{eq:central-charge} and \eqref{eq:levels}. We can then set a modular ansatz for the elliptic genus of this string as 
\begin{align}\label{eq:ellP2-ansatz}
    Z_n(\tau, \epsilon) = \frac{\mathcal{N}_n(\tau, \epsilon)}{\eta(\tau)^{36n} \prod_{s=1}^n \varphi_2(\tau,s\epsilon)} \ .
\end{align}
Here, we use a shorthand notation $ \varphi_2 \equiv \varphi_{-2,1} $ and $ \varphi_0 \equiv \varphi_{0,1} $ for the Jacobi forms. The numerator $ \mathcal{N}_n(\tau,\epsilon) $ is a weak Jacobi form of weight $ 16n $ and index $n(n-1)(n+4)/3$, and takes the form \eqref{eq:numerator-nogauge}. This property fixes the numerator up to a finite number of unknown coefficients. For instance, there are 2 and 17 unknown constants in the 1-string and 2-string modular ansatz.

The requirement that the 2-string elliptic genus must be factorized into the GV-invariant form determines 14 coefficients out of 17. Moreover, the condition that a decompactification of the elliptic fiber leads to a local Calabi-Yau threefold $ \mathcal{O}(-3) \to \mathbb{P}^2 $, which is often called the local $ \mathbb{P}^2 $, gives rise to additional constraints. M-theory compactified on this local Calabi-Yau threefold after the decompactification defines a 5d rank 1 SCFT. The GV-invariants of this 5d SCFT can be computed using various methods such as local mirror symmetry \cite{Chiang:1999tz, Klemm:1999gm}, topological vertex \cite{Aganagic:2003db, Iqbal:2012mt} and blowup equation \cite{Huang:2017mis, Kim:2020hhh}. The decompactification limit corresponds to $ \tau \to i\infty $ limit in the partition function. In this limit, the elliptic genera should be expanded, for example, as 
\begin{align}
        Z_1 = -\frac{3y}{(1-y)^2} \frac{1}{q^{3/2}} + \mathcal{O}(q^{-1/2}) \, , \quad
        Z_2 = \frac{6y}{(1-y)^2} \frac{1}{q^3} + \mathcal{O}(q^{-2}) \, .
\end{align}
Using this, we can determine one coefficient in both $ \mathcal{N}_1 $ and $ \mathcal{N}_2 $. 
The remaining unknown coefficients are determined by the genus zero GV-invariants given in Table~\ref{table:ellP2} which we compute using the mirror technique. 

As a result, we find
\begin{align}
    \mathcal{N}_1 = -\frac{31}{48} E_4^4 - \frac{113}{48} E_4 E_6^2 \, ,
\end{align}
and
\begin{align}
    \mathcal{N}_2 &= \frac{31 E_4^{10} \varphi_2(\epsilon)^4}{1146617856}+\frac{1995541 E_4^7 E_6^2 \varphi_2(\epsilon)^4}{382205952}+\frac{246421 E_4^4 E_6^4 \varphi_2(\epsilon)^4}{14155776}+\frac{4908413 E_4 E_6^6 \varphi_2(\epsilon)^4}{1146617856} \nonumber \\
    &\quad -\frac{250129 E_4^8 E_6 \varphi_2(\epsilon)^3 \varphi_0(\epsilon)}{31850496} -\frac{420143 E_4^5 E_6^3 \varphi_2(\epsilon)^3 \varphi_0(\epsilon)}{15925248}-\frac{181393 E_4^2 E_6^5 \varphi_2(\epsilon)^3 \varphi_0(\epsilon)}{31850496} \nonumber \\
    &\quad +\frac{36469 E_4^9 \varphi_2(\epsilon)^2 \varphi_0(\epsilon)^2}{1146617856}+\frac{28319 E_4^6 E_6^2 \varphi_2(\epsilon)^2 \varphi_0(\epsilon)^2}{42467328} -\frac{274339 E_4^3 E_6^4 \varphi_2(\epsilon)^2 \varphi_0(\epsilon)^2}{382205952} \nonumber \\
    &\quad +\frac{21935 E_6^6 \varphi_2(\epsilon)^2 \varphi_0(\epsilon)^2}{1146617856}+\frac{208991 E_4^7 E_6 \varphi_2(\epsilon) \varphi_0(\epsilon)^3}{95551488}+\frac{377953 E_4^4 E_6^3 \varphi_2(\epsilon) \varphi_0(\epsilon)^3}{47775744} \nonumber \\
    &\quad +\frac{196319 E_4 E_6^5 \varphi_2(\epsilon) \varphi_0(\epsilon)^3}{95551488}+\frac{961 E_4^8 \varphi_0(\epsilon)^4}{23887872}+\frac{3503 E_4^5 E_6^2 \varphi_0(\epsilon)^4}{11943936}+\frac{12769 E_4^2 E_6^4 \varphi_0(\epsilon)^4}{23887872} ,
\end{align}
for the 1-string and 2-string elliptic genera.


\subsubsection{\texorpdfstring{$ B=\mathbb{P}^2 $ with $ SU(2) $}{B=P2 with SU(2)} gauge symmetry}

Our next examples are the 6d theories arising from elliptic CY 3-folds over a base $ B=\mathbb{P}^2 $ case with $ SU(2) $ gauge symmetry supported on a degree $d$ curve $ d \cdot \ell $, where $ \ell $ is the curve class of $ \mathbb{P}^2 $ with self-intersection number $ \ell^2=1 $. When the theory has no hypermultiplet whose representation under the $ SU(2) $ gauge symmetry is larger than the adjoint representation $ \mathbf{3} $, the anomaly cancellation condition \eqref{eq:6dconstraints} is satisfied with $ T=0 $ and
\begin{align}
    a = -3 \, , \quad
    b = d \, , \quad
    \Omega = 1 \, , \quad
    n_{\mathbf{3}} = \frac{(d-1)(d-2)}{2} \, , \quad
    n_{\mathbf{2}} = 2d(12-d) \, .
\end{align}
Note that $n_{\mathbf{3}}, n_{\mathbf{2}}$ agree with \eqref{P2_3} and \eqref{P2_2} respectively. 
Since the number of fundamental hypermultiplets cannot be negative, the degree $ d $ is bounded by $ d \leq 12 $ \cite{Kumar:2010am}. Here we consider $ d \leq 8 $ cases which are described by the following toric data:
\begin{align}\label{eq:ellP2-gauge-toric}
    \begin{array}{rrrr|rrr}
        \multicolumn{4}{c|}{\nabla} & l^1 & l^2 & l^3 \\ \hline
        0 & 0 & 0 & -1 & 0 & 1 & 0 \\ 
        0 & 0 & -1 & 0 & 1 & 0 & 0 \\
        1 & 0 & 1 & 2 & 0 & 0 & 1 \\
        0 & 1 & 1 & 2 & 0 & 0 & 1 \\
        -1 & -1 & 4-d & 5-d & 0 & 0 & 1 \\
        0 & 0 & 2 & 3 & -1 & 1 & -3 \\
        0 & 0 & 1 & 1 & 3 & -2 & d
    \end{array}
\end{align}
Here, $ l^{\alpha=1,2,3} $ are three Mori cone generators. We note that the Mori-cone generators need to be changed when $ d=7 $ and $ d=8 $, which will be discussed in more detail later in this section. 

The hodge numbers of the Calabi-Yau hypersurface are $ (h^{1,1}, h^{2,1}) = (3, 3d^2-45d+273) $. The reflexive polytope $ \nabla $ is a fibered polytope with a base surface $ \mathbb{P}^2 $ similar to the model considered in section~\ref{subsubsec:ellP2}. Let us label the fiber coordinates $ x $ and $ y $ for the divisors associated to the rays $ (0,0,-1,0) $ and $ (0,0,0,-1) $, respectively, and base coordinates $ b_1,b_2,b_3 $ related to the rays $ (1,0,1,2) $, $ (0,1,1,2) $ and $ (-1,-1,4-d,5-d) $.
The hypersurface equation \eqref{hypersurface_eq} is not factorized into the standard Tate form \eqref{Tate} in this case, but the coefficient of $ x^3 $ term depends on the base coordinates. Nevertheless, it still defines a genus-1 curve and can be brought into the Weierstrass form as \cite{Braun:2011ux}. For an elliptic curve of the form
\begin{align}\label{elliptic curve quartic}
    \sum_{i=0}^4 \sum_{j=0}^2 r_{ij} x^i y^j =0 \quad \text{where } r_{ij}=0 \text{ if } j>\frac{4-i}{2} \, ,
\end{align}
and its discriminant
\begin{align}
    \sum_{i=0}^4 s_i x^i = \left( \sum_{i=0}^2 r_{i1} x^i \right)^2 - 4 r_{02} \sum_{i=0}^4 r_{i0} x^i \, ,
\end{align}
with respect to $ y $, the associated Weierstrass form $ y^2 = x^3 + fx + g $ is given by
\begin{align}
    f = -\frac{1}{4} (s_0 s_4 + 3s_2^2 - 4s_1s_3) \, , \quad
    g = -\frac{1}{4} (s_0 s_3^2 + s_1^2 s_4 - s_0 s_2 s_4 - 2s_1 s_2 s_3 + s_2^3) \, .
\end{align}
The discriminant of the Weierstrass form computed from the anti-canonical hypersurface of the toric data \eqref{eq:ellP2-gauge-toric} is
\begin{align}
    \Delta = 4f^3 + 27g^2 = \Delta_0\Delta_1^2  \, , \quad
    \Delta_1 = \sum_{i=0}^d \sum_{j=0}^{d-i} c_{ij} b_1^i b_2^j b_3^{d-i-j} \, ,
\end{align}
where $ \Delta_0 $ is a degree $ 36-2d $ polynomial in the base coordinates. The vanishing order of $ \{f, g, \Delta\} $ are $ \{0,0,2\} $ on the curve $  \{ \Delta_1 = 0 \} $, implying that there is a singular elliptic fiber of Kodaira type $ I_2 $ along the degree $ d $ curve. The toric data \eqref{eq:ellP2-gauge-toric} therefore describes the models we are seeking.

We now analyze the GV-invariants and the elliptic genera of strings in these models. The perturbative partition function is given by
\begin{align}
    \begin{aligned}
        Z_{\mathrm{pert}} &= \PE\bigg[\frac{3y(y+y^{-1})}{(1-y)^2} \frac{1}{1-q} + \frac{2y}{(1-y)^2} \Big(e^{4\pi i \phi} + 1 + q e^{-4\pi i \phi}\Big) \frac{1}{1-q} \\
        &\qquad - \frac{y}{(1-y)^2} \Big( n_{\mathbf{3}} \big(e^{4\pi i \phi}+1+e^{-4\pi i \phi}\big) + n_{\mathbf{2}} \big(e^{2\pi i \phi}+e^{-2\pi i \phi}\big) + n_{\mathbf{1}} \Big) \frac{1+q}{1-q}\bigg] \, ,
    \end{aligned}
\end{align}
where the first line comes from the gravity and $ SU(2) $ vector multiplet contributions, and the second line is the contribution from the hypermultiplets. $ n_{\mathbf{1}} $ is the number of neutral hypermultiplets determined by the anomaly cancellation condition \eqref{eq:6dconstraints}. The central charges and levels for the $ SU(2)_l $ symmetry and $ SU(2) $ gauge symmetry for the string of charge $ Q=n $ are
\begin{align}\label{eq:P2-su2-central}
    c_L = 3n^2 + 27n + 6 \, , \quad
    c_R = 3n^2 + 9n + 6 \, , \quad
    k_l = \frac{n(n-3)}{2} \, , \quad
    k_{SU(2)} = nd \, .
\end{align}
This determines the modular property of the elliptic genus. We write the modular ansatz as
\begin{align}\label{eq:ellP2-gauge-ansatz}
    Z_n(\tau, \epsilon, \phi) = \frac{1}{\eta(\tau)^{36n}} \frac{\mathcal{N}_n(\tau, \epsilon, \phi)}{\prod_{s=1}^n \varphi_2(\tau, s\epsilon) \cdot \prod_{s=1}^\kappa \prod_{l=0}^{s-1} \varphi_2(\tau, (s-1-2l)\epsilon+2\phi)} \, ,
\end{align}
where $ \kappa = \lfloor n/d \rfloor $, $ \phi $ is the $ SU(2) $ gauge holonomy, and we include the boson zero mode contributions from the $SU(2)$ instantons in the denominator. 

We remark some general properties of the elliptic genera \eqref{eq:ellP2-gauge-ansatz}. First, they must be factorized into GV-invariant form \eqref{eq:GV-condition2}. Second, since the decompactification limit of the elliptic fiber yields a local CY 3-fold embedding a $ \mathbb{P}^2 $, the elliptic genera should reproduce the GV-invariants of the local CY3 in the associated limit. Lastly, these theories can be higgsed to the theory without gauge symmetry considered in section~\ref{subsubsec:ellP2}. Thus the limit $ \phi \to 0 $ of \eqref{eq:ellP2-gauge-ansatz}, which corresponds to the Higgsing at the level of elliptic genera, should reduce to the elliptic genera given in \eqref{eq:ellP2-ansatz}. See \cite{Duan:2020imo} for the related discussions on Higgsing, elliptic genera and Jacobi forms in 6d SCFTs. We shall use these conditions and genus zero GV-invariants computed from the toric data to fix the unknown coefficients in the modular ansatz \eqref{eq:ellP2-gauge-ansatz}.

\paragraph{$ d=1 $ case} This theory has 22 fundamental and 232 neutral hypermultiplets. Its hodge numbers and Euler characteristic are $ (h^{1,1},h^{2,1},\chi)=(3,231,-456) $. The non-zero triple intersection numbers of divisors $ J_i $ dual to Mori cone generators are
\begin{gather}
    \begin{gathered}
        \kappa_{111} = 50 \, , \quad
        \kappa_{112} = 80 \, , \quad
        \kappa_{113} = 10 \, , \quad
        \kappa_{122} = 128 \, , \\
        \kappa_{123} = 16 \, , \quad
        \kappa_{133} = 2 \, , \quad
        \kappa_{222} = 203 \, , \quad
        \kappa_{223} = 25 \, , \quad
        \kappa_{233} = 3 \, .
    \end{gathered}
\end{gather}

The genus zero GV-invariants of the theory computed from the Mori cone and the triple intersection numbers are given in Table~\ref{table:P2b1-gv0}, where $ (n_1,n_2,n_3) $ represents the lattice point $ \sum l^\alpha n_\alpha $ of the Mori cone.
\begin{table}[t]
    \scriptsize
    \centering
    \begin{tabular}{c|c||cccc||cccc||cccc}
        & & \multicolumn{12}{c}{$ n_2-n_1 $} \\ \hline
        & & $ 0 $ & $ 1 $ & $ 2 $ & $ 3 $ & $ 0 $ & $ 1 $ & $ 2 $ & $ 3 $ & $ 0 $ & $ 1 $ & $ 2 $ & $ 3 $ \\ \hline
        \parbox[t]{1.5ex}{\multirow{8}{*}{\rotatebox[origin=c]{90}{$ n_1 $}}} & $ 0 $ & & $ -2 $ & & & $ 3 $ & $ 4 $ & $ 3 $ & $ 5 $ & $ -6 $ & $ -10 $ & $ -12 $ & $ -12 $ \\
        & $ 1 $ & $ 44 $ & $ 44 $ & & & & $ -88 $ & $ -88 $ & $ -176 $ & & $ 220 $ & $ 352 $ & $ 396 $ \\
        & $ 2 $ & $ -2 $  & $ 456 $ & $ -2 $ & & & $ -912 $ & $ 42 $ & $ 2862 $ & & $ 2280 $ & $ -164 $ & $ -1620 $ \\
        & $ 3 $ & & $ 44 $ & $ 44 $ & & & $ -88 $ & $ 19800 $ & $ -27368 $ & & $ 220 $ & $ -79288 $ & $ -78144 $ \\
        & $ 4 $ & & $ -2 $ & $ 456 $ & $ -2 $ & & $ 4 $ & $ 103856 $ & $ 249155 $ & & $ -10 $ & $ -416336 $ & $ 658656 $ \\
        & $ 5 $ & & & $ 44 $ & $ 44 $ & & & $ 19800 $ & $ 37236392 $ & & & $ -79288 $ & $ 13220460 $ \\
        & $ 6 $ & & & $ -2 $ & $ 456 $ & & & $ 42 $ & $ 128625156 $ & & & $ -164 $ & $ 47211048 $ \\
        & $ 7 $ & & & & $ 44 $ & & & $ -88 $ & $ 39333544 $ & & & $ 352 $ & $ 13220460 $ \\ \hline
        & & \multicolumn{4}{c||}{$ n_3=0 $} & \multicolumn{4}{c||}{$ n_3=1 $} & \multicolumn{4}{c}{$ n_3=2 $}
    \end{tabular}
    \caption{Genus zero GV-invariant for $ d=1 $ case, where $ (n_1,n_2,n_3) $ labels the lattice point $ \sum l^\alpha n_\alpha $ of the Mori cone.} \label{table:P2b1-gv0}
\end{table}
We note that the reflexive polytope $ \nabla $ in \eqref{eq:ellP2-gauge-toric} for $ d=1 $ differs from the polytope considered in \cite{Huang:2018vup}. However, two Calabi-Yau hypersurfaces from two polytopes have same hodge numbers, triple intersection numbers and second Chern classes (or equivalently the first Pontryagin class), which makes them equivalent Calabi-Yau manifolds according to the theorem of Wall \cite{Wall:1966}. We also checked that they have same genus zero GV-invariants. 

The modular ansatz for elliptic genera of strings has additional denominator factors arising from the $SU(2)$ instanton zero modes. For instance, from \eqref{eq:ellP2-gauge-ansatz}, the 1-string elliptic genus is given by
\begin{align}
    Z_1 = \frac{1}{\eta(\tau)^{36}} \frac{\mathcal{N}_1(\tau,\epsilon,\phi)}{\varphi_2(\tau,\epsilon) \varphi_2(\tau,2\phi)} \, .
\end{align}
We compute the numerator of the 1-string modular ansatz as
\begin{align}
    \begin{aligned}
    \mathcal{N}_1 &= \frac{E_4^6 \varphi_2(\phi)^5}{23328}+\frac{325 E_4^3 E_6^2 \varphi_2(\phi)^5}{373248}+\frac{91 E_6^4 \varphi_2(\phi)^5}{373248}-\frac{137 E_4^4 E_6 \varphi_2(\phi)^4 \varphi_0(\phi)}{82944} \\
    &\quad -\frac{103 E_4 E_6^3 \varphi_2(\phi)^4 \varphi_0(\phi)}{82944}+\frac{89 E_4^5 \varphi_2(\phi)^3 \varphi_0(\phi)^2}{248832}+\frac{343 E_4^2 E_6^2 \varphi_2(\phi)^3 \varphi_0(\phi)^2}{248832} \\
    &\quad +\frac{115 E_4^3 E_6 \varphi_2(\phi)^2 \varphi_0(\phi)^3}{248832}+\frac{29 E_6^3 \varphi_2(\phi)^2 \varphi_0(\phi)^3}{248832}-\frac{31 E_4^4 \varphi_2(\phi) \varphi_0(\phi)^4}{248832} \\
    &\quad -\frac{113 E_4 E_6^2 \varphi_2(\phi) \varphi_0(\phi)^4}{248832} \,,
    \end{aligned}
\end{align}
using the conditions stated above and the genus zero GV-invariants.

\paragraph{$ d=2 $ case}
The theory has 40 fundamental and 196 neutral hypermultiplets. Its hodge numbers and Euler characteristic are $ (h^{1,1},h^{2,1},\chi) = (3,195,-384) $, and the non-zero triple intersection numbers of divisors dual to Mori cone curves are
\begin{gather}
    \begin{gathered}
        \kappa_{111} = 32 \, , \quad
        \kappa_{112} = 56 \, , \quad
        \kappa_{113} = 8 \, , \quad
        \kappa_{122} = 98 \, , \\
        \kappa_{123} = 14 \, , \quad
        \kappa_{133} = 2 \, , \quad
        \kappa_{222} = 167 \, , \quad
        \kappa_{223} = 23 \, , \quad
        \kappa_{233} = 3 \, .
    \end{gathered}
\end{gather}

The genus zero GV-invariants computed from the Mori cone \eqref{eq:ellP2-gauge-toric} with $ d=2 $ and the triple intersection numbers are given in Table~\ref{table:P2b2-gv0}, where $ (n_1,n_2,n_3) $ represents the lattice point $ \sum l^\alpha n_\alpha $ of the Mori cone generated by $ \{l^1,l^2,l^3\} $.
\begin{table}[t]
    \scriptsize
    \centering
    \begin{tabular}{c|c||cccc||cccc}
        & & \multicolumn{8}{c}{$ n_2-n_1 $} \\ \hline
        & & $ 0 $ & $ 1 $ & $ 2 $ & $ 3 $ & $ 0 $ & $ 1 $ & $ 2 $ & $ 3 $ \\ \hline
        \parbox[t]{1.5ex}{\multirow{8}{*}{\rotatebox[origin=c]{90}{$ n_1 $}}} & $ 0 $ & & $ -2 $ & & & $ 3 $ & $ 4 $ & $ 3 $ & \\
        & $ 1 $ & $ 80 $ & $ 80 $ & & & & $ -160 $ & $ -160 $ & \\
        & $ 2 $ & $ -2 $ & $ 384 $ & $ -2 $ & & & $ -768 $ & $ 2400 $ & $ -768 $ \\
        & $ 3 $ & & $ 80 $ & $ 80 $ & & & $ -160 $ & $ 30240 $ & $ 30240 $  \\
        & $ 4 $ & & $ -2 $ & $ 384 $ & $ -2 $ & & $ 4 $ & $ 78404 $ & $ 6816840 $  \\
        & $ 5 $ & & & $ 80 $ & $ 80 $ & & & $ 30240 $ & $ 48267168 $  \\
        & $ 6 $ & & & $ -2 $ & $ 384 $ & & & $ 2400 $ & $ 93844224 $ \\
        & $ 7 $ & & & & $ 80 $ & & & $ -160 $ & $ 48267168 $ \\ \hline
        & & \multicolumn{4}{c||}{$ n_3=0 $} & \multicolumn{4}{c}{$ n_3=1 $}  \\ \hline \hline
        & & \multicolumn{8}{c}{$ n_2-n_1 $} \\ \hline
        & & $ 0 $ & $ 1 $ & $ 2 $ & $ 3 $ & $ 0 $ & $ 1 $ & $ 2 $ & $ 3 $ \\ \hline
        \parbox[t]{1.5ex}{\multirow{8}{*}{\rotatebox[origin=c]{90}{$ n_1 $}}} & $ 0 $ & $ -6 $ & $ -10 $ & $ -12 $ & $ -12 $ & $ 27 $ & $ 64 $ & $ 91 $ & $ 108 $ \\
        & $ 1 $ & & $ 400 $ & $ 640 $ & $ 720 $ & & $ -2560 $ & $ -5600 $ & $ -7680 $ \\
        & $ 2 $ & & $ 1920 $ & $ -9596 $ & $ -15552 $ & & $ -12288 $ & $ 84448 $ & $ 192000 $ \\
        & $ 3 $ & & $ 400 $ & $ -121120 $ & $ 66720 $ & & $ -2560 $ & $ 1062880 $ & $ -822240 $ \\
        & $ 4 $ & & $ -10 $ & $ -314384 $ & $ 3102264 $ & & $ 64 $ & $ 2768332 $ & $ -37640696 $ \\
        & $ 5 $ & & & $ -121120 $ & $ 17492400 $ & & & $ 1062880 $ & $ -213052480 $ \\
        & $ 6 $ & & & $ -9596 $ & $ 33517440 $ & & & $ 84448 $ & $ -410721024 $  \\
        & $ 7 $ & & & $ 640 $ & $ 17492400 $ & & & $ -5600 $ & $ -213052480 $  \\ \hline
        & & \multicolumn{4}{c||}{$ n_3=2 $} & \multicolumn{4}{c}{$ n_3=3 $} 
    \end{tabular}
    \caption{Genus zero GV-invariants for $ d=2 $} \label{table:P2b2-gv0}
\end{table}
In this case, since the $SU(2)$ gauge symmetry is supported on a degree-two curve, i.e. $d=2$, the 2d worldsheet CFT involves $k$-instanton zero modes if the string charge is $Q=2k$ or $Q=2k+1$. For example, the 1-string elliptic genus does not contain instanton zero mode contributions, whereas the 2-string elliptic genus contains 1-instanton contribution. Therefore, we can set the modular ansatz for the 1-string and 2-string elliptic genera as
\begin{align}
    Z_1(\tau, \epsilon, \phi) = \frac{1}{\eta(\tau)^{36}} \frac{\mathcal{N}_1(\tau, \epsilon, \phi)}{\varphi_2(\tau,\epsilon)} \, , \quad
    Z_2(\tau, \epsilon, \phi) = \frac{1}{\eta(\tau)^{72}} \frac{\mathcal{N}_2(\tau,\epsilon,\phi)}{\varphi_2(\tau,\epsilon) \varphi_2(\tau,2\epsilon) \varphi_2(\tau,2\phi) } \, .
\end{align}
We fix unknown coefficients in $ \mathcal{N}_1 $ and $ \mathcal{N}_2 $ using the GV-invariant condition and the genus zero GV-invariants computed from the mirror symmetry. The numerator for 1-string is
\begin{align}
    \begin{aligned}
        \mathcal{N}_1 &= -\frac{31 E_4^4 \varphi _0(\phi ){}^2}{6912}-\frac{113 E_4 E_6^2 \varphi_0(\phi )^2}{6912}+\frac{115 E_4^3 E_6 \varphi _0(\phi ) \varphi _2(\phi)}{3456} \\
        &\quad +\frac{29 E_6^3 \varphi _0(\phi ) \varphi _2(\phi )}{3456}-\frac{47E_4^5 \varphi _2(\phi )^2}{6912}-\frac{97 E_4^2 E_6^2 \varphi _2(\phi)^2}{6912} \, .
    \end{aligned}
\end{align}
The explicit expression of the 2-string elliptic genus is given in Appendix~\ref{appendix:ellP2-gauge-result}.

For 3-strings, we can write the modular ansatz as
\begin{align}
    Z_3(\tau, \epsilon, \phi) = \frac{1}{\eta(\tau)^{108}} \frac{\mathcal{N}_3(\tau,\epsilon,\phi)}{\varphi_2(\tau,\epsilon) \varphi_2(\tau,2\epsilon) \varphi_2(\tau,3\epsilon) \varphi_2(\tau,2\phi) } \, ,
\end{align}
which contains in the denominator the bosonic zero mode contribution for 1-instanton.
In this case, the GV-invariant condition and the genus zero GV-invariants are not enough to fix all the unknown coefficients in $ \mathcal{N}_3 $. We find that to completely determine the numerator in the 3-instanton ansatz, we need additional information from the genus one free energy. The input data for evaluating the genus one free energy \eqref{eq:genus1} are the period vectors and mirror maps obtained from the solutions of the Picard-Fuchs system, the divisor integrals of the second Chern class given by
\begin{align}
    \int c_2 \wedge J_1 = 128 \, , \quad
    \int c_2 \wedge J_2 = 230 \, , \quad
    \int c_2 \wedge J_3 = 36 \, ,
\end{align}
where $ J_{1,2,3} $ are divisor classes dual to Mori cone curves associated to $ l^{1,2,3} $, as well as the discriminants of the Picard-Fuchs operators which serve as denominators of the Yukawa couplings \eqref{eq:Yukawa} as \eqref{eq:Yukawa-rational}. The complete expressions of the Yukawa couplings are rather complicated, so we only give the discriminants in Appendix~\ref{appendix:ellP2-gauge-result}. One can reconstruct the Yukawa couplings from the discriminants and series solutions \eqref{eq:Period-series} of the Picard-Fuchs system. The genus one GV-invariants we computed from the free energy using \eqref{eq:F1} are summarized in Table~\ref{table:P2b2-gv1}.
\begin{table}[t]
    \scriptsize
    \centering
    \begin{tabular}{c|c||cccc||cccc}
        & & \multicolumn{8}{c}{$ n_2-n_1 $} \\ \hline
        & & $ 0 $ & $ 1 $ & $ 2 $ & $ 3 $ & $ 0 $ & $ 1 $ & $ 2 $ & $ 3 $ \\ \hline
        \parbox[t]{1.5ex}{\multirow{8}{*}{\rotatebox[origin=c]{90}{$ n_1 $}}} & $ 0 $ & & & & & & & & \\
        & $ 1 $ & & & & & & & & \\
        & $ 2 $ & & $ 3 $ & & & & $ -6 $ & $ -8 $ & $ -6 $ \\
        & $ 3 $ & & & & & & & $ 320 $ & $ 320 $  \\
        & $ 4 $ & & & $ 3 $ & & & & $ 1518 $ & $ -4824 $ \\
        & $ 5 $ & & & & & & & $ 320 $ & $ -59520 $ \\
        & $ 6 $ & & & & $ 3 $ & & & $ -8 $ & $ -152224 $ \\
        & $ 7 $ & & & & & & & & $ -59520 $ \\ \hline
        & & \multicolumn{4}{c||}{$ n_3=0 $} & \multicolumn{4}{c}{$ n_3=1 $}  \\ \hline \hline
        & & \multicolumn{8}{c}{$ n_2-n_1 $} \\ \hline
        & & $ 0 $ & $ 1 $ & $ 2 $ & $ 3 $ & $ 0 $ & $ 1 $ & $ 2 $ & $ 3 $ \\ \hline
        \parbox[t]{1.5ex}{\multirow{8}{*}{\rotatebox[origin=c]{90}{$ n_1 $}}} & $ 0 $ & & & & & $ -10 $ & $ -18 $ & $ -24 $ & $ -28 $ \\
        & $ 1 $ & & & & & & $ 720 $ & $ 1280 $ & $ 1680 $ \\
        & $ 2 $ & & $ 15 $ & $ 32 $ & $ 39 $ & & $ 3360 $ & $ -19452 $ & $ -36672 $ \\
        & $ 3 $ & & & $ -1280 $ & $ -2400 $ & & $ 720 $ & $ -231840 $ & $ 180640 $ \\
        & $ 4 $ & & & $ -6078 $ & $ 36336 $ & & $ -18 $ & $ -579226 $ & $ 6860448 $ \\
        & $ 5 $ & & & $ -1280 $ & $ 449760 $ & & & $ -231840 $ & $ 36114480 $ \\
        & $ 6 $ & & & $ 32 $ & $ 1158888 $ & & & $ -19452 $ & $ 66037890 $  \\
        & $ 7 $ & & & & $ 449760 $ & & & $ 1280 $ & $ 36114480 $  \\ \hline
        & & \multicolumn{4}{c||}{$ n_3=2 $} & \multicolumn{4}{c}{$ n_3=3 $} 
    \end{tabular}
    \caption{Genus one GV-invariants for $ d=2 $} \label{table:P2b2-gv1}
\end{table}
The explicit expression of $ \mathcal{N}_3 $ in the 3-string ansatz is given in Appendix~\ref{appendix:ellP2-gauge-result}.

\paragraph{$ d=3 $ case}
In this case, the theory has 1 adjoint, 54 fundamental and 165 neutral hypeermultiplets. Its topological numbers are $ (h^{1,1},h^{2,1},\chi)=(3,165,-324) $, and the non-zero triple intersection numbers of divisors dual to the Mori cone generators are
\begin{gather}
    \begin{gathered}
        \kappa_{111} = 18 \, , \quad
        \kappa_{112} = 36 \, , \quad
        \kappa_{113} = 6 \, , \quad
        \kappa_{122} = 72 \, , \\
        \kappa_{123} = 12 \, , \quad
        \kappa_{133} = 2 \, , \quad
        \kappa_{222} = 135 \, , \quad
        \kappa_{223} = 21 \, , \quad
        \kappa_{233} = 3 \, . \quad
    \end{gathered}
\end{gather}

The genus zero GV-invariants from the triple intersection numbers and Mori cone curves are given in Table~\ref{table:P2b3-gv0}, where $ (n_1,n_2,n_3) $ labels the lattice point $ \sum l^\alpha n_\alpha $ in the Mori cone. In this model, 1-string and 2-strings are not instanton strings, so the modular ansatz for their elliptic genera does not involve extra denominator factors from $ SU(2) $ instanton zero modes. 

We compute the numerator in the modular ansatz for the 1-string elliptic genus as
\begin{align}
    \mathcal{N}_1 &= \frac{103 E_4^4 E_6 \varphi_2(\phi )^3}{82944}+\frac{41 E_4 E_6^3\varphi_2(\phi )^3}{82944}-\frac{43 E_4^5 \varphi_2(\phi )^2 \varphi_0(\phi)}{27648}-\frac{101 E_4^2 E_6^2 \varphi_2(\phi )^2 \varphi_0(\phi)}{27648} \nonumber \\
    &\quad +\frac{115 E_4^3 E_6 \varphi_2(\phi ) \varphi_0(\phi)^2}{27648}+\frac{29 E_6^3 \varphi_2(\phi ) \varphi_0(\phi)^2}{27648}-\frac{31 E_4^4 \varphi_0(\phi )^3}{82944}-\frac{113 E_4 E_6^2\varphi_0(\phi )^3}{82944} \, .
\end{align}
The 2-string elliptic genus is computed in Appendix~\ref{appendix:ellP2-gauge-result}.

\begin{table}[t]
    \scriptsize
    \centering
    \begin{tabular}{c|c||cccc||cccc||cccc}
        & & \multicolumn{12}{c}{$ n_2-n_1 $} \\ \hline
        & & $ 0 $ & $ 1 $ & $ 2 $ & $ 3 $ & $ 0 $ & $ 1 $ & $ 2 $ & $ 3 $ & $ 0 $ & $ 1 $ & $ 2 $ & $ 3 $ \\ \hline
        \parbox[t]{1.5ex}{\multirow{8}{*}{\rotatebox[origin=c]{90}{$ n_1 $}}} & $ 0 $ & & & & & $ 3 $ & & & $ 3 $ & $ -6 $ \\
        & $ 1 $ & $ 108 $ & $ 108 $ & & & & $ -216 $ & & $ -216 $ & & $ 540 $  \\
        & $ 2 $ & & $ 324 $ & & & & $ -648 $ & $ 5778 $ & $ 5778 $ & & $ 1620 $ & $ -23112 $ \\
        & $ 3 $ & & $ 108 $ & $ 108 $ & & & $ -216 $ & $ 34560 $ & $ 1164624 $ & & $ 540 $ & $ -138456 $ & $ 612468 $ \\
        & $ 4 $ & & & $ 324 $ & & & & $ 62694 $ & $ 13355901 $ & & & $ -251424 $ & $ 5476248 $ \\
        & $ 5 $ & & & $ 108 $ & $ 108 $ & & & $ 34560 $ & $ 49699224 $ & & & $ -138456 $ & $ 17988804 $ \\
        & $ 6 $ & & & & $ 324 $ & & & $ 5778 $ & $ 75620556 $ & & & $ -23112 $ & $ 26655480 $ \\
        & $ 7 $ & & & & $ 108 $ & & & & $ 49699224 $ & & & & $ 17988804 $ \\ \hline
        & & \multicolumn{4}{c||}{$ n_3=0 $} & \multicolumn{4}{c||}{$ n_3=1 $} & \multicolumn{4}{c}{$ n_3=2 $}
    \end{tabular}
    \caption{Genus zero GV-invariants for $ d=3 $} \label{table:P2b3-gv0}
\end{table}

\paragraph{$ d=4 $ case}
This theory has 3 adjoint, 64 fundamental and 139 neutral hypermultiplets. The topological numbers are $ (h^{1,1},h^{2,1},\chi) = (3,141,-276) $. The non-zero triple intersection numbers of divisors dual to the Mori cone generator are
\begin{gather}
    \begin{gathered}
        \kappa_{111} = 8 \, , \quad
        \kappa_{112} = 20 \, , \quad
        \kappa_{113} = 4 \, , \quad
        \kappa_{122} = 50 \, , \\
        \kappa_{123} = 10 \, , \quad
        \kappa_{133} = 2 \, , \quad
        \kappa_{222} = 107 \, , \quad
        \kappa_{223} = 19 \, , \quad
        \kappa_{233} = 3 \, , \quad
    \end{gathered}
\end{gather}
The genus zero GV-invariants computed using the mirror technique are given in Table~\ref{table:P2b4-gv0}, where $ (n_1,n_2,n_3) $ labels the lattice point $ \sum l^\alpha n_\alpha $ in the Mori cone. 

We compute the numerator in the modular ansatz for the 1-string elliptic genus as
\begin{align}
    \mathcal{N}_1 &= -\frac{37 E_4^6 \varphi_2(\phi )^4}{2985984}-\frac{355 E_4^3 E_6^2\varphi_2(\phi )^4}{2985984}-\frac{5 E_6^4 \varphi_2(\phi)^4}{373248}+\frac{11 E_4^4 E_6 \varphi_2(\phi )^3 \varphi_0(\phi)}{27648} \nonumber \\
    &\quad +\frac{5 E_4 E_6^3 \varphi_2(\phi )^3 \varphi_0(\phi)}{27648}-\frac{125 E_4^5 \varphi_2(\phi )^2 \varphi_0(\phi)^2}{497664}-\frac{307 E_4^2 E_6^2 \varphi_2(\phi )^2 \varphi_0(\phi)^2}{497664} \\
    &\quad +\frac{115 E_4^3 E_6 \varphi_2(\phi ) \varphi_0(\phi)^3}{248832}+\frac{29 E_6^3 \varphi_2(\phi ) \varphi_0(\phi)^3}{248832}-\frac{31 E_4^4 \varphi_0(\phi )^4}{995328}-\frac{113 E_4 E_6^2\varphi_0(\phi )^4}{995328} \, . \nonumber
\end{align}
The 2-string result is given in Appendix~\ref{appendix:ellP2-gauge-result}.

\begin{table}[t]
    \scriptsize
    \centering
    \begin{tabular}{c|c|cccc|cccc|cccc}
        & & \multicolumn{12}{c}{$ n_2-n_1 $} \\ \hline
        & & $ 0 $ & $ 1 $ & $ 2 $ & $ 3 $ & $ 0 $ & $ 1 $ & $ 2 $ & $ 3 $ & $ 0 $ & $ 1 $ & $ 2 $ & $ 3 $ \\ \hline
        \parbox[t]{1.5ex}{\multirow{8}{*}{\rotatebox[origin=c]{90}{$ n_1 $}}} & $ 0 $ & & $ 4 $ & & & $ 3 $ & $ -8 $ & $ 6 $ & $ -8 $ & $ -6 $ & $ 20 $ & $ -24 $ & $ 12 $  \\
        & $ 1 $ & $ 128 $ & $ 128 $ & & & & $ -256 $ & $ 512 $ & $ 512 $ & & $ 640 $ & $ -2048 $ & $ 2304 $  \\
        & $ 2 $ & $ 4 $ & $ 276 $ & $ 4 $ & & & $ -552 $ & $ 9216 $ & $ 182368 $ & & $ 1380 $ & $ -36872 $ & $ 102312 $ \\
        & $ 3 $ & & $ 128 $ & $ 128 $ & & & $ -256 $ & $ 35328 $ & $ 3120896 $ & & $ 640 $ & $ -141568 $ & $ 1438080 $ \\
        & $ 4 $ & & $ 4 $ & $ 276 $ & $ 4 $ & & $ -8 $ & $ 53246 $ & $ 18166584 $ & & $ 20 $ & $ -213536 $ & $ 7159068 $ \\
        & $ 5 $ & & & $ 128 $ & $ 128 $ & & & $ 35328 $ & $ 47904000 $ & & & $ -141568 $ & $ 17267328 $ \\
        & $ 6 $ & & & $ 4 $ & $ 276 $ & & & $ 9216 $ & $ 65322480 $ & & & $ -36872 $ & $ 22872312 $ \\
        & $ 7 $ & & & & $ 128 $ & & & $ 512 $ & $ 47904000 $ & & & $ -2048 $ & $ 17267328 $ \\ \hline
        & & \multicolumn{4}{c|}{$ n_3=0 $} & \multicolumn{4}{c|}{$ n_3=1 $} & \multicolumn{4}{c}{$ n_3=2 $}
    \end{tabular}
    \caption{Genus zero GV-invariants for $ d=4 $} \label{table:P2b4-gv0}
\end{table}

\paragraph{$ d=5 $ case}
This theory has 6 adjoint, 70 fundamental and 118 neutral hypermultiplets. The topological numbers are $ (h^{1,1},h^{2,1},\chi) = (3,123,-240) $. The nonzero triple intersection numbers of divisors dual to Mori cone generators are
\begin{gather}
    \begin{gathered}
        \kappa_{111} = 2 \, , \quad
        \kappa_{112} = 8 \, , \quad
        \kappa_{113} = 2 \, , \quad
        \kappa_{122} = 32 \, , \\
        \kappa_{123} = 8 \, , \quad
        \kappa_{133} = 2 \, , \quad
        \kappa_{222} = 83 \, , \quad
        \kappa_{223} = 17 \, , \quad
        \kappa_{233} = 3 \, .
    \end{gathered}
\end{gather}
The genus zero GV-invariant computed from the toric data are given in Table~\ref{table:P2b5-gv0}, where $ (n_1,n_2,n_3) $ represents lattice points $ \sum l^\alpha n_\alpha $ in the Mori cone generated $ \{l^1,l^2,l^3\} $. 

We compute the numerator in the modular ansatz for the 1-string elliptic genus as
\begin{align}
    \mathcal{N}_1 &= \frac{23 E_4^5 E_6 \varphi_2(\phi )^5}{11943936}+\frac{121 E_4^2 E_6^3\varphi_2(\phi )^5}{11943936}-\frac{55 E_4^6 \varphi_2(\phi )^4\varphi_0(\phi )}{11943936}-\frac{65 E_4^3 E_6^2 \varphi_2(\phi )^4\varphi_0(\phi )}{1327104} \nonumber \\
    &\quad -\frac{5 E_6^4 \varphi_2(\phi )^4 \varphi_0(\phi)}{746496}+\frac{485 E_4^4 E_6 \varphi_2(\phi )^3 \varphi_0(\phi)^2}{5971968}+\frac{235 E_4 E_6^3 \varphi_2(\phi )^3 \varphi_0(\phi)^2}{5971968} \nonumber \\
    &\quad -\frac{205 E_4^5 \varphi_2(\phi )^2 \varphi_0(\phi)^3}{5971968}-\frac{515 E_4^2 E_6^2 \varphi_2(\phi )^2 \varphi_0(\phi)^3}{5971968}+\frac{575 E_4^3 E_6 \varphi_2(\phi ) \varphi_0(\phi)^4}{11943936} \nonumber \\
    &\quad +\frac{145 E_6^3 \varphi_2(\phi ) \varphi_0(\phi)^4}{11943936}-\frac{31 E_4^4 \varphi_0(\phi )^5}{11943936}-\frac{113 E_4E_6^2 \varphi_0(\phi )^5}{11943936} \, .
\end{align}
We summarize the 2-string result in Appendix~\ref{appendix:ellP2-gauge-result}.

\begin{table}[t]
    \scriptsize
    \centering
    \begin{tabular}{c|c|cccc|cccc|cccc}
        & & \multicolumn{12}{c}{$ n_2-n_1 $} \\ \hline
        & & $ 0 $ & $ 1 $ & $ 2 $ & $ 3 $ & $ 0 $ & $ 1 $ & $ 2 $ & $ 3 $ & $ 0 $ & $ 1 $ & $ 2 $ & $ 3 $ \\ \hline
        \parbox[t]{1.5ex}{\multirow{8}{*}{\rotatebox[origin=c]{90}{$ n_1 $}}} & $ 0 $ & & $ 10 $ & & & $ 3 $ & $ -20 $ & $ 45 $ & $ 45 $ & $ -6 $ & $ 50 $ & $ -180 $ & $ 360 $  \\
        & $ 1 $ & $ 140 $ & $ 140 $ & & & & $ -280 $ & $ 1400 $ & $ 23904 $ & & $ 700 $ & $ -5600 $ & $ 18900 $  \\
        & $ 2 $ & $ 10 $ & $ 240 $ & $ 10 $ & & & $ -480 $ & $ 12090 $ & $ 631230 $ & & $ 1200 $ & $  -48380$ & $ 323100 $ \\
        & $ 3 $ & & $ 140 $ & $ 140 $ & & & $ -280 $ & $ 34440 $ & $ 5388840 $ & & $ 700 $ & $ -138040 $ & $ 2336040 $ \\
        & $ 4 $ & & $ 10 $ & $ 240 $ & $ 10 $ & & $ -20 $ & $ 47420 $ & $ 21171435 $ & & $ 50 $ & $ -190160 $ & $ 8162820 $ \\
        & $ 5 $ & & & $ 140 $ & $ 140 $ & & & $ 34440 $ & $ 45601800 $ & & & $ -138040 $ & $ 16345980 $ \\
        & $ 6 $ & & & $ 10 $ & $ 240 $ & & & $ 12090 $ & $ 58436676 $ & & & $ -48380 $ & $ 20436120 $ \\
        & $ 7 $ & & & & $ 140 $ & & & $ 1400 $ & $ 45601800 $ & & & $ -5600 $ & $ 16345980 $ \\ \hline
        & & \multicolumn{4}{c|}{$ n_3=0 $} & \multicolumn{4}{c|}{$ n_3=1 $} & \multicolumn{4}{c}{$ n_3=2 $}
    \end{tabular}
    \caption{Genus zero GV-invariants for $ d=5 $} \label{table:P2b5-gv0}
\end{table}

\paragraph{$ d=6 $ case}
This theory has 10 adjoint, 72 fundamental and 102 neutral hypermultiplets. The Hodge numbers and Euler characteristics are $ (h^{1,1},h^{2,1},\chi) = (3,111,-216) $. The non-zero triple intersection numbers of divisors dual to Mori cone generators are
\begin{align}
    \kappa_{122} = 18 \, , \quad
    \kappa_{123} = 6 \, , \quad
    \kappa_{133} = 2 \, , \quad
    \kappa_{222} = 63 \, , \quad
    \kappa_{223} = 15 \, , \quad
    \kappa_{233} = 3 \, .
\end{align}
We present genus zero GV-invariants computed using toric data and mirror symmetry technique in Table~\ref{table:P2b6-gv0}, where $ (n_1,n_2,n_3) $ labels the lattice point $ \sum l^\alpha n_\alpha $ in the Mori cone. 

We compute the numerator in the modular ansatz for the 1-string elliptic genus as
\begin{align}
    \begin{aligned}
        \mathcal{N}_1 &= \frac{E_4^7 \varphi_2(\phi )^6}{15925248}-\frac{49 E_4^4 E_6^2 \varphi_2(\phi )^6}{143327232}-\frac{13 E_4 E_6^4 \varphi_2(\phi )^6}{17915904}+\frac{19 E_4^5 E_6 \varphi_2(\phi )^5 \varphi_0(\phi )}{23887872} \\
        &\quad +\frac{125 E_4^2 E_6^3 \varphi_2(\phi )^5 \varphi_0(\phi )}{23887872}-\frac{17 E_4^6 \varphi_2(\phi )^4 \varphi_0(\phi )^2}{15925248}-\frac{581 E_4^3 E_6^2 \varphi_2(\phi )^4 \varphi_0(\phi )^2}{47775744} \\
        &\quad -\frac{11 E_6^4 \varphi_2(\phi )^4 \varphi_0(\phi )^2}{5971968}+\frac{479 E_4^4 E_6 \varphi_2(\phi )^3 \varphi_0(\phi )^3}{35831808}+\frac{241 E_4 E_6^3 \varphi_2(\phi )^3 \varphi_0(\phi )^3}{35831808} \\
        &\quad -\frac{203 E_4^5 \varphi_2(\phi )^2 \varphi_0(\phi )^4}{47775744}-\frac{517 E_4^2 E_6^2 \varphi_2(\phi )^2 \varphi_0(\phi )^4}{47775744}+\frac{115 E_4^3 E_6 \varphi_2(\phi ) \varphi_0(\phi )^5}{23887872} \\
        &\quad +\frac{29 E_6^3 \varphi_2(\phi ) \varphi_0(\phi )^5}{23887872}-\frac{31 E_4^4 \varphi_0(\phi )^6}{143327232}-\frac{113 E_4 E_6^2 \varphi_0(\phi )^6}{143327232} \, .
    \end{aligned}
\end{align}
We summarize the 2-string result in Appendix~\ref{appendix:ellP2-gauge-result}.

\begin{table}[t]
    \scriptsize
    \centering
    \begin{tabular}{c|c|cccc|cccc|cccc}
        & & \multicolumn{12}{c}{$ n_2-n_1 $} \\ \hline
        & & $ 0 $ & $ 1 $ & $ 2 $ & $ 3 $ & $ 0 $ & $ 1 $ & $ 2 $ & $ 3 $ & $ 0 $ & $ 1 $ & $ 2 $ & $ 3 $ \\ \hline
        \parbox[t]{1.5ex}{\multirow{8}{*}{\rotatebox[origin=c]{90}{$ n_1 $}}} & $ 0 $ & & $ 18 $ & & & $ 3 $ & $ -36 $ & $ 153 $ & $ 2256 $ & $ -6 $ & $ 90 $ & $ -612 $ & $ 2448 $  \\
        & $ 1 $ & $ 144 $ & $ 144 $ & & & & $ -288 $ & $ 2592 $ & $ 105984 $ & & $ 720 $ & $ -10368 $ & $ 66096 $  \\
        & $ 2 $ & $ 18 $ & $ 216 $ & $ 18 $ & & & $ -432 $ & $ 14112 $ & $ 1356336 $ & & $ 1080 $ & $ -56484 $ & $ 651240 $ \\
        & $ 3 $ & & $ 144 $ & $ 144 $ & & & $ -288 $ & $ 33120 $ & $ 7517856 $ & & $ 720 $ & $ -132768 $ & $ 3145536 $ \\
        & $ 4 $ & & $ 18 $ & $ 216 $ & $ 18 $ & & $ -36 $ & $ 43416 $ & $ 22973544 $ & & $ 90 $ & $ -174096 $ & $ 8712036 $ \\
        & $ 5 $ & & & $ 144 $ & $ 144 $ & & & $ 33120 $ & $ 43396128 $ & & & $ -132768 $ & $ 15493680 $ \\
        & $ 6 $ & & & $ 18 $ & $ 216 $ & & & $ 14112 $ & $ 53366976 $ & & & $ -56484 $ & $ 18668448 $ \\
        & $ 7 $ & & & & $ 144 $ & & & $ 2592 $ & $ 43396128 $ & & & $ -10368 $ & $ 15493680 $ \\ \hline
        & & \multicolumn{4}{c|}{$ n_3=0 $} & \multicolumn{4}{c|}{$ n_3=1 $} & \multicolumn{4}{c}{$ n_3=2 $}
    \end{tabular}
    \caption{Genus zero GV-invariants for $ d=6 $} \label{table:P2b6-gv0}
\end{table}

\paragraph{$ d=7 $ case}
This theory has 15 adjoint, 70 fundamental and 91 neutral hypermultiplets. The topological numbers are $ (h^{1,1},h^{2,1},\chi) = (3,105,-204) $. The 4d reflexive polytope for the toric construction of this theory is the same in \eqref{eq:ellP2-gauge-toric} for $ d=7 $. However, the Mori cone becomes non-simplicial in this case. There are four Mori cone generators $ l^{1,2,3,4} $. The first three are identical to the vectors provided in \eqref{eq:ellP2-gauge-toric}, while the last one is represented as $ l^4 = (2,-1,1,1,1,0,0) $. These generators satisfy $ l^4 = -l^1 + 2l^2 + l^3 $ and define a 3-dimensional cone. The triple intersection numbers for three divisors $ J_{1,2,3} $ dual to three Mori cone curves associated with $ l^{1,2,3} $ are
\begin{gather}
    \begin{gathered}
        \kappa_{111} = 2 \, , \quad
        \kappa_{112} = -4 \, , \quad
        \kappa_{113} = -2 \, , \quad
        \kappa_{122} = 8 \, , \\
        \kappa_{123} = 4 \, , \quad
        \kappa_{133} = 2 \, , \quad
        \kappa_{222} = 47 \, , \quad
        \kappa_{223} = 13 \, , \quad
        \kappa_{233} = 3 \, ,
    \end{gathered}
\end{gather}
where the others are zero. The genus zero GV-invariant computed from the toric information is given in Table~\ref{table:P2b7-gv0}, where $ \sum_{k=1}^3 l^k n_k $ represents the lattice points of the Mori cone. Since the Mori cone is non-simplicial, we have included some negative values of $ n_1 $.

We compute the numerator in the modular ansatz for the 1-string elliptic genus as
\begin{align}
    \mathcal{N}_1 &= -\frac{59 E_4^6 E_6 \varphi_2(\phi )^7}{5159780352}+\frac{187 E_4^3 E_6^3 \varphi_2(\phi )^7}{5159780352}+\frac{19 E_6^5 \varphi_2(\phi )^7}{322486272}+\frac{91 E_4^7 \varphi_2(\phi )^6 \varphi_0(\phi )}{1719926784} \nonumber \\
    &\quad -\frac{371 E_4^4 E_6^2 \varphi_2(\phi )^6 \varphi_0(\phi )}{1719926784}-\frac{91 E_4 E_6^4 \varphi_2(\phi )^6 \varphi_0(\phi )}{214990848}+\frac{343 E_4^5 E_6 \varphi_2(\phi )^5 \varphi_0(\phi )^2}{1719926784} \nonumber \\
    &\quad +\frac{2681 E_4^2 E_6^3 \varphi_2(\phi )^5 \varphi_0(\phi )^2}{1719926784}-\frac{1015 E_4^6 \varphi_2(\phi )^4 \varphi_0(\phi )^3}{5159780352}-\frac{12145 E_4^3 E_6^2 \varphi_2(\phi )^4 \varphi_0(\phi )^3}{5159780352} \nonumber \\
    &\quad -\frac{245 E_6^4 \varphi_2(\phi )^4 \varphi_0(\phi )^3}{644972544}+\frac{3325 E_4^4 E_6 \varphi_2(\phi )^3 \varphi_0(\phi )^4}{1719926784}+\frac{1715 E_4 E_6^3 \varphi_2(\phi )^3 \varphi_0(\phi )^4}{1719926784} \nonumber \\
    &\quad -\frac{847 E_4^5 \varphi_2(\phi )^2 \varphi_0(\phi )^5}{1719926784}-\frac{2177 E_4^2 E_6^2 \varphi_2(\phi )^2 \varphi_0(\phi )^5}{1719926784}+\frac{805 E_4^3 E_6 \varphi_2(\phi ) \varphi_0(\phi )^6}{1719926784} \nonumber \\
    &\quad +\frac{203 E_6^3 \varphi_2(\phi ) \varphi_0(\phi )^6}{1719926784}-\frac{31 E_4^4 \varphi_0(\phi )^7}{1719926784}-\frac{113 E_4 E_6^2 \varphi_0(\phi )^7}{1719926784} \, .
\end{align}
The 2-string result is summarized in Appendix~\ref{appendix:ellP2-gauge-result}.

\begin{table}[t]
    \scriptsize
    \centering
    \begin{tabular}{c|c|cccc|cccc|cccc}
        & & \multicolumn{12}{c}{$ n_2-n_1 $} \\ \hline
        & & $ 0 $ & $ 1 $ & $ 2 $ & $ 3 $ & $ 0 $ & $ 1 $ & $ 2 $ & $ 3 $ & $ 0 $ & $ 1 $ & $ 2 $ & $ 3 $ \\ \hline
        \parbox[t]{1.5ex}{\multirow{9}{*}{\rotatebox[origin=c]{90}{$ n_1 $}}} & $ -1 $ & & & & & & & & $ 56 $ &  \\
        & $ 0 $ & & $ 28 $ & & & $ 3 $ & $ -56 $ & $ 378 $ & $ 12747 $ & $ -6 $ & $ 140 $ & $ -1512 $ & $ 9828 $  \\
        & $ 1 $ & $ 140 $ & $ 140 $ & & & & $ -280 $ & $ 3920 $ & $ 286048 $ & & $ 700 $ & $ -15680 $ & $ 158760 $  \\
        & $ 2 $ & $ 28 $ & $ 204 $ & $ 28 $ & & & $ -408 $ & $ 15330 $ & $ 2253314 $ & & $ 1020 $ & $ -61376 $ & $ 1039248 $ \\
        & $ 3 $ & & $ 140 $ & $ 140 $ & & & $ -280 $ & $ 31920 $ & $ 9341976 $ & & $ 700 $ & $ -127960 $ & $ 3806460 $ \\
        & $ 4 $ & & $ 28 $ & $ 204 $ & $ 28 $ & & $ -56 $ & $ 40274 $ & $ 24037573 $ & & $ 140 $ & $ -161504 $ & $ 9000684 $ \\
        & $ 5 $ & & & $ 140 $ & $ 140 $ & & & $ 31920 $ & $ 41386464 $ & & & $ -127960 $ & $ 14735700 $ \\
        & $ 6 $ & & & $ 28 $ & $ 204 $ & & & $ 15330 $ & $ 49434828 $ & & & $ -61376 $ & $ 17309160 $ \\
        & $ 7 $ & & & & $ 140 $ & & & $ 3920 $ & $ 41386464 $ & & & $ -15680 $ & $ 14735700 $ \\ \hline
        & & \multicolumn{4}{c|}{$ n_3=0 $} & \multicolumn{4}{c|}{$ n_3=1 $} & \multicolumn{4}{c}{$ n_3=2 $}
    \end{tabular}
    \caption{Genus zero GV-invariants for $ d=7 $} \label{table:P2b7-gv0}
\end{table}

\paragraph{$ d=8 $ case}
This theory has 21 adjoint, 64 fundamental and 85 neutral hypermultiplets and the topological numbers are $ (h^{1,1},h^{2,1},\chi) = (3,105,-204) $. Similar to the case when $ d=7 $, the lattice polytope $ \nabla $ for the Calabi-Yau hypersurface corresponds to $ d=8 $ of \eqref{eq:ellP2-gauge-toric}; however the Mori cone is non-simplicial. There are four Mori cone generators $ l^{\alpha=1,2,3,4} $, where $ l^{1,2,3} $ are given in \eqref{eq:ellP2-gauge-toric} and $ l^4 = (1,-2,1,1,1,0,0) $ satisfying $ l^4 = -2l^1 + l^2 + l^3 $. The triple intersection numbers of divisors dual to the Mori cone curve associated to $ l^{1,2,3} $ are
\begin{gather}
    \begin{gathered}
        \kappa_{111} = 8 \, , \quad
        \kappa_{112} = -4 \, , \quad
        \kappa_{113} = -4 \, , \quad
        \kappa_{122} = 2 \, , \\
        \kappa_{123} = 2 \, , \quad
        \kappa_{133} = 2 \, , \quad
        \kappa_{222} = 35 \, , \quad
        \kappa_{223} = 11 \, , \quad
        \kappa_{233} = 3 \, .
    \end{gathered}
\end{gather}
We present the genus zero GV-invariants from the toric data in Table~\ref{table:P2b8-gv0}, where $ (n_1,n_2,n_3) $ represents the lattice point $ l^1 n_1 + l^2 n_2 + l^3 n_3 $ of the Mori cone generated by $ \{l^1,l^2,l^3,l^4\} $. The Mori cone is non-simplicial, and consequently we need to sum over some negative $ n_1 $ in \eqref{eq:fundamental-period}. 

We compute the numerator in the modular ansatz for the 1-string elliptic genus as
\begin{align}
    \begin{aligned}
        \mathcal{N}_1 &= -\frac{37 E_4^8 \varphi_2(\phi )^8}{6879707136}+\frac{367 E_4^5 E_6^2 \varphi_2(\phi )^8}{20639121408}-\frac{25 E_4^2 E_6^4 \varphi_2(\phi )^8}{1289945088}-\frac{29 E_4^6 E_6 \varphi_2(\phi )^7 \varphi_0(\phi )}{2579890176} \\
        &\quad +\frac{77 E_4^3 E_6^3 \varphi_2(\phi )^7 \varphi_0(\phi )}{2579890176}+\frac{E_6^5 \varphi_2(\phi )^7 \varphi_0(\phi )}{26873856}+\frac{37 E_4^7 \varphi_2(\phi )^6 \varphi_0(\phi )^2}{1719926784} \\
        &\quad -\frac{391 E_4^4 E_6^2 \varphi_2(\phi )^6 \varphi_0(\phi )^2}{5159780352}-\frac{91 E_4 E_6^4 \varphi_2(\phi )^6 \varphi_0(\phi )^2}{644972544}+\frac{101 E_4^5 E_6 \varphi_2(\phi )^5 \varphi_0(\phi )^3}{2579890176} \\
        &\quad +\frac{907 E_4^2 E_6^3 \varphi_2(\phi )^5 \varphi_0(\phi )^3}{2579890176}-\frac{325 E_4^6 \varphi_2(\phi )^4 \varphi_0(\phi )^4}{10319560704}-\frac{1345 E_4^3 E_6^2 \varphi_2(\phi )^4 \varphi_0(\phi )^4}{3439853568} \\
        &\quad -\frac{85 E_6^4 \varphi_2(\phi )^4 \varphi_0(\phi )^4}{1289945088}+\frac{661 E_4^4 E_6 \varphi_2(\phi )^3 \varphi_0(\phi )^5}{2579890176}+\frac{347 E_4 E_6^3 \varphi_2(\phi )^3 \varphi_0(\phi )^5}{2579890176} \\
        &\quad -\frac{281 E_4^5 \varphi_2(\phi )^2 \varphi_0(\phi )^6}{5159780352}-\frac{727 E_4^2 E_6^2 \varphi_2(\phi )^2 \varphi_0(\phi )^6}{5159780352}+\frac{115 E_4^3 E_6 \varphi_2(\phi ) \varphi_0(\phi )^7}{2579890176} \\
        &\quad +\frac{29 E_6^3 \varphi_2(\phi ) \varphi_0(\phi )^7}{2579890176}-\frac{31 E_4^4 \varphi_0(\phi )^8}{20639121408}-\frac{113 E_4 E_6^2 \varphi_0(\phi )^8}{20639121408} \, .
    \end{aligned}
\end{align}
The 2-string result is given in Appendix~\ref{appendix:ellP2-gauge-result}.

\begin{table}[t]
    \scriptsize
    \centering
    \begin{tabular}{c|c|cccc|cccc|cccc}
        & & \multicolumn{12}{c}{$ n_2-n_1 $} \\ \hline
        & & $ 0 $ & $ 1 $ & $ 2 $ & $ 3 $ & $ 0 $ & $ 1 $ & $ 2 $ & $ 3 $ & $ 0 $ & $ 1 $ & $ 2 $ & $ 3 $ \\ \hline
        \parbox[t]{1.5ex}{\multirow{10}{*}{\rotatebox[origin=c]{90}{$ n_1 $}}} & $ -2 $ & & & & & & & & $ -4 $ &  \\
        & $ -1 $ & & & & & & & & $ 256 $ &  \\
        & $ 0 $ & & $ 40 $ & & & $ 3 $ & $ -80 $ & $ 780 $ & $ 46224 $ & $ -6 $ & $ 200 $ & $ -3120 $ & $ 29640 $  \\
        & $ 1 $ & $ 128 $ & $ 128 $ & & & & $ -256 $ & $ 5120 $ & $ 572672 $ & & $ 640 $ & $ -20480 $ & $ 299520 $  \\
        & $ 2 $ & $ 40 $ & $ 204 $ & $ 40 $ & & & $ -408 $ & $ 16128 $ & $ 3211024 $ & & $ 1020 $ & $ -64592 $ & $ 1433520 $ \\
        & $ 3 $ & & $ 128 $ & $ 128 $ & & & $ -256 $ & $ 30720 $ & $ 10856448 $ & & $ 640 $ & $ -123136 $ & $ 4334208 $ \\
        & $ 4 $ & & $ 40 $ & $ 204 $ & $ 40 $ & & $ -80 $ & $ 37874 $ & $ 24626000 $ & & $ 200 $ & $ -151904 $ & $ 9127368 $ \\
        & $ 5 $ & & & $ 128 $ & $ 128 $ & & & $ 30720 $ & $ 39596032 $ & & & $ -123136 $ & $ 14073984 $ \\
        & $ 6 $ & & & $ 40 $ & $ 204 $ & & & $ 16128 $ & $ 46253880 $ & & & $ -64592 $ & $ 16214040 $ \\
        & $ 7 $ & & & & $ 128 $ & & & $ 2592 $ & $ 39596032 $ & & & $ -20480 $ & $ 14073984 $ \\ \hline
        & & \multicolumn{4}{c|}{$ n_3=0 $} & \multicolumn{4}{c|}{$ n_3=1 $} & \multicolumn{4}{c}{$ n_3=2 $}
    \end{tabular}
    \caption{Genus zero GV-invariants for $ d=8 $} \label{table:P2b8-gv0}
\end{table}

\subsubsection{\texorpdfstring{$ B=\mathbb{P}^2 $ with $ SU(3) $}{B=P2 with SU(3)} gauge symmetry} \label{subsubsec:P2SU3}

We now consider the 6d supergravity theories with $SU(3)$ gauge symmetry realized by elliptic 3-folds over a base $ B=\mathbb{P}^2 $ with $ SU(3) $ gauge algebra living on a degree 1 curve $\ell$ in the base. This theory is anomaly free with 24 fundamental hypermultiplets and
\begin{align}
    a = -3 \, , \quad
    b = 1 \, , \quad
    \Omega = 1 \, ,
\end{align}
together with $ 209 $ neutral hypermultiplets. To compute the GV-invariants of this theory, we consider a Calabi-Yau hypersurface constructed by the following toric data:
\begin{align}\label{eq:ellP2-SU3-toric}
    \begin{array}{rrrr|rrrr}
        \multicolumn{4}{c|}{\nabla} & l^1 & l^2 & l^3 & l^4 \\ \hline
        0 & 0 & 0 & -1 & 1 & 0 & 0 & 0  \\
        0 & 0 & -1 & 0 & 0 & 1 & 0 & 0  \\
        1 & 0 & 2 & 3 & 0 & 1 & -2 & 1 \\
        1 & 0 & 1 & 2 & 1 & -2 & 1 & 0  \\
        1 & 0 & 1 & 1 & -1 & 1 & 1 & 0  \\
        0 & 1 & 2 & 3 & 0 & 0 & 0 & 1  \\
        -1 & -1 & 2 & 3 & 0 & 0 & 0 & 1  \\
        0 & 0 & 2 & 3 & 0 & 0 & 1 & -3
    \end{array}
\end{align}
This toric data defines an elliptically fibered Calabi-Yau threefold with hodge numbers $ (h^{1,1},h^{2,1}) = (4,208) $. We can construct the Weierstrass form of the elliptic fibration using same method in the previous subsection. One then finds a Kodaira type $ I_3 $ singularity over the degree 1 curve of $ \mathbb{P}^2 $. The non-zero triple intersection numbers of divisors dual to the Mori cone generators are
\begin{gather}
    \begin{gathered}
        \kappa_{111} = 147\, , \ 
        \kappa_{112} = 105\, , \ 
        \kappa_{113} = 63\, , \ 
        \kappa_{114} = 21\, , \ 
        \kappa_{122} = 75\, , \
        \kappa_{123} = 45\, , \\
        \kappa_{124} = 15\, , \ 
        \kappa_{133} = 27\, , \ 
        \kappa_{134} = 9\, , \ 
        \kappa_{144} = 3\, , \ 
        \kappa_{222} = 50\, , \ 
        \kappa_{223} = 30\, , \\
        \kappa_{224} = 10\, , \ 
        \kappa_{233} = 18\, , \ 
        \kappa_{234} = 6\, , \ 
        \kappa_{244} = 2\, , \ 
        \kappa_{333} = 9\, , \ 
        \kappa_{334} = 3\, , \ 
        \kappa_{344} = 1 \, .
    \end{gathered}
\end{gather}

Let us identify $ l^4 $ as the curve class $\ell$ in $ \mathbb{P}^2 $ and the K\"ahler parameters associated with $ \{l^1,l^2,l^3\} $ as $ \{\phi_1-\phi_2, -\phi_1+2\phi_2, \tau-\phi_1-\phi_2\} $, where $ \phi_1 $ and $ \phi_2 $ are holonomies for the $ SU(3) $ gauge symmetry. 
We can compute the genus zero GV-invariants of this theory by using the Mori cone generators and triple intersection numbers. The result agrees with the perturbative partition function of the $SU(3)$ gauge theory given by
\begin{align}
    \begin{aligned}
        Z_{\mathrm{pert}} &= \PE\bigg[ \frac{3y(y+y^{-1})}{(1-y)^2} \frac{1}{1-q} + \frac{2y}{(1-y)^2} \Big( \sum_{\alpha \in \Delta^+} \big(e^{2\pi i \alpha \cdot \phi} + q e^{-2\pi i \alpha \cdot \phi}\big) + 2\Big) \frac{1}{1-q} \\
        &\qquad \qquad - \frac{y}{(1-y)^2} \Big(12 \sum_{w \in \mathbf{F}} e^{2\pi i w \cdot \phi} + 12\sum_{w \in \overline{\mathbf{F}}} e^{2\pi i w\cdot\phi} + 209\Big) \frac{1+q}{1-q} \bigg] \, ,
    \end{aligned}
\end{align}
where the first line represents the gravity and vector multiplet contributions and the second line comes from the fundamental and neutral hypermultiplets. Here, $ \Delta^+ $ is the set of positive roots of $ SU(3) $, and $ \mathbf{F} $, $ \overline{\mathbf{F}} $ denote fundamental and antifundamental representations.

Next, let us look at the BPS strings and their elliptic genera. The central charges $ c_L $, $ c_R $ and the level $ k_l $ in the worldsheet CFT are the same as those given in \eqref{eq:P2-su2-central} for the string with charge $ Q=n $, and the level for the $ SU(3) $ gauge symmetry is $ k_{SU(3)} = n $. The modular ansatz for the elliptic genus of this string is
\begin{align}
    Z_n(\tau, \epsilon, \phi_1, \phi_2) = \frac{1}{\eta(\tau)^{36n}} \frac{\mathcal{N}_n(\tau, \epsilon, \phi_1, \phi_2)}{\prod_{s=1}^n \varphi_2(\tau, s\epsilon) \prod_{e \in \Delta^+} \prod_{l=0}^{s-1} \varphi_2(\tau, (s-1-2 l)\epsilon+e\cdot\phi)} \, .
\end{align}
The numerator $ \mathcal{N}_n $ is generated by the Eisenstein series $ E_4(\tau) $, $ E_6(\tau) $, the weak Jacobi forms $ \varphi_2(\tau,\epsilon) $, $ \varphi_0(\tau,\epsilon) $ for the $ SU(2)_l $ symmetry, and the $ SU(3) $ Weyl invariant Jacobi forms $ \varphi_0^{A_2}(\tau,\phi_1,\phi_2) $, $ \varphi_2^{A_2}(\tau,\phi_1,\phi_2) $, $ \varphi_3^{A_2}(\tau,\phi_1,\phi_2) $ defined in Appendix~\ref{appendix:modular}. The Jacobi forms $ \varphi_k^{A_2} $ have weight $ -k $ and index $ 1 $ for the  $ SU(3) $ symmetry. We fix the numerator for 1-string in this ansatz as
\begin{align}
    \mathcal{N}_1 &= -\frac{17 E_4^6 E_6 (\varphi^{A_2}_3)^6 \varphi^{A_2}_2}{2654208}+\frac{E_4^3 E_6^3 (\varphi^{A_2}_3)^6 \varphi^{A_2}_2}{221184}+\frac{5 E_6^5 (\varphi^{A_2}_3)^6  \varphi^{A_2}_2}{2654208} \\
    &\ -\frac{85 E_4^7 (\varphi^{A_2}_3)^4  (\varphi^{A_2}_2)^3}{23887872} -\frac{977 E_4^4 E_6^2 (\varphi^{A_2}_3)^4  (\varphi^{A_2}_2)^3}{11943936} -\frac{553 E_4 E_6^4 (\varphi^{A_2}_3)^4  (\varphi^{A_2}_2)^3}{23887872} \nonumber \\
    &\ -\frac{325 E_4^5 E_6 (\varphi^{A_2}_3)^2  (\varphi^{A_2}_2)^5}{3981312} -\frac{107 E_4^2 E_6^3 (\varphi^{A_2}_3)^2  (\varphi^{A_2}_2)^5}{3981312}-\frac{E_4^6 (\varphi^{A_2}_2)^7}{23328}-\frac{325 E_4^3 E_6^2 (\varphi^{A_2}_2)^7}{373248} \nonumber \\
    &\ -\frac{91 E_6^4 (\varphi^{A_2}_2)^7}{373248} -\frac{7 E_4^7 (\varphi^{A_2}_3)^6 \varphi^{A_2}_0}{1990656}-\frac{19 E_4^4 E_6^2  (\varphi^{A_2}_3)^6 \varphi^{A_2}_0}{1990656}+\frac{13 E_4 E_6^4 (\varphi^{A_2}_3)^6 \varphi^{A_2}_0}{995328} \nonumber \\
    &\ -\frac{43 E_4^5 E_6 (\varphi^{A_2}_3)^4 (\varphi^{A_2}_2)^2 \varphi^{A_2}_0}{110592} -\frac{29 E_4^2 E_6^3 (\varphi^{A_2}_3)^4 (\varphi^{A_2}_2)^2 \varphi^{A_2}_0}{110592}-\frac{595 E_4^6 (\varphi^{A_2}_3)^2 (\varphi^{A_2}_2)^4 \varphi^{A_2}_0}{5971968} \nonumber \\
    &\ -\frac{8869 E_4^3 E_6^2 (\varphi^{A_2}_3)^2 (\varphi^{A_2}_2)^4 \varphi^{A_2}_0}{5971968}-\frac{275 E_6^4 (\varphi^{A_2}_3)^2 (\varphi^{A_2}_2)^4 \varphi^{A_2}_0}{746496} -\frac{137 E_4^4 E_6 (\varphi^{A_2}_2)^6 \varphi^{A_2}_0}{41472} \nonumber \\
    &\ -\frac{103 E_4 E_6^3 (\varphi^{A_2}_2)^6 \varphi^{A_2}_0}{41472}-\frac{41 E_4^6 (\varphi^{A_2}_3)^4 \varphi^{A_2}_2 (\varphi^{A_2}_0)^2}{221184}-\frac{113 E_4^3 E_6^2 (\varphi^{A_2}_3)^4 \varphi^{A_2}_2 (\varphi^{A_2}_0)^2}{110592} \nonumber \\
    &\ -\frac{7 E_6^4 (\varphi^{A_2}_3)^4  \varphi^{A_2}_2 (\varphi^{A_2}_0)^2}{73728} -\frac{2123 E_4^4 E_6 (\varphi^{A_2}_3)^2  (\varphi^{A_2}_2)^3 (\varphi^{A_2}_0)^2}{497664} \nonumber \\
    &\ -\frac{1765 E_4 E_6^3 (\varphi^{A_2}_3)^2  (\varphi^{A_2}_2)^3 (\varphi^{A_2}_0)^2}{497664}-\frac{89 E_4^5 (\varphi^{A_2}_2)^5 (\varphi^{A_2}_0)^2}{62208} -\frac{343 E_4^2 E_6^2 (\varphi^{A_2}_2)^5 (\varphi^{A_2}_0)^2}{62208} \nonumber \\
    &\ -\frac{17 E_4^4 E_6 (\varphi^{A_2}_3)^4 (\varphi^{A_2}_0)^3}{82944}-\frac{55 E_4 E_6^3 (\varphi^{A_2}_3)^4 (\varphi^{A_2}_0)^3}{82944}-\frac{149 E_4^5  (\varphi^{A_2}_3)^2 (\varphi^{A_2}_2)^2 (\varphi^{A_2}_0)^3}{82944} \nonumber \\
    &\ -\frac{571 E_4^2 E_6^2  (\varphi^{A_2}_3)^2 (\varphi^{A_2}_2)^2 (\varphi^{A_2}_0)^3}{82944} +\frac{115 E_4^3 E_6  (\varphi^{A_2}_2)^4 (\varphi^{A_2}_0)^3}{31104}+\frac{29 E_6^3 (\varphi^{A_2}_2)^4 (\varphi^{A_2}_0)^3}{31104} \nonumber \\
    &\ +\frac{115 E_4^3 E_6 (\varphi^{A_2}_3)^2 \varphi^{A_2}_2 (\varphi^{A_2}_0)^4}{27648} +\frac{29 E_6^3 (\varphi^{A_2}_3)^2 \varphi^{A_2}_2 (\varphi^{A_2}_0)^4}{27648}+\frac{31 E_4^4 (\varphi^{A_2}_2)^3 (\varphi^{A_2}_0)^4}{15552} \nonumber \\
    &\ +\frac{113 E_4 E_6^2 (\varphi^{A_2}_2)^3 (\varphi^{A_2}_0)^4}{15552}+\frac{31 E_4^4  (\varphi^{A_2}_3)^2 (\varphi^{A_2}_0)^5}{13824}+\frac{113 E_4 E_6^2 (\varphi^{A_2}_3)^2 (\varphi^{A_2}_0)^5}{13824} \, , \nonumber
\end{align}
by comparing the ansatz against the genus zero GV-invariants from the mirror computation.

\subsubsection{\texorpdfstring{$ B=\mathbb{F}_1 $ with $ SU(2) $}{B=F1 with SU(2)} gauge symmetry} \label{subsubsec:F1SU2}

The supergravity theories with $T=1$ can be engineered in F-theory compactifications on elliptic CY 3-folds over a  Hirzebruch surface $ \mathbb{F}_n $. We shall discuss the $SU(2)$ gauge theories with only fundamental and adjoint hypermultiplets arising from $\mathbb{F}_1$ base. There are two independent curve classes in $ \mathbb{F}_1 $, which are the base curve $ e $ and the fiber curve $ f $ satisfying $ e^2=-1 $ and $ f^2=0 $. We label tensor charge $ Q $ of the BPS string wrapping $ e $ and $ f $ as $ (0,1) $ and $ (1,-1) $, respectively. Now, consider the $ SU(2) $ gauge symmetry supported on a $ b=(d,0) $ curve whose self-intersection number is $ d^2 $. The low energy gravity theory is anomaly free with
\begin{align}
    a = (-3,1) \, , \
    b = (d,0) \, , \
    \Omega = \operatorname{diag}(1,-1) \, , \
    n_{\mathbf{3}} = \frac{(d-1)(d-2)}{2} \, , \
    n_{\mathbf{2}} = 2d(12-d) \, .
\end{align}
The numbers $ n_{\mathbf{3}},  n_{\mathbf{2}}$ are consistent with \eqref{F1_3} and \eqref{F1_2} respectively.

The perturbative part of the partition function of this theory is given by
\begin{align}
    Z_{\mathrm{pert}} &= \PE\bigg[ \frac{3y(y+y^{-1})}{(1-y)^2} \frac{1}{1-q} + \frac{y(y+y^{-1})}{(1-y)^2} \frac{1}{1-q} + \frac{2y}{(1-y)^2} \Big(e^{4\pi i \phi} + 1 + q e^{-4\pi i \phi}\Big) \frac{1}{1-q} \nonumber \\
    &\qquad - \frac{y}{(1-y)^2} \Big( n_{\mathbf{3}} \big(e^{4\pi i \phi} + 1 + e^{-4\pi i \phi}\big) + n_{\mathbf{2}} \big(e^{2\pi i \phi} + e^{-2\pi i \phi}\big) + n_{\mathbf{1}}\Big) \frac{1+q}{1-q} \bigg]  \, ,
\end{align}
where $ \phi $ is the holonomy for the $ SU(2) $ gauge symmetry.
Here, the first line involves the contributions from the gravity, tensor and vector multiplets, respectively. The second line corresponds to the contributions from the adjoint and the fundamental hypermultiplets.

The theories with $ d \leq 6 $ can be constructed by Calabi-Yau hypersurfaces with the toric data given by
\begin{align}\label{eq:ellF1-gauge-toric}
    \begin{array}{rrrr|rrrrrr}
        \multicolumn{4}{c|}{\nabla} & l^1 & l^2 & l^3 & l^4 & \tilde{l}^4 & l^5 \\ \hline
        0 & 0 & 0 & -1 & 0 & 1 & 0 & 0 & 0 & 1 \\ 
        0 & 0 & -1 & 0 & 0 & 0 & 1 & 0 & -2 & -1 \\
        1 & 0 & 1 & 2 & 0 & 0 & 0 & 1 & 1 & 1 \\
        0 & 1 & 2 & 3 & 1 & 0 & 0 & 0 & 0 & 0 \\
        -1 & 0 & 3-d & 4-d & -1 & 0 & 0 & 1 & 1 & 1 \\
        -1 & -1 & 3-d & 4-d & 1 & 0 & 0 & 0 & 0 & 0 \\
        0 & 0 & 2 & 3 & -1 & 1 & -1 & -2 & 0 & 0 \\
        0 & 0 & 1 & 1 & 0 & -2 & 3 & d & 0 & 0
    \end{array}
\end{align}
The lattice polytope $ \nabla $ for $ d=0 $ describes the elliptic fibration over $ \mathbb{F}_1 $ without gauge symmetry as discussed in \cite{Klemm:2012sx}. The 3-fold in this case has the hodge numbers $ (h^{1,1},h^{2,1})=(3,243) $ and the Euler characteristic $ \chi=-480 $. The Mori cone of $ 1 \leq d \leq 4 $ cases are generated by $ \{l^1,l^2,l^3,l^4\} $ as studied in \cite{Cota:2020zse}. When $ d=5 $, the Mori cone becomes non-simplicial and is generated by $ \{l^1,l^2,l^3,l^4,l^5\} $. For the Calabi-Yau hypersurface with $ d=6 $, the Mori-cone is simplicial but is generated by $ \{l^1,l^2,l^3,\tilde{l}^4\} $. We can compute the genus zero GV-invariants from the toric data together with the triple intersection numbers of divisors which we discuss below.

The BPS string with tensor charge $ Q=(k_1,k_2) $ in these theories have the central charges as
\begin{alignat}{2}\label{eq:F1-su2-central}
    \begin{aligned}
        &c_L = k_1^2 - k_2^2 + 9(3k_1+k_2) + 6 \, , \quad
        &&c_R = k_1^2 - k_2^2 + 3(3k_1+k_2) + 6 \, , \\
        &k_l = \frac{1}{2}(k_1^2 + k_2^2 - 3k_1 - k_2) \, , \quad
        &&k_{SU(2)} = k_1 d \, .
    \end{aligned}
\end{alignat}
From this, we can set the modular ansatz for the elliptic genera of the strings as
\begin{align}\label{eq:ellF1-ansatz}
    \begin{aligned}
        Z_{(k_1,k_2)} &= \frac{1}{\eta(\tau)^{12(3k_1+k_2)}} \frac{\mathcal{N}_{(k_1,k_2)}(\tau, \epsilon, \phi)}{\prod_{s_1=1}^{k_1} \varphi_2(\tau, s_1 \epsilon) \cdot \prod_{s_2=1}^{k_1+k_2} \varphi_2(\tau, s_2\epsilon)} \\
        &\qquad \cdot \frac{1}{\prod_{s_1=1}^{\kappa} \prod_{l=0}^{s_1-1} \varphi_2(\tau, (s_1-1-2l)\epsilon+2\phi)} \, ,
    \end{aligned}
\end{align}
with $k_1\ge 0$, $k_2\ge -k_1$, and $\kappa= \lfloor {\rm min}(k_1,k_1+k_2)/d \rfloor$.

Before presenting the results for each $ d $, we discuss some general properties of the elliptic genus \eqref{eq:ellF1-ansatz}. The strings with charge $ Q=(0,k) $ are not charged under the $ SU(2) $ gauge symmetry. Their elliptic genera are same with the (massless) E-string theory whose modular ansatz is studied in \cite{Gu:2017ccq}. For instance, there is only one unknown coefficient in $ Q=(0,1) $ modular ansatz, and it is determined by providing one genus zero GV-invariants:
\begin{align}
    \mathcal{N}_{(0,1)} = -E_4 \quad \text{for every } d \, .
\end{align}
On the other hand, the elliptic genera of strings with charge $ Q=(k,-k) $, which are D3-branes wrapping a fiber curve of the Hirzebruch surface multiple times, have an index $ k $ for  the $ SU(2) $ gauge symmetry. Notably, once  the elliptic genus $ Z_{(1,-1)} $ is determined, the higher string elliptic genus with charge $ Q=(k,-k) $ for $ k>1 $ are generated by the Hecke transformation of $ Z_{(1,-1)} $ as follows:
\begin{align}
    1 + \sum_{k=1}^\infty w^k Z_{(k,-k)}(\tau, \epsilon, \phi) = \exp(\sum_{k=1}^\infty \frac{w^k}{k} \sum_{ad=k} \sum_{b(\operatorname{mod}d)} Z_{(1,-1)}\Big(\frac{a\tau+b}{d}, a\epsilon, a\phi\Big) ) \, ,
\end{align}
where $ w $ is the fugacity for the winding numbers of the strings. This indicates that these strings do not form a bound state at higher string number. In fact, they are the heterotic strings. We will discuss this property in more details in the next section. 

The elliptic genus $ Z_{(1,-1)} $ has been calculated for $ 1 \leq d \leq 4 $ in \cite{Cota:2020zse}. In the remaining part of this section, we review these calculations and discuss extensions for $ d=5,6 $ cases.

\paragraph{$ d=1 $ case} This theory contains 22 fundamental and 203 neutral hypermultiplets. The Calabi-Yau threefold has $ (h^{1,1},h^{2,1},\chi) = (4,202,-396) $. The non-zero triple intersection numbers of divisors dual to the four Mori cone generators are
\begin{gather}
    \begin{gathered}
        \kappa_{122} = 16 \, , \
        \kappa_{123} = 10 \, , \
        \kappa_{124} = 3 \, , \
        \kappa_{133} = 6 \, , \
        \kappa_{134} = 2 \, , \
        \kappa_{222} = 176 \, , \
        \kappa_{223} = 110 \, , \\
        \kappa_{224} = 25 \, , \
        \kappa_{233} = 68 \, , \
        \kappa_{234} = 16 \, , \
        \kappa_{244} = 3 \, , \
        \kappa_{333} = 42 \, , \
        \kappa_{334} = 10 \, , \
        \kappa_{344} = 2 \, .
    \end{gathered}
\end{gather}
Using the mirror results, one can compute the elliptic genus of the string with charge $ Q=(1,-1) $ with the numerator in the ansatz as
\begin{align}
    \mathcal{N}_{(1,-1)} &= -\frac{13}{144} E_4^3 \varphi_2(\phi )-\frac{11}{144} E_6^2 \varphi_2(\phi )+\frac{1}{6} E_4 E_6 \varphi_0(\phi ) \, .
\end{align}
The elliptic genus for $ Q=(1,0) $ is given in Appendix~\ref{appendix:F1-SU2}.

\paragraph{$ d=2 $ case} This theory has 40 fundamental and 167 neutral hypermultiplets. The topological numbers of the Calabi-Yau 3-fold are $ (h^{1,1},h^{2,1},\chi) = (4,166,324) $. The non-zero triple intersection numbers of divisors dual to Mori cone generators are
\begin{gather}
    \begin{gathered}
        \kappa_{122} = 14 \, ,\
        \kappa_{123} = 8 \, ,\
        \kappa_{124} = 3 \, ,\
        \kappa_{133} = 4 \, ,\
        \kappa_{134} = 2 \, ,\
        \kappa_{222} = 140 \, ,\
        \kappa_{223} = 80 \, ,\\
        \kappa_{224} = 23 \, ,\
        \kappa_{233} = 44 \, ,\
        \kappa_{234} = 14 \, ,\
        \kappa_{244} = 3 \, ,\
        \kappa_{333} = 24 \, ,\
        \kappa_{334} = 8 \, ,\
        \kappa_{344} = 2 \, .
    \end{gathered}
\end{gather}
Using the mirror results, one can compute the elliptic genus of the string with charge $ Q=(1,-1) $ with the numerator in the ansatz as
\begin{align}
    \mathcal{N}_{(1,-1)} &= \frac{E_4^2 E_6 \varphi_2(\phi )^2}{72} -\frac{13E_4^3 \varphi_2(\phi ) \varphi_0(\phi )}{864} -\frac{11E_6^2 \varphi_2(\phi ) \varphi_0(\phi )}{864} +\frac{E_4 E_6 \varphi_0(\phi )^2}{72} \, .
\end{align}
The result for $ Q=(1,0) $ is given in Appendix~\ref{appendix:F1-SU2}.

\paragraph{$ d=3 $ case} This theory has 1 adjoint, 54 fundamental and 136 neutral hypermultiplets. The hodge numbers and Euler characteristics are $ (h^{1,1},h^{2,1},\chi) = (4,136,-264) $. The non-trivial triple intersection numbers of divisors dual to the Mori cone generators are
\begin{gather}
    \begin{gathered}
        \kappa_{122} = 12 \, ,\ 
        \kappa_{123} = 6 \, ,\ 
        \kappa_{124} = 3 \, ,\ 
        \kappa_{133} = 2 \, ,\ 
        \kappa_{134} = 2 \, ,\ 
        \kappa_{222} = 108 \, ,\ 
        \kappa_{223} = 54 \, ,\\
        \kappa_{224} = 21 \, ,\ 
        \kappa_{233} = 24 \, ,\ 
        \kappa_{234} = 12 \, ,\ 
        \kappa_{244} = 3 \, ,\ 
        \kappa_{333} = 10 \, ,\ 
        \kappa_{334} = 6 \, ,\ 
        \kappa_{344} = 2 \, .
    \end{gathered}
\end{gather}
Using the mirror results, one can compute the elliptic genus of the string with charge $ Q=(1,-1) $ with the numerator in the ansatz as
\begin{align}
    \begin{aligned}
        \mathcal{N}_{(1,-1)} &= -\frac{E_4^4 \varphi_2(\phi )^3}{2304}-\frac{5 E_4 E_6^2 \varphi_2(\phi )^3}{6912}+\frac{E_4^2 E_6 \varphi_2(\phi )^2 \varphi_0(\phi )}{288} \\
        &\quad -\frac{13 E_4^3 \varphi_2(\phi ) \varphi_0(\phi )^2}{6912}-\frac{11 E_6^2 \varphi_2(\phi ) \varphi_0(\phi )^2}{6912}+\frac{1}{864} E_4 E_6 \varphi_0(\phi )^3 \, .
    \end{aligned}
\end{align}
The result for $ Q=(1,0) $ string is given in Appendix~\ref{appendix:F1-SU2}.

\paragraph{$ d=4 $ case} This theory has 3 adjoint, 64 fundamental and 110 neutral hypermultiplets. The hodge numbers and Euler characteristics of the 3-fold are $ (h^{1,1},h^{2,1},\chi)=(4,112,-216) $. The triple intersection numbers of divisors dual to the Mori cone generators are
\begin{gather}
    \begin{gathered}
        \kappa_{122} = 10 \, ,\
        \kappa_{123} = 4 \, ,\
        \kappa_{124} = 3 \, ,\
        \kappa_{134} = 2 \, ,\
        \kappa_{222} = 80 \, ,\
        \kappa_{223} = 32 \, ,\\
        \kappa_{224} = 19 \, ,\
        \kappa_{233} = 8 \, ,\
        \kappa_{234} = 10 \, ,\
        \kappa_{244} = 3 \, ,\
        \kappa_{334} = 4 \, ,\
        \kappa_{344} = 2 \, .
    \end{gathered}
\end{gather}
where the others are zero. Using the mirror results, one can compute the elliptic genus of the string with charge $ Q=(1,-1) $ with the numerator in the ansatz as
\begin{align}
    \mathcal{N}_{(1,-1)} &= \frac{E_4^3 E_6 \varphi_2(\phi )^4}{15552}+\frac{E_6^3 \varphi_2(\phi )^4}{31104}-\frac{E_4^4 \varphi_2(\phi )^3 \varphi_0(\phi )}{6912}-\frac{5 E_4 E_6^2 \varphi_2(\phi )^3 \varphi_0(\phi )}{20736} \\
    &\quad +\frac{E_4^2 E_6 \varphi_2(\phi )^2 \varphi_0(\phi )^2}{1728}-\frac{13 E_4^3 \varphi_2(\phi ) \varphi_0(\phi )^3}{62208}-\frac{11 E_6^2 \varphi_2(\phi ) \varphi_0(\phi )^3}{62208}+\frac{E_4 E_6 \varphi_0(\phi )^4}{10368} \, . \nonumber
\end{align}
The result for $ Q=(1,0) $ string is given in Appendix~\ref{appendix:F1-SU2}.

\paragraph{$ d=5 $ case} This theory contains 6 adjoint, 70 fundamental and 89 neutral hypermultiplets. The hodge numbers and Euler characteristics of the associated 3-fold are $ (h^{1,1},h^{2,1}) = (4,94,-180) $. The Mori cone is non-simplicial four-dimensional cone with 5 generators satisfying $ l^5 = l^2 - l^3 + l^4 $. The non-zero triple intersection numbers of divisors dual to four Mori cone curves associated to $ l^{1,2,3,4} $ are
\begin{gather}
    \begin{gathered}
        \kappa_{122} = 8 \, ,\ 
        \kappa_{123} = 2 \, ,\ 
        \kappa_{124} = 3 \, ,\ 
        \kappa_{133} = -2 \, ,\ 
        \kappa_{134} = 2 \, ,\ 
        \kappa_{222} = 56 \, ,\ 
        \kappa_{223} = 14 \, ,\\
        \kappa_{224} = 17 \, ,\ 
        \kappa_{233} = -4 \, ,\ 
        \kappa_{234} = 8 \, ,\ 
        \kappa_{244} = 3 \, ,\ 
        \kappa_{333} = -6 \, ,\ 
        \kappa_{334} = 2 \, ,\ 
        \kappa_{344} = 2 \, .
    \end{gathered}
\end{gather}
We provide some leading genus zero GV-invariants from the non-simplicial Mori cone in Table~\ref{table:F1b5-gv0}, where $ (n_1,n_2,n_3,n_4) $ labels the lattice point $ \sum_{k=1}^4 l^k n_k $ of the Mori cone. Here, we need to consider negative values of $ n_3 $ since the Mori cone is non-simplicial.
\begin{table}[t]
    \scriptsize
    \centering
    \begin{tabular}{c|c|cccc|cccc|cccc}
        & & \multicolumn{12}{c}{$ n_2-n_3 $} \\ \hline
        & & $ 0 $ & $ 1 $ & $ 2 $ & $ 3 $ & $ 0 $ & $ 1 $ & $ 2 $ & $ 3 $ & $ 0 $ & $ 1 $ & $ 2 $ & $ 3 $ \\ \hline
        \parbox[t]{1.5ex}{\multirow{7}{*}{\rotatebox[origin=c]{90}{$ n_3 $}}} & $ -1 $ & & & & & & & & & & & $ 28 $ &  \\
        & $ 0 $ & & $ 10 $ & & & $ 1 $ & & & & $ -2 $ & $ 10 $ & $ 960 $ & $ 960 $ \\
        & $ 1 $ & $ 140 $ & $ 140 $ & & & & & & & & $ 140 $ & $ 8540 $ & $ 23880 $ \\
        & $ 2 $ & $ 10 $ & $ 180 $ & $ 10 $ & & & $ 252 $ & & & & $ 180 $ &$ 30460 $ & $ 188160 $  \\
        & $ 3 $ & & $ 140 $ & $ 140 $ & & & & & & & $ 140 $ & $ 62440 $ & $ 781780 $  \\
        & $ 4 $ & & $ 10 $ & $ 180 $ & $ 10 $ & & & $ 5130 $ & & & $ 10 $ & $ 78032 $ & $ 2009520 $ \\
        & $ 5 $ & & & $ 140 $ & $ 140 $ & & & & & & & $ 62440 $ & $ 3459780 $  \\ \hline
        & & \multicolumn{4}{c|}{$ (n_1,n_4)=(0,0) $} & \multicolumn{4}{c|}{$ (n_1,n_4)=(1,0) $} & \multicolumn{4}{c}{$ (n_1,n_4)=(0,1) $} \\ \hline \hline
        & & \multicolumn{12}{c}{$ n_2-n_3 $} \\ \hline
        & & $ 0 $ & $ 1 $ & $ 2 $ & $ 3 $ & $ 0 $ & $ 1 $ & $ 2 $ & $ 3 $ & $ 0 $ & $ 1 $ & $ 2 $ & $ 3 $ \\ \hline
        \parbox[t]{1.5ex}{\multirow{6}{*}{\rotatebox[origin=c]{90}{$ n_3 $}}}& $ 0 $ & $ 3 $ & $ -20 $ & $ 45 $ & $ 45 $ & & & & & & & & $ 960 $ \\
        & $ 1 $ & & $ -280 $ & $ 1400 $ & $ 22224 $ & & & & & & & & $ 23880 $ \\
        & $ 2 $ & & $ -360 $ & $ 11490 $ & $ 573630 $ & & & & & & & $ 10 $ & $ 188160 $ \\
        & $ 3 $ & & $ -280 $ & $ 26040 $ & $ 4876440 $ & & & & & & & $ 140 $ & $ 781780 $ \\
        & $ 4 $ & & $ -20 $ & $ 40220 $ & $ 19325835 $ & & & $ -9252 $ & & & & $ 180 $ & $ 2009520 $ \\
        & $ 5 $ & & & $ 26040 $ & $ 41603400 $ & & & & & & & $ 140 $ & $ 3459780 $ \\ \hline
        & & \multicolumn{4}{c|}{$ (n_1,n_4)=(1,1) $} & \multicolumn{4}{c|}{$ (n_1,n_4)=(2,0) $} & \multicolumn{4}{c}{$ (n_1,n_4)=(0,2) $}
    \end{tabular}
    \caption{Genus zero GV-invariants for $ d=5 $} \label{table:F1b5-gv0}
\end{table}
We identify the Mori cone curves associated with $ l_1 $ and $ l_4 $ as $ e $ and $ f $ curve of the base $ \mathbb{F}_1 $. The numerator in the modular ansatz for $ Q=(1,-1) $ string can be completely fixed by the genus zero free energy as
\begin{align}
    \begin{aligned}
    \mathcal{N}_{(1,-1)} &= \frac{E_4^5 \varphi_2(\phi )^5}{995328}-\frac{E_4^2 E_6^2 \varphi_2(\phi )^5}{110592}+\frac{5 E_4^3 E_6 \varphi_2(\phi )^4 \varphi_0(\phi )}{186624}+\frac{5 E_6^3 \varphi_2(\phi )^4 \varphi_0(\phi )}{373248} \\
    &\quad -\frac{5 E_4^4 \varphi_2(\phi )^3 \varphi_0(\phi )^2}{165888}-\frac{25 E_4 E_6^2 \varphi_2(\phi )^3 \varphi_0(\phi )^2}{497664}+\frac{5 E_4^2 E_6 \varphi_2(\phi )^2 \varphi_0(\phi )^3}{62208} \\
    &\quad -\frac{65 E_4^3 \varphi_2(\phi ) \varphi_0(\phi )^4}{2985984}-\frac{55 E_6^2 \varphi_2(\phi ) \varphi_0(\phi )^4}{2985984}+\frac{E_4 E_6 \varphi_0(\phi )^5}{124416} \, .
    \end{aligned}
\end{align}
The result for $ Q=(1,0) $ string is given in Appendix~\ref{appendix:F1-SU2}.

\paragraph{$ d=6 $ case} This theory has 10 adjoint, 72 fundamental and 73 neutral hypermultiplets. The hodge numbers and Euler characteristics of the CY 3-fold are $ (h^{1,1},h^{2,1},\chi) = (4,82,-156) $. The Mori cone is generated by $ \{l^1,l^2,l^3,\tilde{l}^4\} $ in \eqref{eq:ellF1-gauge-toric}. The non-trivial triple intersection numbers dual to the four Mori cone curves are
\begin{gather}
    \begin{gathered}
        \kappa_{122} = 6 \, ,\ 
        \kappa_{123} = 6 \, ,\ 
        \kappa_{124} = 3 \, ,\ 
        \kappa_{133} = 4 \, ,\ 
        \kappa_{134} = 2 \, ,\ 
        \kappa_{222} = 36 \, ,\ 
        \kappa_{223} = 30 \, ,\\
        \kappa_{224} = 15 \, ,\ 
        \kappa_{233} = 24 \, ,\ 
        \kappa_{234} = 12 \, ,\ 
        \kappa_{244} = 3 \, ,\ 
        \kappa_{333} = 16 \, ,\ 
        \kappa_{334} = 8 \, ,\ 
        \kappa_{344} = 2 \, .
    \end{gathered}
\end{gather}
The genus zero GV-invariants computed from the data are given in Table~\ref{table:F1b6-gv0}. In the table, $ (n_1,n_2,n_3,n_4) $ labels the lattice point $ l^1 n_1 + l^2 n_2 + l^3 n_3 + \tilde{l}^4 n_4 $ of the Mori cone. Note that $ 2l_3+\tilde{l}_4 $ is $ l_4 $ of \eqref{eq:ellF1-gauge-toric} with $ d=6 $, and the 2-cycles associated with $ (l_1,l_4) $ are identified with $ (e,f) $ curve of the base $ \mathbb{F}_1 $. 
\begin{table}[t]
    \scriptsize
    \centering
    \begin{tabular}{c|c|cccc|cccc|cccc}
        & & \multicolumn{12}{c}{$ n_2-n_3+2n_4 $} \\ \hline
        & & $ 0 $ & $ 1 $ & $ 2 $ & $ 3 $ & $ 0 $ & $ 1 $ & $ 2 $ & $ 3 $ & $ 0 $ & $ 1 $ & $ 2 $ & $ 3 $ \\ \hline
        & $ -2 $ & & & & & & & & & & & $ 2 $ \\
        \parbox[t]{1.7ex}{\multirow{7}{*}{\rotatebox[origin=c]{90}{$ n_3-2n_4 $}}} & $ -1 $ & & & & & & & & & & & $ 144 $ & $ 144 $ \\
        & $ 0 $ & & $ 18 $ & & & $ 1 $ & & & & $ -2 $ & $ 18 $ & $ 2142 $ & $ 5660 $ \\
        & $ 1 $ & $ 144 $ & $ 144 $ & & & & & & & & $ 144 $ & $ 11680 $ & $ 59472 $ \\
        & $ 2 $ & $ 18 $ & $ 156 $ & $ 18 $ & & & $ 252 $ & & & & $ 156 $ & $ 32526 $ & $ 298980 $   \\
        & $ 3 $ & & $ 144 $ & $ 144 $ & & & & & & & $ 144 $ & $ 59472 $ & $ 955168 $  \\
        & $ 4 $ & & $ 18 $ & $ 156 $ & $ 18 $ & & & $ 5130 $ & & & $ 18 $ & $ 70956 $ & $ 2068596 $  \\
        & $ 5 $ & & & $ 144 $ & $ 144 $ & & & & & & & $ 59472 $ & $ 3259008 $  \\ \hline
        & & \multicolumn{4}{c|}{$ (n_1,n_4)=(0,0) $} & \multicolumn{4}{c|}{$ (n_1,n_4)=(1,0) $} & \multicolumn{4}{c}{$ (n_1,n_4)=(0,1) $} \\ \hline \hline
        & & \multicolumn{12}{c}{$ n_2-n_3+2n_4 $} \\ \hline
        & & $ 0 $ & $ 1 $ & $ 2 $ & $ 3 $ & $ 0 $ & $ 1 $ & $ 2 $ & $ 3 $ & $ 0 $ & $ 1 $ & $ 2 $ & $ 3 $ \\ \hline
        \parbox[t]{1.7ex}{\multirow{7}{*}{\rotatebox[origin=c]{90}{$ n_3-2n_4 $}}} & $ -1 $ & & & & & & & & & & & & $ 144 $ \\
        & $ 0 $ & $ 3 $ & $ -36 $ & $ 153 $ & $ 2136 $ & & & & & & & & $ 5660 $ \\
        & $ 1 $ & & $ -288 $ & $ 2592 $ & $ 97344 $ & & & & & & & & $ 59472 $ \\
        & $ 2 $ & & $ -312 $ & $ 13032 $ & $ 1227816 $ & & & & & & & $ 18 $ & $ 298980 $ \\
        & $ 3 $ & & $ -288 $ & $ 24480 $ & $ 6817056 $ & & & & & & & $ 144 $ & $ 955168 $ \\
        & $ 4 $ & & $ -36 $ & $ 37656 $ & $ 20989584 $ & & & $ -9252 $ & & & & $ 156 $ & $ 2068596 $ \\
        & $ 5 $ & & & $ 24480 $ & $ 39568608 $ & & & & & & & $ 144 $ & $ 3259008 $ \\ \hline
        & & \multicolumn{4}{c|}{$ (n_1,n_4)=(1,1) $} & \multicolumn{4}{c|}{$ (n_1,n_4)=(2,0) $} & \multicolumn{4}{c}{$ (n_1,n_4)=(0,2) $}
    \end{tabular}
    \caption{Genus zero GV-invariants for $ d=6 $} \label{table:F1b6-gv0}
\end{table}
From this data, we can compute the elliptic genus of the string with charge $ Q=(1,-1) $. The numerator factor in the modular ansatz is 
\begin{align}
    \mathcal{N}_{(1,-1)} &= \frac{E_4 E_6^3 \varphi_2(\phi )^6}{1492992}+\frac{E_4^5 \varphi_2(\phi )^5 \varphi_0(\phi )}{1990656}-\frac{E_4^2 E_6^2 \varphi_2(\phi )^5 \varphi_0(\phi )}{221184}+\frac{5 E_4^3 E_6 \varphi_2(\phi )^4 \varphi_0(\phi )^2}{746496} \nonumber \\
    &\ +\frac{5 E_6^3 \varphi_2(\phi )^4 \varphi_0(\phi )^2}{1492992}-\frac{5 E_4^4 \varphi_2(\phi )^3 \varphi_0(\phi )^3}{995328}-\frac{25 E_4 E_6^2 \varphi_2(\phi )^3 \varphi_0(\phi )^3}{2985984} \\
    &\ +\frac{5 E_4^2 E_6 \varphi_2(\phi )^2 \varphi_0(\phi )^4}{497664}-\frac{13 E_4^3 \varphi_2(\phi ) \varphi_0(\phi )^5}{5971968}-\frac{11 E_6^2 \varphi_2(\phi ) \varphi_0(\phi )^5}{5971968}+\frac{E_4 E_6 \varphi_0(\phi )^6}{1492992} \, . \nonumber
\end{align}
The result  for $ Q=(1,0) $ string is given in Appendix~\ref{appendix:F1-SU2}.

\subsubsection{\texorpdfstring{$ B=\mathbb{F}_1 $ with $ SU(2)\times SU(2) $}{B=F1 with SU(2)×SU(2)} gauge symmetry} \label{subsubsec:F1-SU2SU2}

Another interesting example is the supergravity with $T=1$ and $SU(2)^2$ gauge symmetry arising from F-theory compactified on an elliptic 3-fold over a base $ B = \mathbb{F}_1 $ with two $ SU(2) $ gauge algebras living on $ e $ and $ f $ curves in the base, respectively. This theory is anomaly free with 8 fundamentals for the first $SU(2)$, 14 fundamentals for the second $SU(2)$, one bi-fundamental, and 202 neutral hypermultiplets, together with
\begin{align}
    a = (-3,1) \, , \
    b_1 = (0,1) \, , \
    b_2 = (1,-1) \, , \
    \Omega = \operatorname{diag}(1,-1) \, .
\end{align}

The perturbative partition function of this theory is given by
\begin{align}
    Z_{\mathrm{pert}} &= \PE\bigg[\frac{4y(y+y^{-1})}{(1-y)^2} \frac{1}{1-q} + \frac{2y}{(1-y)^2} \Big( e^{2\pi i (2\phi_1+2\phi_2)} + 2 + q e^{-2\pi i(2\phi_1+2\phi_2)} \Big) \frac{1}{1-q} \nonumber \\
    &\qquad- \frac{y}{(1-y)^2} \Big( e^{2\pi i(\pm \phi_1 \pm \phi_2)} + 8e^{\pm 2\pi i \phi_1} + 14e^{\pm 2\pi i \phi_2} + 202 \Big) \frac{1+q}{1-q}\bigg] \, ,
\end{align}
where $ \phi_1 $ and $ \phi_2 $ are two gauge holonomies for $ SU(2) \times SU(2) $ gauge algebras. The first line contains the gravity, tensor and vector multiplet contributions, and the second line is from the hypermultiplet contributions. The Calabi-Yau hypersurface can be constructed from the following toric data:
\begin{align}
    \begin{array}{rrrr|rrrrr}
        \multicolumn{4}{c|}{\nabla} & l^1 & l^2 & l^3 & l^4 & l^5 \\ \hline
        0 & 0 & 0 & -1 & 0 & 1 & 0 & 0 & 1 \\
        0 & 0 & -1 & 0 & 0 & 0 & 0 & 0 & 1 \\
        1 & 0 & 2 & 3 & 0 & 0 & 1 & 0 & 0 \\
        -1 & 0 & 2 & 3 & -1 & -2 & 1 & 1 & 0 \\
        0 & -1 & 2 & 3 & 1 & 0 & 0 & -1 & 1 \\
        -1 & 1 & 2 & 3 & 1 & 0 & 0 & 0 & 0 \\
        -1 & 0 & 1 & 2 & 0 & 2 & 0 & -1 & 0 \\
        0 & -1 & 1 & 2 & 0 & 0 & 0 & 1 & -1 \\
        0 & 0 & 2 & 3 & -1 & 1 & -2 & 0 & 0
    \end{array}
\end{align}
The Hodge numbers and Euler characteristics of the hypersurface are $ (h^{1,1},h^{2,1},\chi)=(5,201,-392) $. The hypersurface equation \eqref{hypersurface_eq} from the toric data can be brought into Weierstrass form with singular elliptic fibers of Kodaira type $ I_2 $ living on $ e $ and $ f $ curves in the $ \mathbb{F}_1 $ base. The non-zero triple intersection numbers of divisors dual to Mori cone generators are
\begin{gather}
    \begin{gathered}
        \kappa_{122} = 2 \, , \
        \kappa_{123} = 4 \, , \
        \kappa_{124} = 4 \, , \
        \kappa_{125} = 1 \, , \
        \kappa_{133} = 6 \, , \
        \kappa_{134} = 6 \, , \
        \kappa_{135} = 2 \, , \\
        \kappa_{144} = 6 \, , \
        \kappa_{145} = 2 \, , \
        \kappa_{222} = 8 \, , \
        \kappa_{223} = 16 \, , \
        \kappa_{224} = 16 \, , \
        \kappa_{225} = 3 \, , \
        \kappa_{233} = 30 \, , \\
        \kappa_{234} = 30 \, , \
        \kappa_{235} = 6 \, , \
        \kappa_{244} = 26 \, , \
        \kappa_{245} = 6 \, , \
        \kappa_{255} = 1 \, , \
        \kappa_{333} = 54 \, , \
        \kappa_{334} = 54 \, , \\
        \kappa_{335} = 12 \, , \
        \kappa_{344} = 48 \, , \
        \kappa_{345} = 12 \, , \
        \kappa_{355} = 2 \, , \
        \kappa_{444} = 42 \, , \
        \kappa_{445} = 10 \, , \
        \kappa_{455} = 2 \, .
    \end{gathered}
\end{gather}
We can compute the genus zero GV-invariants from the toric data, and identify 2-cycles in $\mathbb{F}_1$ and the K\"ahler parameters associated to $ l^\alpha $ as follows: $ l^1 $ and $ l^3 $ are identified with $ e $ and $ f $ curves, and $ \{l^2, l^4, l^5\} $ are related to the 2-cycles with volumes $ \{\tau-2\phi_1, \phi_1-\phi_2, \phi_2\} $.

Let us now compute the elliptic genera of BPS strings using the modular ansatz and the mirror results. The BPS string with tensor charge $ Q=(k_1,k_2)$ has the following central charges:
\begin{alignat}{2}
    \begin{aligned}
        &c_L = k_1^2 - k_2^2 + 9(3k_1+k_2) + 6 \, , \quad
        &&c_R = k_1^2 - k_2^2 + 3(3k_1+k_2) + 6 \, , \\
        &k_l = \frac{1}{2}(k_1^2 + k_2^2 - 3k_1 - k_2) \, , \quad
        &&k_{SU(2)_1} = -k_2 \, , \qquad
        k_{SU(2)_2} = k_1 + k_2 \, ,
    \end{aligned}
\end{alignat}
with $k_1\ge0$ and $k_2\ge -k_1$.
We can write the modular ansatz for the elliptic genus of this string as
\begin{align}
    \begin{aligned}
        Z_{(k_1,k_2)} &= \frac{1}{\eta(\tau)^{12(3k_1+k_2)}} \frac{\mathcal{N}_{(k_1,k_2)}(\tau, \epsilon, \phi)}{\prod_{s_1=1}^{k_1} \varphi_2(\tau, s_1 \epsilon) \cdot \prod_{s_2=1}^{k_1+k_2} \varphi_2(\tau, s_2\epsilon)} \\
        &\qquad \cdot \frac{1}{\prod_{i=1}^2 \prod_{s_i=1}^{\kappa_i} \prod_{l=0}^{s_i-1} \varphi_2(\tau, (s_i-1-2l)\epsilon+2\phi_i)} \, ,
    \end{aligned}
\end{align}
with $\kappa_1=k_1+k_2$ and $\kappa_2=k_1$.

For the string with charge $ Q=(0,1) $, its elliptic genus has indices $ \{-1,-1,1\} $ for the $SU(2)$ symmetries with holonomies $ \{\epsilon, \phi_1, \phi_2\} $, respectively. Using this together with the genus zero GV-invariants computed from the toric data, we can fix the numerator $ \mathcal{N}_{(0,1)} $ in the modular ansatz as
\begin{align}
    \mathcal{N}_{(0,1)} &= -\frac{E_4 E_6 \varphi_2(\phi _1)^3 \varphi_2(\phi _2)}{216}+\frac{E_4^2 \varphi_2(\phi _1)^2 \varphi_0(\phi _1) \varphi_2(\phi _2)}{432} +\frac{E_6 \varphi_2(\phi _1) \varphi_0(\phi _1)^2 \varphi_2(\phi _2)}{432} \nonumber \\
    & +\frac{E_4^2 \varphi_2(\phi _1)^3 \varphi_0(\phi _2)}{216} -\frac{E_6 \varphi_2(\phi _1)^2 \varphi_0(\phi _1) \varphi_0(\phi _2)}{432} -\frac{E_4 \varphi_2(\phi _1) \varphi_0(\phi _1)^2 \varphi_0(\phi _2)}{432} .
\end{align}
For the string with $ Q=(1,-1) $, its elliptic genus has indices $ \{-1,1,0\} $ for $ \{\epsilon, \phi_1, \phi_2\} $. We again fix the numerator in the modular ansatz using the genus zero GV-invariants. The result is given by
\begin{align}
    \begin{aligned}
        \mathcal{N}_{(1,-1)} &= -\frac{13 E_4^3 E_6 \varphi_2(\phi _1) \varphi_2(\phi _2)^4}{31104}-\frac{11 E_6^3 \varphi_2(\phi _1) \varphi_2(\phi _2)^4}{31104}+\frac{E_4^4 \varphi_0(\phi _1) \varphi_2(\phi _2)^4}{7776} \\
        &\quad +\frac{5 E_4 E_6^2 \varphi_0(\phi _1) \varphi_2(\phi _2)^4}{7776}+\frac{31 E_4^4 \varphi_2(\phi _1) \varphi_2(\phi _2)^3 \varphi_0(\phi _2)}{62208} \\
        &\quad +\frac{41 E_4 E_6^2 \varphi_2(\phi _1) \varphi_2(\phi _2)^3 \varphi_0(\phi _2)}{62208}-\frac{E_4^2 E_6 \varphi_0(\phi _1) \varphi_2(\phi _2)^3 \varphi_0(\phi _2)}{864} \\
        &\quad -\frac{E_4^3 \varphi_0(\phi _1) \varphi_2(\phi _2)^2 \varphi_0(\phi _2)^2}{31104}+\frac{E_6^2 \varphi_0(\phi _1) \varphi_2(\phi _2)^2 \varphi_0(\phi _2)^2}{31104} \\
        &\quad -\frac{11 E_4^3 \varphi_2(\phi _1) \varphi_2(\phi _2) \varphi_0(\phi _2)^3}{62208}-\frac{13 E_6^2 \varphi_2(\phi _1) \varphi_2(\phi _2) \varphi_0(\phi _2)^3}{62208} \\
        &\quad +\frac{E_4 E_6 \varphi_0(\phi _1) \varphi_2(\phi _2) \varphi_0(\phi _2)^3}{2592} \, .
    \end{aligned}
\end{align}
We note that the elliptic genera of strings with charge $ Q=(k,-k) $ are not generated by the Hecke transformation of the elliptic genus of a single string coming from a D3 wrapped on a fiber curve $f$. Thus, this is not a heterotic string in this case, which is expected since there is an $SU(2)$ gauge symmetry supported on the fiber curve. Some higher string elliptic genera are summarized in Appendix~\ref{appendix:F1-SU2SU2}.

\section{Emergent Strings and Swampland Conjectures}
\label{sec:emergent-string}

We shall now discuss the Swampland Conjectures associated with BPS strings. In previous studies \cite{Lee:2018urn,Lee:2019xtm}, the authors performed a comprehensive analysis of weak coupling limits in the moduli space of F-theory compactifications to six dimensions. This analysis gave rise to the Emergent String Conjecture, which states that every infinite-distance limit in the moduli space is either a decompactification limit, where a tower of Kaluza-Klein excitations become light, or an emergent string limit, characterized by the emergence of an asymptotically tensionless critical string. In the limit, the effective field theory breaks down due to a tower of infinitely many light states, which is in accordance with the Swampland Distance Conjecture proposed in \cite{Ooguri:2006in}. 

The Emergent String Conjecture has been rigorously tested for F-theory models also in \cite{Lee:2018urn,Lee:2019xtm,Lee:2019wij}. Specifically, when considering the weak coupling limits of 2-form gauge fields while keeping the gravitational coupling finite, these limits lie at infinite distance in the tensor moduli space. In such a limit, there is always a shrinking rational cycle $C_0$ with zero self-intersection in the K\"ahler base of the elliptic fibration. Moreover, when a D3-brane wraps around this curve, it provides a weakly-coupled tensionless string. By F-theory-heterotic duality, this string is identified with the critical heterotic string compactified on a certain K3 surface.

In the presence of an Abelian gauge symmetry, the limit under consideration corresponds to a weak gauge coupling limit. The emergent critical string in the limit is always charged under the gauge symmetry.  The study of the elliptic genus of this  heterotic string at weak gauge coupling limit plays a crucial role in proving the sublattice Weak Gravity Conjecture (sWGC), proposed in \cite{Heidenreich:2015nta,Heidenreich:2016aqi}, with respect to the gauge symmetry.
More precisely, as we discussed above, the elliptic genera in the worldsheet CFTs encode the BPS protected states of strings interacting with the bulk supergravity theory in six-dimensions. 
These elliptic genera take the form of meromorphic Jacobi forms, and their charge lattice structure is significantly constrained by modular invariance. 
For the emergent heterotic string, its elliptic genus captures (non-BPS) charged particle excitations in the perturbative heterotic string regime, and the masses of these particle states become exponentially light at infinite distance in moduli space. By utilizing the theory of weak Jacobi forms applied to the elliptic genus of the heterotic string, we can explicitly determine both the charges and the masses of these particles. Consequently, this enables us to concretely identify superextremal particle states that populate a sublattice of the charge lattice.
This study, in turn, provides a mathematical proof of the sublattice Weak Gravity Conjecture for 6d F-theory on elliptically fibered Calabi-Yau 3-folds with an Abelian gauge symmetry \cite{Lee:2018urn}.

In this section, we will investigate strings and their BPS spectra at infinite distance in the moduli space of specific 6-dimensional supergravity theories that do not have an F-theory realization, and provide evidences to show that these theories indeed satisfy the Swampland Conjectures. Additionally, we will discuss an interesting infinite-distance limit in the moduli space where the asymptotically tensionless string carrying a charge under the gauge symmetry is not a critical string but rather a non-critical little string. Such little strings are characteristic objects of 6d little string theories that arise in the gravity decoupling limit.

\subsection{Swampland conjectures in non-geometric theories}

Let us consider the theory with $T=1$ and $SU(2)$ gauge group coupled to one triple-symmetric and 84 fundamental hypermultiplets, i.e. $1\times {\bf 4}+84\times {\bf 2}$ hypermultiplets. This theory satisfies the anomaly cancellation conditions in \eqref{eq:6dconstraints} with
\begin{align}\label{eq:charge-vector-b=5}
	a = (-3,1) \ , \quad b = (5,0) \ , \quad \Omega = {\rm diag}(1,-1) \ ,
\end{align}
and $75$ neutral hypermultiplets, where $b$ is the charge vector supporting the $SU(2)$ gauge algebra. 

We will argue that a geometric realization of this theory in F-theory is not possible.
First, suppose that the theory has a geometric construction. In such a case, it can be represented by an elliptic CY 3-fold over a Hirzebruch surface $\mathbb{F}_1$ with $SU(2)$ gauge symmetry on a curve $b=5h$ with $h\subset \mathbb{F}_1$ and $h^2=1$. It is important to note that a geometric transition always exists from $\mathbb{F}_1$ to $\mathbb{P}^2$ by blowing down the exceptional curve in $\mathbb{F}_1$. This transition results in a change in the field theory description from one tensor to 29 neutral hypermultiplets without affecting the gauge group and the charged hypermultiplets. After this transition, we arrive at an F-theory model on a base $\mathbb{P}^2$ with an $SU(2)$ gauge symmetry supported on a degree 5 curve within $\mathbb{P}^2$, i.e. $b=5\ell$ with $\ell\subset \mathbb{P}^2$ and $\ell^2=1$.  This F-theory model is coupled to $1\times {\bf 4}+84\times {\bf 2}$ hypermultiplets.

However, it was shown in \cite{Klevers:2017aku} that a half hypermultiplet in the triple-symmetric representation of $SU(2)$ can be realized in F-theory only when the gauge group lives on a divisor with a triple point singularity, and a degree 5 curve in $\mathbb{P}^2$  can have at most one triple point singularity. This implies that the theory under consideration cannot accommodate one full hypermultiplet in the triple-symmetric representation which requires two triple point singularities.\footnote{A potential loophole in this argument was presented in \cite{Raghuram:2020vxm}. The quintic curve on $\mathbb{P}^2$ that supports the triple point singularities actually reduces to the union of a line and a quartic curve. When this happens, the $I_2$ singularity along the curve leads to an enhanced gauge algebra $\mathfrak{su}(2)\times \mathfrak{su}(2)$. Therefore, the geometric model that realizes the $\mathfrak{su}(2)$ gauge algebra on $b=5\ell$ with a full triple symmetric hypermultiplet exhibits an automatic symmetry enhancement to the $\mathfrak{su}(2)\times \mathfrak{su}(2)$ gauge algebra.} Therefore, we can conclude that the original theory with $T=1$ and one full hypermultiplet in the triple-symmetric representation of $SU(2)$ gauge group cannot have a geometric realization in F-theory, and thus it is a non-geometric theory.

Now, we will examine several Swampland Conjectures in this non-geometric theory. To do so, we will investigate infinite distance limits in the moduli space of the tensor branch. We first note that this theory can be Higgsed to a geometric theory realized in F-theory on $\mathbb{F}_1$ without gauge symmetry and charged matters. The tensor moduli space in the Higgsed theory is described by the geometry of the Hirzebruch surface $\mathbb{F}_1$. The BPS strings in the 6d theory come from D3-branes wrapped around holomorphic curves in the base $\mathbb{F}_1$. The charge lattice of such BPS strings are generated by two basis curves $e=(0,1)$ with $e^2=-1$ and $f=(1,-1)$ with $f^2=0$. The only infinite distance limit in the moduli space of $\mathbb{F}_1$, while keeping the gravity dynamical with the condition $M^4_{\rm Pl} =\frac{1}{2} J\cdot J =1$, is
\begin{align}\label{eq:distance}
  J\cdot h = J\cdot (1,0) \ \sim \ t  \ , \qquad J\cdot f = J\cdot (1,-1) \ \sim \ t^{-1}\ , \qquad t \ \rightarrow\ \infty,
\end{align}
where $h=(1,0)$ denotes the base curve with $h^2=1$. In this limit, the volume of a fiber curve vanishes, i.e. ${\rm vol}(f)\sim J\cdot f \rightarrow 0$, and thus the D3-brane wrapping this fiber curve becomes tensionless and weakly coupled, which agrees with the Emergent String Conjecture. This asymptotically tensionless string turns out to be the critical heterotic string compactified on a K3 surface in a duality frame, as investigated in \cite{Lee:2018urn}.

As discussed in \cite{Tarazi:2021duw}, the Higgsing process does not affect the tensor multiplets and the string charge lattice $\Gamma$ because it involves only the vector multiplet and the hypermultiplets. This implies that the moduli space of the tensor branch and the charge lattice of BPS strings remain unchanged both before and after the Higgsing procedure. This also means that the charge lattice of BPS strings in the non-geometric theory, before Higgsing, is spanned by the charges $(0,1)$ and $(1,-1)$. Thus, the original non-geometric theory exhibits the same infinite distance limit  described in \eqref{eq:distance}. Notably, this limit corresponds to a weak gauge coupling limit
\begin{align}
  \frac{1}{g_{{\rm YM}}^2} = J\cdot b = J\cdot (5,0)\ \rightarrow\ \infty \ .
\end{align}

The Emergent String Conjecture predicts an asymptotically tensionless string in this weak coupling limit \cite{Lee:2019wij}. Indeed, the string with tensor charge $Q=(1,-1)$, with its tension expressed as $T_{(1,-1)}= J\cdot Q\sim t^{-1}$, becomes tensionless in the limit with \eqref{eq:distance}.  This tensionless string is evidently present in the string charge lattice $\Gamma$ in the non-geometric theory, because it exists in the Higgsed theory and the process of Higgsing does not affect the charge lattice.

We now assert that the tensionless string arising in the weak coupling limit \eqref{eq:distance} of the non-geometric theory is, in fact, the heterotic string on a K3 surface. The evidences for this claim are as follows. Firstly, the central charges of the tensionless string match those of the heterotic string. Specifically, the string with charge $Q=(1,-1)$ has
\begin{align}\label{eq:b=5-central}
  Q\cdot Q = 0 \ , \quad c_L = 24 \ , \quad c_R = 12 \  , \quad k_l = -1 \ ,
\end{align}
which precisely agree with the properties of the heterotic string.
Secondly, this string after the Higgsing reduces to the string coming from a D3-brane wrapping a fiber curve $f$ in $\mathbb{F}_1$. This is indeed a weakly coupled, perturbative heterotic string, as discussed above. Lastly, as we will show shortly, the elliptic genera for $n$ strings with charge $nQ=(n,-n)$, can be derived from the Hecke transformation of the elliptic genus for a single string with $n=1$. This is a distinguished feature of the critical heterotic strings, reflecting the fact that they do not form bound states \cite{Dijkgraaf:1996xw}. Altogether, these evidences strongly support our claim that the string with charge $(1,-1)$ corresponds to the heterotic string in 6d.

The last evidence can be easily proven as follows. We first prove that the solution to the modular ansatz of the form \eqref{eq:ansatz-old} for the elliptic genus of $n$ strings with charge $nQ=(n,-n)$ is unique by employing the argument given in Sec. 5.1.1 in \cite{Gu:2017ccq}. Let us define $ \mathcal{Z}_{kQ} $ as
\begin{align}
    1+\sum_{k=1}^\infty w^k Z_{kQ} = \PE\left[\sum_{k=1}^\infty w^k \mathcal{Z}_{kQ}\right] \, .
\end{align}\
Suppose that the elliptic genera of charges $ kQ = (k,-k) $ with $ k < n $ are fully fixed, and that there are two distinct sets of coefficients $ \{c_i\} $ and $ \{\tilde{c}_i\} $ in the modular ansatz \eqref{eq:numerator-nogauge} of $ Z_{nQ} $ which satisfy the GV-invariant condition \eqref{eq:GV-condition2}. The difference between two modular ans\"atze
\begin{align}
    Z_{nQ}(c_i) - Z_{nQ}(\tilde{c}_i) = \mathcal{Z}_{nQ}(c_i) - \mathcal{Z}_{nQ}(\tilde{c}_i)
\end{align}
is of the form \eqref{eq:ansatz-old} that has the same modular properties with $ Z_{nQ} $, and can be expressed as the form within the plethystic exponential in \eqref{eq:GV-condition2}. The `lemma' in \cite{Gu:2017ccq} is that Jacobi forms which have such properties can have at most a single $ \varphi_{-2,1}(\tau,\epsilon) $ factor in the denominator. Thus, for a weak Jacobi form $ \mathcal{N}_{nQ}^{\mathrm{diff}} $, the difference of two ans\"atze is given by
\begin{align}
    Z_{nQ}(c_i) - Z_{nQ}(\tilde{c}_i) = \frac{1}{\eta(\tau)^{24n} } \frac{\mathcal{N}_{nQ}^{\mathrm{diff}}(\tau, \epsilon, \phi)}{\varphi_2(\tau, \epsilon)} \, .
\end{align}

Now note that the elliptic genus $ Z_{nQ} $ has an index $ -n $ with respect to $ \epsilon $, so $ \mathcal{N}_{nQ}^{\mathrm{diff}} $ has an index $ -n+1 $ with respect to $ \epsilon $. However, since there is no weak Jacobi form of a negative index, $ \mathcal{N}_{nQ}^{\mathrm{diff}} $ always vanishes if $ n>1 $. This implies that GV-invariant condition uniquely fixes all the unknown coefficients in the modular ansatz of $ Z_{nQ} $ when $ n>1 $. Using this uniqueness, we conclude that the elliptic genus for the strings with charge $ nQ $ is given by the Hecke transformation of $ Z_Q $ as
\begin{align}\label{eq:Hecke-b5}
    1 + \sum_{n=1}^\infty w^n Z_{nQ}(\tau, \epsilon, \phi) = \exp(\sum_{n=1}^\infty \frac{w^n}{n} \sum_{ad=n} \sum_{b(\operatorname{mod}d)} Z_{Q}\Big(\frac{a\tau+b}{d}, a\epsilon, a\phi\Big) ) \, ,
\end{align}
where $ Q=(1,-1) $. The reasons for this are as follows. First, $ Z_{nQ} $ obtained using this equation exhibits correct modular properties: it has weight zero, index $ -n $ for $ \epsilon $ and $ 5n $ for $ \phi $. This is a property of the Hecke transformation given in Appendix~\ref{appendix:modular}. More detailed study for the Hecke transformation is given in \cite{Eichler:1995}. Second, it is known that the partition function of the heterotic string is generated by the Hecke transformation \cite{Dijkgraaf:1996xw}, so $ Z_{nQ} $ from \eqref{eq:Hecke-b5} is an elliptic genus of a heterotic string. In other words, it has the same denominator structure as \eqref{eq:ansatz-old} and satisfy the GV-invariant condition in \eqref{eq:GV-condition2}. Thus, $ Z_{nQ} $ from \eqref{eq:Hecke-b5} is a candidate for the elliptic genus we are interested in. Due to the uniqueness of the modular ansatz, it must be the correct answer for the elliptic genus of the strings with charge $ nQ=(n,-n) $. This suggests that strings with charge $ Q=(1,-1) $ in our non-geometric theory is a heterotic string.

Let us turn to the computation of the elliptic genus of the string with charge $Q=(1,-1)$. This string has the central charges given in \eqref{eq:b=5-central} with $k_{SU(2)}=5$. Thus we can take the modular ansatz for the elliptic genus as
\begin{align}\label{eq:modular-anzatz-elliptic-genus}
  Z_{(1,-1)}^{b=(5,0)}(\tau,\epsilon,\phi) = \frac{\mathcal{N}_{(10,5)}(\tau,\phi)}{\eta(\tau)^{24}\varphi_{2}(\tau,\epsilon)} \ .
\end{align}
The numerator $\mathcal{N}_{(10,5)}$ is a weak Jacobi form of weight 10 and  index 5 for the $SU(2)$ gauge group. It can be expressed as a polynomial in the modular forms $E_4(\tau), E_6(\tau), \varphi_{2}(\tau,\phi)$, and $\varphi_{0}(\tau,\phi)$ with 10 unknown coefficients. To narrow down these coefficients, we can impose the following two conditions.  First, the ground state in the elliptic genus is an $SU(2)$ singlet. Second, this elliptic genus should reduce to the elliptic genus of the heterotic string on K3 without any gauge symmetry after the Higgsing. By enforcing these conditions, we can determine 6 coefficients. In order to determine the remaining 4 coefficients in the modular ansatz, we will use the claim discussed above, which asserts that this string in the weak gauge coupling is a perturbative heterotic string.

The elliptic genus counts left-moving excitations of the strings, and the excitation level $\ell$ is captured by the power of $q$ in the $q$-expansion. In the case of the heterotic string, the left-moving string excitations, along with right-moving pairs by level-matching, provide particle states in the bulk 6d theory. In particular, the states at $\ell=0$, corresponding to the $q^0$-th order, generate the massless sector in the 6d theory. It then follows that the elliptic genus for the string of $Q=(1,-1)$ at $q^0$-th order, denoted as $n_0$, must capture the chiral index of the massless matter content in the 6d theory \cite{Lee:2018urn} as
\begin{align}
  n_0|_{\mathfrak{q}} = 2(V+3+T-H)|_{\mathfrak{q}}\ ,
\end{align}
where the subscript $\mathfrak{q}$ indicates the restriction to the states with a specific charge $\mathfrak{q}$. This condition imposes significant constraints on the elliptic genus for the heterotic string.  All the remaining coefficients in the modular ansatz are determined by this condition. Remarkably, the conditions for $\mathfrak{q}\ge2$ are sufficient to fix all the unknown coefficients, and as a result, the conditions at $\mathfrak{q}=0$ and $\mathfrak{q}= 1$ are automatically satisfied.

The result after fixing the coefficients is
\begin{align}\label{eq:numerator-b=5}
    \mathcal{N}_{(10,5)}(\tau,\phi) =& -\frac{7E_4^5\varphi_2(\phi)^5+E_4^2E_6^2\varphi_2(\phi)^4}{995328} -\frac{19E_4^4\varphi_{2}(\phi)^3\varphi_{0}(\phi)^2}{497664}+\frac{E_4^3E_6\varphi_2(\phi)^4\varphi_0(\phi)}{23328} \nonumber \\
    &-\frac{(65E_4^3+55E_6^2)\varphi_2(\phi)\varphi_0(\phi)^4}{E_4^4\varphi_{2}(\phi)^3\varphi_{0}(\phi)^2}+\frac{5E_4^2E_6\varphi_2(\phi)^2\varphi_0(\phi)^3}{62208} \nonumber \\
    &-\frac{7E_4E_6^2\varphi_2(\phi)^3\varphi_0(\phi)^2}{62208}+\frac{E_4E_6\varphi_0(\phi)^5}{124416}-\frac{E_6^3\varphi_2(\phi)^4\varphi_0(\phi)}{373248} \ .
\end{align}
Plugging this into the modular ansatz, we compute the elliptic genus which can be expanded as
\begin{align}\label{eq:elliptic-expanded-b=5}
    -&Z_{(1,-1)}^{b=(5,0)}(\tau,\epsilon,\phi)\varphi_{-2,1}(\tau,\epsilon) = -2q^{-1}+(140+170\xi^{\pm1}-2\xi^{\pm2}+2\xi^{\pm3})q^0 \\
    &\qquad \qquad +\left(77336 + 62752\xi^{\pm1}+30604\xi^{\pm2}+8456\xi^{\pm3}+908\xi^{\pm4}+56\xi^{\pm5}\right)q +\mathcal{O}(q^2) \ ,\nonumber
\end{align}
where $\xi = e^{2\pi i \phi}$ and $\xi^{\pm n}\equiv \xi^{n}+\xi^{-n}$. So we can uniquely determine the elliptic genus for the string with charge $(1,-1)$, which is consistent with the structure of the  critical heterotic string on K3.

The left-moving excitations captured by the elliptic genus, when properly paired with right-moving excitations, realize actual charged (but non-BPS) particles states in the 6d supergravity. In the tensionless limit, these states provide a tower of asymptotically massless charged states. This results in the break down of the low-energy effective theory, which is expected since the tensionless limit lies at infinite distance in the moduli space. 

In \cite{Lee:2018urn,Lee:2018spm}, the Sublattice Weak Gravity Conjecture (sLWGC) has been rigorously verified in F-theory on elliptically fibered Calabi-Yau 3-folds by examining the spectrum of charged states originating from the emergent heterotic string at the weak gauge coupling limit. The sLWGC postulates that any $U(1)$ gauge theory coupled to gravity must have a sublattice in the charge lattice such that for any site in the sublattice, there exists a superextremal particle-like state satisfying the condition that its charge-to-mass ratio surpasses that of a extremal Reissner-Nordstr\"om black hole \cite{Heidenreich:2016aqi}. This is a refined version of the Weak Gravity Conjecture first proposed in \cite{Arkani-Hamed:2006emk}. The WGC for 6d $\mathcal{N}=(1,0)$ supergravity, accounting for contributions from massless scalar fields, leads to the extremality bound \cite{Lee:2018urn,Lee:2018spm},
\begin{align}\label{eq:weak-gravity}
  \mathfrak{q}^2\, g_{\rm YM}^2 \ge \frac{M^2}{M_{\rm pl}^4} \ .
\end{align}

Using the modular properties of the elliptic genus and the theory of Jacobi forms, we can prove sLWGC for 6d F-theory models with Abelian gauge group factors. For this, we first note the fact that any weak Jacobi form of weight $w$ and index $m$ can be expanded as \cite{Eichler:1995} 
\begin{align}\label{eq:modular-jacobi-form}
  \varphi_{w,m}(\tau,\phi) = \sum_{\ell \in \mathbb{Z} \, {\rm mod} \, 2m}h_\ell(\tau)\theta_{m,\ell}(\tau,\phi) \ ,
\end{align}
where $h_\ell(\tau)$ represents a vector-valued weakly-holomorphic modular form of weight $w-1/2$ and
\begin{align}
  \theta_{m,\ell}(\tau,\phi) = \sum_{k\in \mathbb{Z}} q^{(\ell+2mk)^2/4m} \xi^{\ell+2mk} \ 
\end{align}
is the theta function of index $m$ and characteristic $\ell$. The numerator of the elliptic genus of the heterotic string, as demonstrated in equation \eqref{eq:modular-anzatz-elliptic-genus} for our specific case, takes the form of a weak Jacobi form, and therefore it admits an expansion in terms of the theta functions. The sector at $\ell=0$ is of particular interest. This specific sector is particularly intriguing as it pertains to the sublattice associated with the left-moving ground state, which corresponds to the leading coefficient in the expansion $h_0(\tau) = 2 + \mathcal{O}(q)$. So it always has non-trivial contribution to the elliptic genus and thus to the particle spectrum in the 6d supergravity. The excitation level and the charge for the states in this sublattice can be easily extracted from the theta function at $\ell=0$, which are
\begin{align}
  n(k) = mk^2 \ , \quad \mathfrak{q}_k = 2mk \ , \quad k\in \mathbb{Z} \ .
\end{align}
The charged states arising from the string states populating in this sublattice turn out to be the superextremal particles satisfying the extremality bound as \cite{Lee:2018urn}
\begin{align}\label{eq:weak-gravity-bound}
  \mathfrak{q}_k^2 = 4mn(k) = 4m\left(\frac{M_k^2}{4mg_{\rm YM}^2}+1\right) > \frac{M_k^2}{g_{\rm YM}^2} \ ,
\end{align}
with the normalization $M_{\rm Pl}=1$ at a weak gauge coupling limit, e.g. $t\rightarrow \infty$. Here, we used the mass formula for the excitations of the heterotic string, $M_k^2 = 4T(n(k)-1)$. This provides a proof for the Sublattice Weak Gravity Conjecture in the context of F-theory compactification on elliptic Calabi-Yau 3-folds with a single $U(1)$ factor. Also, the analogous proof for F-theory constructions involving multiple Abelian factors is outlined in \cite{Lee:2018spm}.

The extension of this proof to non-Abelian gauge theories can be carried out by employing the same idea.
The sLWGC for non-Abelian gauge fields is a stronger conjecture compared to the conventional Weak Gravity Conjecture mentioned earlier. It asserts that any gauge theory in dimensions $d\ge 5$ with a non-Abelian gauge group $G$ coupled to gravity must have a finite-index Weyl-invariant sublattice $\Gamma_0$ of the weight lattice $\Gamma_G$ such that for every dominant weight ${\bf\lambda}_{\bf R}\in\Gamma_0$, there exists a superextremal particle-like state transforming in the $G$ irrep ${\bf R}$ with highest weight ${\bf \lambda}_{\bf R}$ \cite{Heidenreich:2017sim}. The proof of this conjecture for F-theory compactified on elliptic or genus one fibered Calabi-Yau 3-folds when it has a tensionless charged heterotic string at weak gauge coupling limit was presented in \cite{Cota:2020zse}, which was achieved once again by utilizing the modular properties of Jacobi forms appearing in the elliptic genus of the heterotic string.

These proofs for the sLWGC can also be applied to the non-geometric theory, specifically in the case where $T=1$ and $SU(2)$ gauge group on $b=(5,0)$ is coupled to $1\times {\bf 4}+84\times {\bf 2}$ hypermultiplets. The elliptic genus of the string with charge $(1,-1)$ in this theory can be expressed as \eqref{eq:modular-anzatz-elliptic-genus} and the numerator given in \eqref{eq:numerator-b=5} is a weak Jacobi form. Therefore, by relying on our claim that this string becomes a heterotic string at weak gauge coupling limit and using the property of weak Jacobi forms in \eqref{eq:modular-jacobi-form}, we can demonstrate that this supergravity theory has the superextremal particles, which arise from the tensionless heterotic string, satisfying the extremality condition\footnote{The translation from the bound in \eqref{eq:weak-gravity} to the one in \eqref{eq:weak-gravity-bound} was performed for F-theory models in \cite{Lee:2018urn,Lee:2018spm}. The same bound can be obtained for the theories in this subsection by substituting the K\"ahler form $J$ and the curve classes $C,C_0$ in their F-theory models by those in our theories as $J_{\rm theirs}=J_{\rm ours}$, $C=b$, and $C_0=(1,-1)$.} \eqref{eq:weak-gravity-bound}, and thus satisfies the (non-Abelian) Sublattice Weak Gravity Conjecture. As this theory also satisfies the other Swampland Conjectures like the Distance Conjecture and the Emergent String Conjecture, it seems plausible that this field theory can be consistently coupled to quantum gravity, and thus it resides outside the Swampland.

Another interesting supergravity theory is the $SU(2)$ gauge theory with $T=1$ and  $\frac{3}{2}\times {\bf 4}+1\times {\bf 3}+93\times {\bf 2}+52\times {\bf 1}$ hypermultiplets. This is an anomaly free theory with the anomaly vectors and the intersection form
\begin{align}
  a = (-3,1) \ , \quad b = (6,0) \ , \quad \Omega = {\rm diag}(1,-1) \ .
\end{align}
In contrast to the previous example involving $SU(2)$ gauge group on $b=(5,0)$, we cannot rule out the possibility of this theory being realized in F-theory. The realization of this theory demands three triple point singularities on the curve $b=(6,0)$ to accommodate three triple symmetric half-hypermultiplets. This configuration can be achieved in the F-theory framework \cite{Klevers:2017aku}. However, the explicit construction of the corresponding elliptic Calabi-Yau 3-fold is unknown at present. Thus the geometric method, such as the mirror technique, to investigate this theory is not currently feasible. Nonetheless, the technique developed in this subsection can still be employed to carry out concrete tests for various Swampland Conjectures in this theory.

The moduli space of this theory coincides with that of the previous theory described in \eqref{eq:charge-vector-b=5}, and the charge lattice of BPS strings is also spanned by the basis vectors $(0,1)$ and $(1,-1)$. The weak gauge coupling limit, while gravity is kept dynamical, is located at infinite distance in the moduli space
\begin{align}
  \frac{1}{g^2_{\rm YM}} = J\cdot b = J\cdot (6,0) \  \rightarrow \  \infty \ .
\end{align}
In this limit, the string with the tensor charge $(1,-1)$ again becomes tensionless. The existence of this string is  ensured by its presence in the Higgsed theory, which corresponds to the F-theory model on an elliptic 3-fold over a base $\mathbb{F}_1$. Based on the same reasons accounted for proving the emergence of the heterotic string in the previous example, which relies on its central charges, Higgsing, and special properties of its elliptic genus, we claim that this string is also the heterotic string on a K3. This assertion enables us to calculate the elliptic genus of this tensionless string.

The central charges of the string with charge $(1,-1)$ are given by
\begin{align}\label{eq:b=6-central}
  Q\cdot Q = 0 \ , \quad c_L = 24 \ , \quad c_R = 12 \  , \quad k_l = -1 \  , \quad k_{SU(2)} = 6 \ .
\end{align}
From this, we can take the following ansatz for the elliptic genus of this string:
\begin{align}\label{eq:modular-anzatz-elliptic-genus2}
  Z_{(1,-1)}^{b=(6,0)}(\tau,\epsilon,\phi) = \frac{\mathcal{N}_{(10,6)}(\tau,\phi)}{\eta(\tau)^{24}\varphi_{-2,1}(\tau,\epsilon)} \ .
\end{align}
The numerator $\mathcal{N}_{(10,6)}$ is now a weak Jacobi form of weight $10$ and index $5$ for the $SU(2)$ gauge group. It contains 12 unknown coefficients when expressed in terms of the basis modular forms. By using the properties of the heterotic string at $q^{-1}$ and $q^0$ orders in the $q$-expansion, we can uniquely fix the numerator,
\begin{align}
  \mathcal{N}_{(10,6)}(\tau,\phi) =& -\frac{(5E_4^4+3E_6^2)E_4^2\varphi_2(\phi)^5\varphi_0(\phi)}{1990656}+\frac{(6E_4^4E_6-2E_4E_6^3)\varphi_2(\phi)^6}{5971968}  \\
  &-\frac{(13E_4^3+11E_6^2)\varphi_2(\phi)\varphi_0(\phi)^5}{5971968}-\frac{(9E_4^3+11E_6^2)E_4\varphi_2(\phi)^3\varphi_0(\phi)^3}{1492992} \nonumber \\
  & +\frac{(29E_4^3+E_6^2)E_6\varphi_2(\phi)^4\varphi_0(\phi)^2}{2985984}+\frac{5E_4^2E_6\varphi_2(\phi)^2\varphi_0(\phi)^4}{497664}+\frac{E_4E_6\varphi_0(\phi)^6}{1492992}\nonumber \ .
\end{align}
Substituting this numerator into the modular ansatz, the elliptic genus is expanded as
\begin{align}\label{eq:elliptic-expanded-b=6}
    -&Z_{(1,-1)}^{b=(6,0)}(\tau,\epsilon,\phi)\varphi_{-2,1}(\tau,\epsilon) = -2q^{-1}+(96+189\xi^{\pm1}+3\xi^{\pm3})q^0 \\
    & \quad +\left(69156 + 60858\xi^{\pm1}+31752\xi^{\pm2}+12169\xi^{\pm3}+1890\xi^{\pm4}+189\xi^{\pm5}+8\xi^{\pm6}\right)q +\mathcal{O}(q^2) \, . \nonumber
\end{align}
The result is consistent with the structure of the elliptic genus of the heterotic string on K3: all coefficients present in the $q$-expansion are integers, the chiral index for the massless states are correctly captured, and when $\xi$ is set to $1$, this is Higgsed to the elliptic genus of the heterotic string associated to a rational fiber curve in a base $\mathbb{F}_1$ in an F-theory. 

In this context,  even without an explicit F-theory construction for this theory at present, it is still possible to examine certain Swampland Conjectures. The characteristics of the string with charge $(1,-1)$ including the consistent elliptic genus computed above offer compelling evidence that supports the emergence of a tensionless heterotic string at weak gauge coupling. By noting that the elliptic genus of the heterotic string encodes the particle states in the 6d supergravity as well as their charge-to-mass ratios, which can be derived from the special feature of weak Jacobi forms discussed in \eqref{eq:modular-jacobi-form}, we can indeed confirm that this theory satisfies the non-Abelian Sublattice Weak Gravity Conjecture.

\subsection{Emergent little strings}

In certain supergravity theories, infinite distance limits in the moduli space correspond to strong gauge coupling limits. In this subsection, we will highlight one such instance. As we will see, the theory also has asymptotically tensionless strings in a neighborhood of the infinite distance limit, consistent with the Emergent String Conjecture. However, it is important to note that the tensionless strings, which are charged under the gauge group, are not the heterotic string in this case. Instead, they are non-critical strings at strong coupling in a 6d little string theory (LST) that interacts with gravity.

Consider the 6d supergravity with $T=1$ and $SU(2)$ gauge group coupled to $16\times {\bf 2} + 215 \times {\bf 1}$ hypermultiplets. This theory is anomaly free with the charge vectors and the intersection form,
\begin{align}
  a = (-3,1) \ , \quad b = (1,-1) \ , \quad \Omega = {\rm diag}(1,-1) \ .
\end{align}
This theory can be engineered in F-theory compactification on an elliptically fibered 3-fold with $(h^{1,1},h^{2,1})=(4,214)$. The base surface of the CY$_3$ is a Hirzebruch surface $\mathbb{F}_1$ and the fiber curve $f\subset \mathbb{F}_1$ with $f^2=0$ supports an $SU(2)$ gauge algebra. This model is a Calabi-Yau hypersurface and its toric data is given as follows:
\begin{align}
    \begin{array}{rrrr|rrrr}
        \multicolumn{4}{c|}{\nabla} & l^1 & l^2 & l^3 & l^4 \\ \hline
        0 & 0 & 0 & -1 & 0 & 1 & 0 & 1 \\
        0 & 0 & -1 & 0 & 0 & 0 & 0 & 1 \\
        -1 & -1 & 2 & 3 & 0 & 0 & 1 & 0 \\
        0 & 1 & 2 & 3 & 1 & 0 & 0 & 0 \\
        0 & -1 & 2 & 3 & 1 & 0 & -1 & 0 \\
        1 & 0 & 1 & 2 & 0 & 2 & 0 & -1 \\
        1 & 0 & 2 & 3 & 0 & -2 & 1 & 1 \\
        0 & 0 & 2 & 3 & -2 & 1 & -1 & 0
    \end{array}
\end{align}
The Mori cone curve associated with $ l^1 $ and $ l^3 $ are fiber $ f $ and base $ e $ classes of the Hirzebruch surface $ \mathbb{F}_1 $, respectively, and $ l^2+2l^4 $ is related to the elliptic fiber.

The only infinite distance limit in the moduli space, while the volume of $\mathbb{F}_1$ stays finite, is the infinite volume limit of the base curve $h$, i.e. ${\rm Vol}(h)\rightarrow \infty$. In the limit, the fiber curve $f$ shrinks to zero size, i.e. ${\rm Vol}(f)\rightarrow 0$. This implies that the $SU(2)$ gauge coupling diverges because its inverse is controlled by the volume of the fiber curve. Thus, this limit represents the strong coupling regime in the moduli space for the $SU(2)$ gauge group.

Let us consider D3-branes wrapped around the fiber curve $f$ with zero self-intersection. Since the volume of the wrapped curve approaches zero in the infinite distance limit, these wrapped D3-branes yield tensionless strings in the supergravity theory. It turns out that there are two distinct strings which become tensionless as the gauge coupling diverges. 

The first one is the critical heterotic string, which coincides with the emergent strings previously highlighted in \cite{Lee:2018urn}. The oscillations of this tensionless heterotic string produces a tower of massless particle states, leading to the breakdown of the effective field theory. This is consistent with expectations from the Distance Conjecture at infinite distance in the moduli space. We note, however, that this string is not charged under the $SU(2)$ gauge field since the wrapped fiber curve does not intersect with the fiber curve that supports the gauge algebra. So, this means that the previous proof of the Weak Gravity Conjecture for F-theory models hosting only Abelian gauge factors in \cite{Lee:2018urn,Lee:2018spm}, which heavily leans on the distinct features of the charged heterotic strings appearing at infinite distance limits, does not apply in this case, even though the theory has an F-theory realization.

The second string, which becomes tensionless in the infinite distance limit, is the instantonic string for the $SU(2)$ gauge group. The worldvolume theory on the instantonic strings contains bosonic zero modes parametrizing the instanton moduli space, and thus they are distinguished from the heterotic string which solely involves  fermionic zero modes apart from the center of mass modes. It is clear that the instantonic strings carry non-trivial charges for the $SU(2)$ gauge symmetry. These strings are instanton solitons in the 6d LST with $SU(2)$ gauge group and 16 fundamental hypermultiplets, which can be realized on a D5-brane in the Type I string theory \cite{Witten:1995gx,Gimon:1996rq}.

We will now demonstrate that the elliptic genus of the second tensionless string matches exactly with that of the non-critical string in the 6d LST. To support this claim, we will directly calculate the elliptic genus of the string arising from a wrapped D3-brane around the fiber curve $f$, which hosts the $SU(2)$ singularity. This string is an instantonic string with
\begin{align}
  c_L = 24 \ , \quad c_R = 12 \  , \quad k_l = -1 \  , \quad k_{SU(2)} = 0 \ .
\end{align} 
Hence, we can take the modular ansatz for the elliptic genus as
\begin{align}\label{eq:elliptic-genus-LST}
  Z_{(1,-1)}^{b=(1,-1)}(\tau,\phi) = \frac{\mathcal{N}_{(8,4)}(\tau,\phi)}{\eta(\tau)^{24}\varphi_{2}(\tau,\epsilon)\varphi_{2}(\tau,2\phi)} \ .
\end{align} 
Here, the denominator factor depending on the $SU(2)$ gauge chemical potential is chosen for the bosonic zero mode of a single $SU(2)$ instanton moduli space. The numerator is a weak Jacobi form of weight 8 and $SU(2)$ index $4$. By comparing it to the genus zero GV-invariant obtained from the mirror analysis, we can uniquely determine the numerator,
\begin{align}
  \mathcal{N}_{(8,4)}(\tau,\phi) =& -\frac{(E_4^4+5E_4E_6^2)\varphi_2(\phi)^4}{648}+\frac{(E_4^3-E_6^2)\varphi_2(\phi)^2\varphi_0(\phi)^2}{2592} \nonumber\\
  &-\frac{E_4E_6(\varphi_0(\phi)^2-3E_4\varphi_2(\phi)^2)\varphi_2(\phi)\varphi_0(\phi)}{216} \ .
\end{align}
Given this result, the elliptic genus \eqref{eq:elliptic-genus-LST} perfectly matches the elliptic genus of a single $SU(2)$ little string which was computed in \cite{Kim:2018gak} using the ADHM construction for this little string in \cite{Johnson:1998yw} and also in \cite{Kim:2023glm} using the blowup equation.\footnote{More precisely, only the states charged under the $SU(2)$ gauge symmetry match, after switching off all the $SO(32)$ flavor fugacities, which we checked numerically and also up to $q^{10}$ order in $q$-expansion. The charge-neutral sector can differ from the LST results, as the partition functions of local theories are defined up to certain factors independent of dynamical parameters like $\phi$.} One can also compute the elliptic genera of strings from  multi-wrapped D3-branes on $f$, and check that they agree with those of the strings at higher instanton numbers in the LST. This verifies that the strings corresponding to D3-branes wrapped around a fiber curve $f$ with an $SU(2)$ singularity are little strings in the 6d LST.

It is also worth noting that when we turn off the $SU(2)$ gauge fugacity (by setting $\phi\rightarrow 0$), this elliptic genus matches that of the heterotic string on K3. This reflects the fact that when we Higgs the $SU(2)$ gauge symmetry, the theory reduces to the 6d F-theory corresponding to a Hirzebruch base $\mathbb{F}_1$ without gauge singularities, and the string in the Higgsed theory is dual to a six-dimensional heterotic string in the emergent string limit. In fact, the string arising from a D3-brane wrapped over a fiber curve away from the $SU(2)$ singularity before the Higgsing coincides with the heterotic string after the Higgsing. This is because the string remains unaffected during the Higgsing process as it does not interact with the $SU(2)$ gauge field. From this perspective, we can say that the elliptic genus in the emergent string limit encodes two spectra: the BPS states of the little strings carrying $SU(2)$ gauge charges and those of the charge-neutral string dual to a 6d heterotic string. In this case, the elliptic genus of the tensionless string cannot capture the $SU(2)$ charged particle states in the bulk 6d supergravity since the emergent heterotic string does not carry any gauge charge. Thus, as mentioned earlier, we cannot confirm the Weak Gravity Conjecture for theories of this kind using the argument  from references \cite{Lee:2018urn,Lee:2018spm} for Abelian gauge groups (also \cite{Cota:2020zse} for non-Abelian groups) which is based on the spectrum of a charged heterotic string.

\section{Conclusions}\label{sec:conclusion}

In this work, we have explored BPS strings and their behavior at infinite-distance limits in the moduli space of 6d $\mathcal{N}=(1,0)$ supergravity theories. Our main tool for this exploration is the exact spectrum of BPS states in 2d worldsheet CFTs living on these strings. In the first part of this work,  we outlined a computational strategy that uses a modular bootstrap approach for calculating the elliptic genera of BPS strings in 6d supergravity.  We introduced a generalized modular ansatz that potentially covers all string configurations, including instantonic strings for 6d gauge groups. We then confirmed the validity of this approach through explicit calculations, using topological strings and mirror symmetry in the context of F-theory compactifications on elliptically fibered CY 3-folds.

In the second part, we examined various Swampland conjectures in 6d supergravity theories, which are not easily described by geometric constructions in F-theory compactifications. These theories have exotic matter representations of the gauge group, specifically the triple symmetric representation of the $SU(2)$ group. Such representations are difficult or even impossible to explicitly realize using Weierstrass models in F-theory. Our investigation strongly suggests that these models give rise to emergent heterotic strings at weak gauge coupling limit, which is located at infinite distance in the moduli space, and the effective field theory breaks down due to a tower of asymptotically massless charged states. This is consistent with both the Swampland Distance and Emergent String Conjectures. Furthermore, the excitations of these strings provide states occupying a sublattice in the charge lattice that meets the criteria of a sublattice version of the Weak Gravity Conjecture. Our results demonstrate that it is possible to rigorously test the Swampland conjectures in certain EFTs without the need for geometric realizations or specific string theory embeddings.

Expanding our approach to explore other non-geometric supergravity theories could be quite interesting.  Specifically, if the theory includes a BPS string whose central charges match those given in \eqref{eq:b=5-central}, we can argue that this string becomes a critical heterotic string in a duality frame. The elliptic genus of this string is strongly constrained by the properties of the heterotic string. So, by combining this knowledge with the modular bootstrap technique, we might be able to exactly calculate the elliptic genus. 
If we find an elliptic genus that is consistent with both the features of the heterotic string and the modular ansatz, as in the example we discussed in Section \ref{sec:emergent-string}, then we could use this to solidly confirm certain Swampland conjectures like the Weak Gravity Conjecture.  This would offer strong evidence that these non-geometric theories have consistent UV-completions and thus being in the landscape. In these cases,  it would also be fascinating to further examine whether these non-geometric theories cannot be realized in any string framework, thereby potentially serving as counterexamples to the String Lamppost Principle (SLP). Conversely, if we cannot find a consistent elliptic genus for the string, it would strongly suggest that the supergravity theory in question is inconsistent when coupled to gravity and hence belongs to the Swampland. Therefore, this analysis could serve as an additional Swampland constraint for 6d supergravity theories.
\bigskip

\acknowledgments
We would like to thank Yuta Hamada, Seung-Joo Lee, Sungwoo Nam, Shing Yan Li and Xin Wang for useful discussion and conversation. HH thanks the hospitality of POSTECH where part of this work was done. The work of HH is supported in part by JSPS KAKENHI Grant Number JP18K13543 and JP23K03396. HK and MK are supported by Samsung Science and Technology Foundation under Project Number SSTF-BA2002-05 and by the National Research Foundation of Korea (NRF) grant funded by the Korea government (MSIT) (2023R1A2C1006542).

\appendix

\section{Modular forms and Jacobi forms}\label{appendix:modular}

In this appendix, we review basic definitions and properties of the modular forms and Jacobi forms.

\subsection{Modular forms}

Let $ \mathcal{H} = \{ z \in \mathbb{C} \mid \imaginary z > 0 \} $ be the upper half plane of the complex plane. \emph{Modular form} is a function $ f : \mathcal{H} \to \mathbb{C} $ satisfying
\begin{align}
    f\left(\frac{a\tau+b}{c\tau+d}\right) = (c\tau+d)^k f(\tau) \, , \quad
    \mqty(a & b \\ c & d) \in \mathrm{SL}(2,\mathbb{Z}) \, ,
\end{align}
for an integer $ k $, which is called the \emph{weight} of the modular form. The modular forms are periodic under $ \tau \to \tau+1 $, and they have Fourier expansion of the form
\begin{align}
    f(\tau) = \sum_{n\in \mathbb{Z}} c(n) q^n \qquad (q = e^{2\pi i \tau}) \, .
\end{align}
$ f $ is called a holomorphic (resp. cusp or weakly holomorphic) modular form if $ c(n)=0 $ unless $ n\geq 0 $ (resp. $ n>0 $ or $ n\geq -N $ for some $ N \geq 0 $). 

An important example of the modular forms is the Eisenstein series
\begin{align}
    E_{2k}(\tau) = \frac{1}{2\zeta(2k)} \sum_{(m,n) \neq (0, 0)} \frac{1}{(m+n\tau)^{2k}}
    = 1 + \frac{(2\pi i)^{2k}}{\zeta(2k) (2k-1)!} \sum_{n=1}^\infty \sigma_{2k-1}(n) q^n \, ,
\end{align}
of weight $ 2k $, where $ \zeta(s) $ is the Riemann zeta function and $ \sigma_k(n) = \sum_{d|n} d^k $ is the divisor function. When $ k>1 $, $ E_{2k}(\tau) $ is a holomorphic modular form, while $ E_2(\tau) $ transforms as
\begin{align}
    E_2\qty(\frac{a\tau+b}{c\tau+d}) = (c\tau+d)^2 E_2(\tau) - \frac{6i}{\pi} c(c\tau + d) \, .
\end{align}
$ E_2 $ is an example of \emph{quasi-modular form}, meaning that it is a holomorphic part of the non-holomorphic function $ \hat{E}_2 = E_2 - 3/(\pi \imaginary \tau) $ which transforms like a modular form of weight 2. The ring of holomorphic modular forms on $ \mathrm{SL}(2,\mathbb{Z}) $ is freely generated by two Eisenstein series $ E_4(\tau) $ and $ E_6(\tau) $, i.e., any holomorphic modular form $ f(\tau) $ of weight $ k $ can be written as $ f = \sum_{4\alpha+6\beta=k} c_{\alpha\beta} E_4^\alpha E_6^{\beta} $ for some constants $ c_{\alpha\beta} \in \mathbb{C} $. For instance, the 24-th power of the Dedekind eta function defined by
\begin{align}
    \eta(\tau) = q^{1/24} \prod_{n=1}^\infty (1-q^n) \, ,
\end{align}
is an example of weight 12 cusp modular form called the modular discriminant, and it can be represented as $ \eta(\tau)^{24} = (E_4(\tau)^3-E_6(\tau)^2)/1728 $.

\subsection{Jacobi forms}

A function $ \varphi_{k,m} : \mathcal{H} \times \mathbb{C} \to \mathbb{C} $ is called a \emph{Jacobi form} if it satisfies two transformation properties
\begin{alignat}{2}
    \varphi_{k,m}\left(\frac{a\tau+b}{c\tau+d}, \frac{z}{c\tau+d}\right) &= (c\tau+d)^k e^{\frac{2\pi i m c z^2}{c\tau+d}} \varphi_{k,m}(\tau,z) \, , \quad
    &&\text{for }\ \mqty(a & b \\ c & d) \in \mathrm{SL}(2,\mathbb{Z}) \, , \\
    \varphi_{k,m}(\tau, z+\lambda\tau + \mu) &= e^{-2\pi i m (\lambda^2\tau + 2\lambda z)} \varphi_{k,m}(\tau,z) \, , \quad
    &&\text{for }\ \lambda, \mu \in \mathbb{Z} \, , \label{eq:jacobi-transf2}
\end{alignat}
where $ k \in \mathbb{Z} $ is called the \emph{weight} and $ m \in \mathbb{Z}_{\geq 0} $ is called the \emph{index} or \emph{level}. It has a Fourier expansion of the form
\begin{align}
    \varphi_{k,m}(\tau,z) = \sum_{n,r} c(n,r) q^n \xi^r \, ,
\end{align}
where $ \xi = e^{2\pi i z} $. The Jacobi form $ \varphi_{k,m} $ is called a holomorphic (resp. cusp or weak) Jacobi form if $ c(n,r)=0 $ unless $ 4nm \geq r^2 $ (resp. $ 4nm>r^2 $ or $ n \geq 0 $). The transformation property \eqref{eq:jacobi-transf2} implies that the Fourier expansion can be written as
\begin{align}\label{eq:jacobi-theta-decomp}
    \varphi_{k,m}(\tau,z) = \sum_{\ell \in \mathbb{Z}} h_\ell(\tau) q^{\ell^2/4m} \xi^\ell = \sum_{\ell \in \mathbb{Z}/2m\mathbb{Z}} h_{\ell}(\tau) \vartheta_{m,\ell}(\tau,z) \, ,
\end{align}
where $ h_{\ell}(\tau) = h_{\ell+2m \lambda}(\tau) $ for $ \lambda\in\mathbb{Z} $ and
\begin{align}
    \vartheta_{m,\ell}(\tau,z) = \sum_{n \equiv \ell (\operatorname{mod}2m)} q^{n^2/4m} \xi^n = \sum_{k \in \mathbb{Z}} q^{(\ell+2mk)^2/4m} \xi^{\ell+2mk}
\end{align}
is the theta function of index $ m $ and characteristic $ \ell $. If $ \varphi_{k,m}(\tau,z) $ is a holomorphic (resp. cusp of weak) Jacobi form, then $ h_{\ell}(\tau) $ is a weight $ k-1/2 $ holomorphic (resp. cusp or weakly holomorphic) vector-valued modular form. The ring of weak Jacobi form is freely generated over the ring of modular forms on $ \mathrm{SL}(2,\mathbb{Z}) $. The generators of the ring are given by
\begin{align}
    \varphi_{-2,1}(\tau,z) = -\frac{\theta_1(\tau,z)^2}{\eta(\tau)^6} \, , \quad
    \varphi_{0,1}(\tau,z) = 4\sum_{i=2}^4 \frac{\theta_i(\tau,z)^2}{\theta_i(\tau,0)^2} \, ,
\end{align}
where
\begin{align}
    \begin{alignedat}{2}\label{Jacobi-theta}
        &\theta_1(\tau, z) = -i \sum_{n \in \mathbb{Z}} (-1)^{n} q^{\frac{1}{2}(n+1/2)^2} \xi^{n+1/2} \, , \quad
        &&\theta_2(\tau, z) = \sum_{n \in \mathbb{Z}} q^{\frac{1}{2}(n+1/2)^2} \xi^{n+1/2} \, , \\
        &\theta_3(\tau, z) = \sum_{n \in \mathbb{Z}} q^{\frac{n^2}{2}} \xi^n \, , \quad
        &&\theta_4(\tau, z) = \sum_{n \in \mathbb{Z}} (-1)^n q^{\frac{n^2}{2}} \xi^n \, ,
    \end{alignedat}
\end{align}
are Jacobi theta functions. Hence, any weak Jacobi form $ \varphi_{k,m} $ of weight $ k $, index $ m $ can be uniquely represented as
\begin{align}
    \varphi_{k,m}(\tau,z) = \underset{a_3+a_4=m, a_i \in \mathbb{Z}_{\geq0}}{\sum_{4a_1+6a_2-2a_3=k}} c_{a_i} E_4(\tau)^{a_1} E_6(\tau)^{a_2} \varphi_{-2,1}(\tau,z)^{a_3} \varphi_{0,1}(\tau,z)^{a_4} \, ,
\end{align}
where $ c_{a_i} \in \mathbb{C} $ are some constants. For a Jacobi form $ \varphi_{k,m} $ of weight $ k $ and index $ m $, one can introduce the \emph{Hecke operator} defined as \cite{Eichler:1995}
\begin{align}
    T_n \varphi_{k,m}(\tau,z) = n^{k-1} \sum_{ad=n} \sum_{b(\operatorname{mod}d)} d^{-k} \varphi_{k,m}\left(\frac{a\tau+b}{d}, az\right) \, ,
\end{align}
which is a weight $ k $, index $ nm $ Jacobi form. This operator is useful for considering the elliptic genera of heterotic strings.

The weak Jacobi forms are invariant under the Weyl group transformation of $ SU(2) $ acting on $ z $. This can be generalized into other Lie algebras. Let $ \mathfrak{g} $ be a simple Lie algebra of rank $ l $, $ \mathfrak{h}_{\mathbb{C}} \cong \mathbb{C}^l $ be the complexification of its Cartan subalgebra, $ W $ be its Weyl group, and $ \langle \cdot, \cdot \rangle $ be the Killing form on $ \mathfrak{h}_{\mathbb{C}} $. A \emph{Weyl invariant Jacobi form} is a function $ \varphi_{k,m} : \mathcal{H} \times \mathfrak{h}_{\mathbb{C}} \to \mathbb{C} $ which has Weyl-invariance $ \varphi_{k,m}(\tau,wz) = \varphi_{k,m}(\tau,z) $ and satisfies two transformation properties
\begin{alignat}{2}
    \varphi_{k,m}\left(\frac{a\tau+b}{c\tau+d}, \frac{z}{c\tau+d}\right) &= (c\tau+d)^k e^{\frac{\pi i m c \langle z, z \rangle}{c\tau+d}} \varphi_{k,m}(\tau,z) \, , \quad
    &&\text{for }\ \mqty(a & b \\ c & d) \in \mathrm{SL}(2,\mathbb{Z}) \, , \\
    \varphi_{k,m}(\tau, z+\lambda\tau + \mu) &= e^{-\pi i m (\langle \lambda, \lambda \rangle \tau + 2\langle \lambda, z \rangle)} \varphi_{k,m}(\tau,z) \, , \quad
    &&\text{for }\ \lambda, \mu \in Q^\vee \, ,
\end{alignat}
where $ Q^\vee $ is the coroot lattice of $ \mathfrak{g} $. The Fourier expansion of the Weyl invariant Jacobi form which is a counterpart of \eqref{eq:jacobi-theta-decomp} is
\begin{align}
    \varphi_{k,m}(\tau,z) = \sum_{\lambda \in P/mQ^\vee} h_{\lambda}(\tau) \vartheta_{m,\lambda}(\tau,z)
\end{align}
where $ P $ is the weight lattice of $ \mathfrak{g} $ and the theta function associated to the integral lattice $ mQ^\vee $ is given by \cite{Kac:1984mq}
\begin{align}
    \vartheta_{m,\lambda}(\tau,z) = \underset{\mu \equiv \lambda (\operatorname{mod} mQ^\vee)}{\sum_{\mu \in P}} q^{\frac{1}{2m} \langle \mu ,\mu \rangle} e^{2\pi i \langle \mu, z \rangle} \, .
\end{align}
The ring of Weyl-invariant Jacobi form is freely generated over the ring of $ \mathrm{SL}(2,\mathbb{Z}) $ modular forms except $ \mathfrak{g}=E_8 $ case \cite{Wirthmuller}. In this paper, we use $ SU(3) $ Weyl invariant Jacobi forms in subsection~\ref{subsubsec:P2SU3}, which is generated by three fundamental Jacobi forms $ \varphi_k^{A_2} $ of weight $ -k $ and index $ 1 $ with $ k=0,2,3 $. They are given by
\begin{align}
    \varphi_k^{A_2}(\tau,\phi_1,\phi_2) = \mathcal{Z}^{3-k} \left. \prod_{j=1}^3 \frac{i\theta_1(\tau,x_j)}{\eta(\tau)^3} \right|_{\sum x_j = 0} \, , \quad (k=0,2,3) \, ,
\end{align}
where
\begin{align}
    \mathcal{Z} = \frac{1}{2\pi i} \left( \sum_{j=1}^3 \frac{\partial}{\partial x_j} + \frac{\pi^2}{3} E_2(\tau) \sum_{j=1}^3 x_j \right)
\end{align}
and $ x_1=\phi_1 $, $ x_2=-\phi_1+\phi_2 $, $ x_3=-\phi_2 $.

\newpage

\section{Expressions of Modular ansatz at higher string orders}

In this appendix, we report the results of the numerator of the ansatz for higher string order.

\subsection{\texorpdfstring{$ B=\mathbb{P}^2 $}{B=P2} with \texorpdfstring{$ SU(2) $}{SU(2)} gauge symmetry} \label{appendix:ellP2-gauge-result}

\paragraph{$ d=2 $ case}
The numerator in the modular ansatz for the 2-string elliptic genus is
\begin{align*}
    \scalebox{0.43}{$\begin{aligned}
            \mathcal{N}_2 &= \frac{66373 E_4^{12} E_6 \varphi_2(\epsilon )^4 \varphi_2(\phi )^8}{15407021574586368}+\frac{14413445 E_4^9 E_6^3 \varphi_2(\epsilon )^4 \varphi_2(\phi )^8}{15407021574586368}+\frac{17891213 E_4^6 E_6^5 \varphi_2(\epsilon )^4 \varphi_2(\phi )^8}{5135673858195456}+\frac{23578127 E_4^3 E_6^7 \varphi_2(\epsilon )^4 \varphi_2(\phi )^8}{15407021574586368}+\frac{6515 E_6^9 \varphi_2(\epsilon )^4 \varphi_2(\phi )^8}{120367356051456}-\frac{E_4^{13} \varphi_2(\epsilon )^3 \varphi_2(\phi )^8 \varphi_0(\epsilon )}{156728328192} -\frac{1750859 E_4^{10} E_6^2 \varphi_2(\epsilon )^3 \varphi_2(\phi )^8 \varphi_0(\epsilon )}{1283918464548864} \\
            &-\frac{274261 E_4^7 E_6^4 \varphi_2(\epsilon )^3 \varphi_2(\phi )^8 \varphi_0(\epsilon )}{53496602689536}-\frac{2960047 E_4^4 E_6^6 \varphi_2(\epsilon )^3 \varphi_2(\phi )^8 \varphi_0(\epsilon )}{1283918464548864}-\frac{72455 E_4 E_6^8 \varphi_2(\epsilon )^3 \varphi_2(\phi )^8 \varphi_0(\epsilon )}{641959232274432}-\frac{308963 E_4^{11} E_6 \varphi_2(\epsilon )^2 \varphi_2(\phi )^8 \varphi_0(\epsilon )^2}{5135673858195456} -\frac{235025 E_4^8 E_6^3 \varphi_2(\epsilon )^2 \varphi_2(\phi )^8 \varphi_0(\epsilon )^2}{1711891286065152}+\frac{18461 E_4^5 E_6^5 \varphi_2(\epsilon )^2 \varphi_2(\phi )^8 \varphi_0(\epsilon )^2}{190210142896128} \\
            &+\frac{515591 E_4^2 E_6^7 \varphi_2(\epsilon )^2 \varphi_2(\phi )^8 \varphi_0(\epsilon )^2}{5135673858195456}+\frac{E_4^{12} \varphi_2(\epsilon ) \varphi_2(\phi )^8 \varphi_0(\epsilon )^3}{470184984576}+\frac{1748245 E_4^9 E_6^2 \varphi_2(\epsilon ) \varphi_2(\phi )^8 \varphi_0(\epsilon )^3}{3851755393646592}+\frac{255593 E_4^6 E_6^4 \varphi_2(\epsilon ) \varphi_2(\phi )^8 \varphi_0(\epsilon )^3}{160489808068608} +\frac{2474641 E_4^3 E_6^6 \varphi_2(\epsilon ) \varphi_2(\phi )^8 \varphi_0(\epsilon )^3}{3851755393646592}+\frac{42817 E_6^8 \varphi_2(\epsilon ) \varphi_2(\phi )^8 \varphi_0(\epsilon )^3}{1925877696823296} \\
            &+\frac{2209 E_4^{10} E_6 \varphi_2(\phi )^8 \varphi_0(\epsilon )^4}{106993205379072}+\frac{4559 E_4^7 E_6^3 \varphi_2(\phi )^8 \varphi_0(\epsilon )^4}{53496602689536}+\frac{9409 E_4^4 E_6^5 \varphi_2(\phi )^8 \varphi_0(\epsilon )^4}{106993205379072}-\frac{31 E_4^{13} \varphi_2(\epsilon )^4 \varphi_2(\phi )^7 \varphi_0(\phi )}{380420285792256} -\frac{22592101 E_4^{10} E_6^2 \varphi_2(\epsilon )^4 \varphi_2(\phi )^7 \varphi_0(\phi )}{10271347716390912}-\frac{55135309 E_4^7 E_6^4 \varphi_2(\epsilon )^4 \varphi_2(\phi )^7 \varphi_0(\phi )}{3423782572130304} \\
            &-\frac{15496247 E_4^4 E_6^6 \varphi_2(\epsilon )^4 \varphi_2(\phi )^7 \varphi_0(\phi )}{1141260857376768}-\frac{93295 E_4 E_6^8 \varphi_2(\epsilon )^4 \varphi_2(\phi )^7 \varphi_0(\phi )}{80244904034304}+\frac{8318819 E_4^{11} E_6 \varphi_2(\epsilon )^3 \varphi_2(\phi )^7 \varphi_0(\epsilon ) \varphi_0(\phi )}{2567836929097728} +\frac{10167271 E_4^8 E_6^3 \varphi_2(\epsilon )^3 \varphi_2(\phi )^7 \varphi_0(\epsilon ) \varphi_0(\phi )}{427972821516288}+\frac{5758763 E_4^5 E_6^5 \varphi_2(\epsilon )^3 \varphi_2(\phi )^7 \varphi_0(\epsilon ) \varphi_0(\phi )}{285315214344192} \\
            &+\frac{297355 E_4^2 E_6^7 \varphi_2(\epsilon )^3 \varphi_2(\phi )^7 \varphi_0(\epsilon ) \varphi_0(\phi )}{160489808068608}+\frac{308963 E_4^{12} \varphi_2(\epsilon )^2 \varphi_2(\phi )^7 \varphi_0(\epsilon )^2 \varphi_0(\phi )}{3423782572130304}+\frac{217883 E_4^9 E_6^2 \varphi_2(\epsilon )^2 \varphi_2(\phi )^7 \varphi_0(\epsilon )^2 \varphi_0(\phi )}{380420285792256} -\frac{124901 E_4^6 E_6^4 \varphi_2(\epsilon )^2 \varphi_2(\phi )^7 \varphi_0(\epsilon )^2 \varphi_0(\phi )}{1141260857376768}-\frac{1839623 E_4^3 E_6^6 \varphi_2(\epsilon )^2 \varphi_2(\phi )^7 \varphi_0(\epsilon )^2 \varphi_0(\phi )}{3423782572130304} \\
            &-\frac{193 E_6^8 \varphi_2(\epsilon )^2 \varphi_2(\phi )^7 \varphi_0(\epsilon )^2 \varphi_0(\phi )}{11888133931008}-\frac{2729239 E_4^{10} E_6 \varphi_2(\epsilon ) \varphi_2(\phi )^7 \varphi_0(\epsilon )^3 \varphi_0(\phi )}{2567836929097728}-\frac{351559 E_4^7 E_6^3 \varphi_2(\epsilon ) \varphi_2(\phi )^7 \varphi_0(\epsilon )^3 \varphi_0(\phi )}{47552535724032} -\frac{5102957 E_4^4 E_6^5 \varphi_2(\epsilon ) \varphi_2(\phi )^7 \varphi_0(\epsilon )^3 \varphi_0(\phi )}{855945643032576}-\frac{162229 E_4 E_6^7 \varphi_2(\epsilon ) \varphi_2(\phi )^7 \varphi_0(\epsilon )^3 \varphi_0(\phi )}{320979616137216} \\
            &-\frac{2209 E_4^{11} \varphi_2(\phi )^7 \varphi_0(\epsilon )^4 \varphi_0(\phi )}{71328803586048}-\frac{35297 E_4^8 E_6^2 \varphi_2(\phi )^7 \varphi_0(\epsilon )^4 \varphi_0(\phi )}{106993205379072}-\frac{128371 E_4^5 E_6^4 \varphi_2(\phi )^7 \varphi_0(\epsilon )^4 \varphi_0(\phi )}{213986410758144} -\frac{2813 E_4^2 E_6^6 \varphi_2(\phi )^7 \varphi_0(\epsilon )^4 \varphi_0(\phi )}{26748301344768}+\frac{994963 E_4^{11} E_6 \varphi_2(\epsilon )^4 \varphi_2(\phi )^6 \varphi_0(\phi )^2}{855945643032576}+\frac{20218399 E_4^8 E_6^3 \varphi_2(\epsilon )^4 \varphi_2(\phi )^6 \varphi_0(\phi )^2}{855945643032576} \\
            &+\frac{33743881 E_4^5 E_6^5 \varphi_2(\epsilon )^4 \varphi_2(\phi )^6 \varphi_0(\phi )^2}{855945643032576}+\frac{2251031 E_4^2 E_6^7 \varphi_2(\epsilon )^4 \varphi_2(\phi )^6 \varphi_0(\phi )^2}{285315214344192} -\frac{497387 E_4^{12} \varphi_2(\epsilon )^3 \varphi_2(\phi )^6 \varphi_0(\epsilon ) \varphi_0(\phi )^2}{285315214344192}-\frac{90052831 E_4^9 E_6^2 \varphi_2(\epsilon )^3 \varphi_2(\phi )^6 \varphi_0(\epsilon ) \varphi_0(\phi )^2}{2567836929097728}-\frac{50039299 E_4^6 E_6^4 \varphi_2(\epsilon )^3 \varphi_2(\phi )^6 \varphi_0(\epsilon ) \varphi_0(\phi )^2}{855945643032576} \\
            &-\frac{1116959 E_4^3 E_6^6 \varphi_2(\epsilon )^3 \varphi_2(\phi )^6 \varphi_0(\epsilon ) \varphi_0(\phi )^2}{95105071448064}+\frac{5911 E_6^8 \varphi_2(\epsilon )^3 \varphi_2(\phi )^6 \varphi_0(\epsilon ) \varphi_0(\phi )^2}{160489808068608} -\frac{452923 E_4^{10} E_6 \varphi_2(\epsilon )^2 \varphi_2(\phi )^6 \varphi_0(\epsilon )^2 \varphi_0(\phi )^2}{855945643032576}-\frac{343399 E_4^7 E_6^3 \varphi_2(\epsilon )^2 \varphi_2(\phi )^6 \varphi_0(\epsilon )^2 \varphi_0(\phi )^2}{855945643032576}+\frac{746111 E_4^4 E_6^5 \varphi_2(\epsilon )^2 \varphi_2(\phi )^6 \varphi_0(\epsilon )^2 \varphi_0(\phi )^2}{855945643032576} \\
            &+\frac{5579 E_4 E_6^7 \varphi_2(\epsilon )^2 \varphi_2(\phi )^6 \varphi_0(\epsilon )^2 \varphi_0(\phi )^2}{95105071448064}+\frac{1430531 E_4^{11} \varphi_2(\epsilon ) \varphi_2(\phi )^6 \varphi_0(\epsilon )^3 \varphi_0(\phi )^2}{2567836929097728} +\frac{9249029 E_4^8 E_6^2 \varphi_2(\epsilon ) \varphi_2(\phi )^6 \varphi_0(\epsilon )^3 \varphi_0(\phi )^2}{855945643032576}+\frac{5033809 E_4^5 E_6^4 \varphi_2(\epsilon ) \varphi_2(\phi )^6 \varphi_0(\epsilon )^3 \varphi_0(\phi )^2}{285315214344192}+\frac{9125653 E_4^2 E_6^6 \varphi_2(\epsilon ) \varphi_2(\phi )^6 \varphi_0(\epsilon )^3 \varphi_0(\phi )^2}{2567836929097728} \\
            &+\frac{2209 E_4^9 E_6 \varphi_2(\phi )^6 \varphi_0(\epsilon )^4 \varphi_0(\phi )^2}{6687075336192}+\frac{36161 E_4^6 E_6^3 \varphi_2(\phi )^6 \varphi_0(\epsilon )^4 \varphi_0(\phi )^2}{26748301344768} +\frac{8185 E_4^3 E_6^5 \varphi_2(\phi )^6 \varphi_0(\epsilon )^4 \varphi_0(\phi )^2}{13374150672384}+\frac{841 E_6^7 \varphi_2(\phi )^6 \varphi_0(\epsilon )^4 \varphi_0(\phi )^2}{26748301344768}+\frac{467 E_4^{12} \varphi_2(\epsilon )^4 \varphi_2(\phi )^5 \varphi_0(\phi )^3}{3423782572130304}-\frac{115454791 E_4^9 E_6^2 \varphi_2(\epsilon )^4 \varphi_2(\phi )^5 \varphi_0(\phi )^3}{10271347716390912} \\
            &-\frac{50271893 E_4^6 E_6^4 \varphi_2(\epsilon )^4 \varphi_2(\phi )^5 \varphi_0(\phi )^3}{1141260857376768} -\frac{66340999 E_4^3 E_6^6 \varphi_2(\epsilon )^4 \varphi_2(\phi )^5 \varphi_0(\phi )^3}{3423782572130304}-\frac{139243 E_6^8 \varphi_2(\epsilon )^4 \varphi_2(\phi )^5 \varphi_0(\phi )^3}{320979616137216}+\frac{43397977 E_4^{10} E_6 \varphi_2(\epsilon )^3 \varphi_2(\phi )^5 \varphi_0(\epsilon ) \varphi_0(\phi )^3}{2567836929097728}+\frac{9365339 E_4^7 E_6^3 \varphi_2(\epsilon )^3 \varphi_2(\phi )^5 \varphi_0(\epsilon ) \varphi_0(\phi )^3}{142657607172096} \\
            &+\frac{24366563 E_4^4 E_6^5 \varphi_2(\epsilon )^3 \varphi_2(\phi )^5 \varphi_0(\epsilon ) \varphi_0(\phi )^3}{855945643032576} +\frac{135379 E_4 E_6^7 \varphi_2(\epsilon )^3 \varphi_2(\phi )^5 \varphi_0(\epsilon ) \varphi_0(\phi )^3}{320979616137216}-\frac{633485 E_4^{11} \varphi_2(\epsilon )^2 \varphi_2(\phi )^5 \varphi_0(\epsilon )^2 \varphi_0(\phi )^3}{10271347716390912}+\frac{941921 E_4^8 E_6^2 \varphi_2(\epsilon )^2 \varphi_2(\phi )^5 \varphi_0(\epsilon )^2 \varphi_0(\phi )^3}{3423782572130304}-\frac{969269 E_4^5 E_6^4 \varphi_2(\epsilon )^2 \varphi_2(\phi )^5 \varphi_0(\epsilon )^2 \varphi_0(\phi )^3}{3423782572130304} \\
            &+\frac{715529 E_4^2 E_6^6 \varphi_2(\epsilon )^2 \varphi_2(\phi )^5 \varphi_0(\epsilon )^2 \varphi_0(\phi )^3}{10271347716390912} -\frac{12931165 E_4^9 E_6 \varphi_2(\epsilon ) \varphi_2(\phi )^5 \varphi_0(\epsilon )^3 \varphi_0(\phi )^3}{2567836929097728}-\frac{8461877 E_4^6 E_6^3 \varphi_2(\epsilon ) \varphi_2(\phi )^5 \varphi_0(\epsilon )^3 \varphi_0(\phi )^3}{427972821516288}-\frac{2539205 E_4^3 E_6^5 \varphi_2(\epsilon ) \varphi_2(\phi )^5 \varphi_0(\epsilon )^3 \varphi_0(\phi )^3}{285315214344192}-\frac{66991 E_6^7 \varphi_2(\epsilon ) \varphi_2(\phi )^5 \varphi_0(\epsilon )^3 \varphi_0(\phi )^3}{320979616137216} \\
            &-\frac{6533 E_4^{10} \varphi_2(\phi )^5 \varphi_0(\epsilon )^4 \varphi_0(\phi )^3}{213986410758144} -\frac{114005 E_4^7 E_6^2 \varphi_2(\phi )^5 \varphi_0(\epsilon )^4 \varphi_0(\phi )^3}{106993205379072}-\frac{247549 E_4^4 E_6^4 \varphi_2(\phi )^5 \varphi_0(\epsilon )^4 \varphi_0(\phi )^3}{213986410758144}-\frac{9077 E_4 E_6^6 \varphi_2(\phi )^5 \varphi_0(\epsilon )^4 \varphi_0(\phi )^3}{53496602689536}+\frac{4294439 E_4^{10} E_6 \varphi_2(\epsilon )^4 \varphi_2(\phi )^4 \varphi_0(\phi )^4}{5135673858195456} \\
            &+\frac{23498077 E_4^7 E_6^3 \varphi_2(\epsilon )^4 \varphi_2(\phi )^4 \varphi_0(\phi )^4}{1711891286065152} +\frac{24603167 E_4^4 E_6^5 \varphi_2(\epsilon )^4 \varphi_2(\phi )^4 \varphi_0(\phi )^4}{1711891286065152}+\frac{5677669 E_4 E_6^7 \varphi_2(\epsilon )^4 \varphi_2(\phi )^4 \varphi_0(\phi )^4}{5135673858195456}-\frac{1610827 E_4^{11} \varphi_2(\epsilon )^3 \varphi_2(\phi )^4 \varphi_0(\epsilon ) \varphi_0(\phi )^4}{1283918464548864}-\frac{2231221 E_4^8 E_6^2 \varphi_2(\epsilon )^3 \varphi_2(\phi )^4 \varphi_0(\epsilon ) \varphi_0(\phi )^4}{106993205379072} \\
            &-\frac{3029807 E_4^5 E_6^4 \varphi_2(\epsilon )^3 \varphi_2(\phi )^4 \varphi_0(\epsilon ) \varphi_0(\phi )^4}{142657607172096} -\frac{788809 E_4^2 E_6^6 \varphi_2(\epsilon )^3 \varphi_2(\phi )^4 \varphi_0(\epsilon ) \varphi_0(\phi )^4}{641959232274432}+\frac{2339677 E_4^9 E_6 \varphi_2(\epsilon )^2 \varphi_2(\phi )^4 \varphi_0(\epsilon )^2 \varphi_0(\phi )^4}{5135673858195456}+\frac{175 E_4^6 E_6^3 \varphi_2(\epsilon )^2 \varphi_2(\phi )^4 \varphi_0(\epsilon )^2 \varphi_0(\phi )^4}{21134460321792}-\frac{787483 E_4^3 E_6^5 \varphi_2(\epsilon )^2 \varphi_2(\phi )^4 \varphi_0(\epsilon )^2 \varphi_0(\phi )^4}{1711891286065152} \\
            &-\frac{19753 E_6^7 \varphi_2(\epsilon )^2 \varphi_2(\phi )^4 \varphi_0(\epsilon )^2 \varphi_0(\phi )^4}{5135673858195456} +\frac{142369 E_4^{10} \varphi_2(\epsilon ) \varphi_2(\phi )^4 \varphi_0(\epsilon )^3 \varphi_0(\phi )^4}{427972821516288}+\frac{319811 E_4^7 E_6^2 \varphi_2(\epsilon ) \varphi_2(\phi )^4 \varphi_0(\epsilon )^3 \varphi_0(\phi )^4}{53496602689536}+\frac{2868973 E_4^4 E_6^4 \varphi_2(\epsilon ) \varphi_2(\phi )^4 \varphi_0(\epsilon )^3 \varphi_0(\phi )^4}{427972821516288}+\frac{4375 E_4 E_6^6 \varphi_2(\epsilon ) \varphi_2(\phi )^4 \varphi_0(\epsilon )^3 \varphi_0(\phi )^4}{7925422620672} \\
            &+\frac{3847 E_4^8 E_6 \varphi_2(\phi )^4 \varphi_0(\epsilon )^4 \varphi_0(\phi )^4}{35664401793024} +\frac{10889 E_4^5 E_6^3 \varphi_2(\phi )^4 \varphi_0(\epsilon )^4 \varphi_0(\phi )^4}{17832200896512}+\frac{8935 E_4^2 E_6^5 \varphi_2(\phi )^4 \varphi_0(\epsilon )^4 \varphi_0(\phi )^4}{35664401793024}-\frac{467 E_4^{11} \varphi_2(\epsilon )^4 \varphi_2(\phi )^3 \varphi_0(\phi )^5}{10271347716390912}+\frac{6451567 E_4^8 E_6^2 \varphi_2(\epsilon )^4 \varphi_2(\phi )^3 \varphi_0(\phi )^5}{3423782572130304}+\frac{6196135 E_4^5 E_6^4 \varphi_2(\epsilon )^4 \varphi_2(\phi )^3 \varphi_0(\phi )^5}{1141260857376768} \\
            &+\frac{17446055 E_4^2 E_6^6 \varphi_2(\epsilon )^4 \varphi_2(\phi )^3 \varphi_0(\phi )^5}{10271347716390912}-\frac{7256579 E_4^9 E_6 \varphi_2(\epsilon )^3 \varphi_2(\phi )^3 \varphi_0(\epsilon ) \varphi_0(\phi )^5}{2567836929097728}-\frac{3366091 E_4^6 E_6^3 \varphi_2(\epsilon )^3 \varphi_2(\phi )^3 \varphi_0(\epsilon ) \varphi_0(\phi )^5}{427972821516288}-\frac{256777 E_4^3 E_6^5 \varphi_2(\epsilon )^3 \varphi_2(\phi )^3 \varphi_0(\epsilon ) \varphi_0(\phi )^5}{95105071448064}+\frac{5911 E_6^7 \varphi_2(\epsilon )^3 \varphi_2(\phi )^3 \varphi_0(\epsilon ) \varphi_0(\phi )^5}{320979616137216} \\
            &-\frac{137 E_4^{10} \varphi_2(\epsilon )^2 \varphi_2(\phi )^3 \varphi_0(\epsilon )^2 \varphi_0(\phi )^5}{1141260857376768}-\frac{1358603 E_4^7 E_6^2 \varphi_2(\epsilon )^2 \varphi_2(\phi )^3 \varphi_0(\epsilon )^2 \varphi_0(\phi )^5}{3423782572130304}+\frac{1322215 E_4^4 E_6^4 \varphi_2(\epsilon )^2 \varphi_2(\phi )^3 \varphi_0(\epsilon )^2 \varphi_0(\phi )^5}{3423782572130304}+\frac{36799 E_4 E_6^6 \varphi_2(\epsilon )^2 \varphi_2(\phi )^3 \varphi_0(\epsilon )^2 \varphi_0(\phi )^5}{3423782572130304}+\frac{776861 E_4^8 E_6 \varphi_2(\epsilon ) \varphi_2(\phi )^3 \varphi_0(\epsilon )^3 \varphi_0(\phi )^5}{855945643032576} \\
            &+\frac{1039267 E_4^5 E_6^3 \varphi_2(\epsilon ) \varphi_2(\phi )^3 \varphi_0(\epsilon )^3 \varphi_0(\phi )^5}{427972821516288}+\frac{628253 E_4^2 E_6^5 \varphi_2(\epsilon ) \varphi_2(\phi )^3 \varphi_0(\epsilon )^3 \varphi_0(\phi )^5}{855945643032576}+\frac{31 E_4^9 \varphi_2(\phi )^3 \varphi_0(\epsilon )^4 \varphi_0(\phi )^5}{213986410758144}+\frac{24259 E_4^6 E_6^2 \varphi_2(\phi )^3 \varphi_0(\epsilon )^4 \varphi_0(\phi )^5}{106993205379072}+\frac{10295 E_4^3 E_6^4 \varphi_2(\phi )^3 \varphi_0(\epsilon )^4 \varphi_0(\phi )^5}{213986410758144} \\
            &+\frac{841 E_6^6 \varphi_2(\phi )^3 \varphi_0(\epsilon )^4 \varphi_0(\phi )^5}{53496602689536}-\frac{4093 E_4^9 E_6 \varphi_2(\epsilon )^4 \varphi_2(\phi )^2 \varphi_0(\phi )^6}{10030613004288}-\frac{1284709 E_4^6 E_6^3 \varphi_2(\epsilon )^4 \varphi_2(\phi )^2 \varphi_0(\phi )^6}{213986410758144}-\frac{191657 E_4^3 E_6^5 \varphi_2(\epsilon )^4 \varphi_2(\phi )^2 \varphi_0(\phi )^6}{35664401793024}-\frac{147887 E_6^7 \varphi_2(\epsilon )^4 \varphi_2(\phi )^2 \varphi_0(\phi )^6}{641959232274432} \\
            &+\frac{1571869 E_4^{10} \varphi_2(\epsilon )^3 \varphi_2(\phi )^2 \varphi_0(\epsilon ) \varphi_0(\phi )^6}{2567836929097728}+\frac{7746575 E_4^7 E_6^2 \varphi_2(\epsilon )^3 \varphi_2(\phi )^2 \varphi_0(\epsilon ) \varphi_0(\phi )^6}{855945643032576}+\frac{6762949 E_4^4 E_6^4 \varphi_2(\epsilon )^3 \varphi_2(\phi )^2 \varphi_0(\epsilon ) \varphi_0(\phi )^6}{855945643032576}+\frac{684647 E_4 E_6^6 \varphi_2(\epsilon )^3 \varphi_2(\phi )^2 \varphi_0(\epsilon ) \varphi_0(\phi )^6}{2567836929097728}-\frac{18907 E_4^8 E_6 \varphi_2(\epsilon )^2 \varphi_2(\phi )^2 \varphi_0(\epsilon )^2 \varphi_0(\phi )^6}{213986410758144} \\
            &+\frac{5659 E_4^5 E_6^3 \varphi_2(\epsilon )^2 \varphi_2(\phi )^2 \varphi_0(\epsilon )^2 \varphi_0(\phi )^6}{106993205379072}+\frac{7589 E_4^2 E_6^5 \varphi_2(\epsilon )^2 \varphi_2(\phi )^2 \varphi_0(\epsilon )^2 \varphi_0(\phi )^6}{213986410758144}-\frac{443885 E_4^9 \varphi_2(\epsilon ) \varphi_2(\phi )^2 \varphi_0(\epsilon )^3 \varphi_0(\phi )^6}{2567836929097728}-\frac{760613 E_4^6 E_6^2 \varphi_2(\epsilon ) \varphi_2(\phi )^2 \varphi_0(\epsilon )^3 \varphi_0(\phi )^6}{285315214344192}-\frac{2121013 E_4^3 E_6^4 \varphi_2(\epsilon ) \varphi_2(\phi )^2 \varphi_0(\epsilon )^3 \varphi_0(\phi )^6}{855945643032576} \\
            &-\frac{282151 E_6^6 \varphi_2(\epsilon ) \varphi_2(\phi )^2 \varphi_0(\epsilon )^3 \varphi_0(\phi )^6}{2567836929097728}-\frac{3565 E_4^7 E_6 \varphi_2(\phi )^2 \varphi_0(\epsilon )^4 \varphi_0(\phi )^6}{53496602689536}-\frac{6947 E_4^4 E_6^3 \varphi_2(\phi )^2 \varphi_0(\epsilon )^4 \varphi_0(\phi )^6}{26748301344768}-\frac{3277 E_4 E_6^5 \varphi_2(\phi )^2 \varphi_0(\epsilon )^4 \varphi_0(\phi )^6}{53496602689536}+\frac{31 E_4^{10} \varphi_2(\epsilon )^4 \varphi_2(\phi ) \varphi_0(\phi )^7}{10271347716390912}+\frac{1995541 E_4^7 E_6^2 \varphi_2(\epsilon )^4 \varphi_2(\phi ) \varphi_0(\phi )^7}{3423782572130304} \\
            &+\frac{246421 E_4^4 E_6^4 \varphi_2(\epsilon )^4 \varphi_2(\phi ) \varphi_0(\phi )^7}{126806761930752}+\frac{4908413 E_4 E_6^6 \varphi_2(\epsilon )^4 \varphi_2(\phi ) \varphi_0(\phi )^7}{10271347716390912}-\frac{250129 E_4^8 E_6 \varphi_2(\epsilon )^3 \varphi_2(\phi ) \varphi_0(\epsilon ) \varphi_0(\phi )^7}{285315214344192}-\frac{420143 E_4^5 E_6^3 \varphi_2(\epsilon )^3 \varphi_2(\phi ) \varphi_0(\epsilon ) \varphi_0(\phi )^7}{142657607172096}-\frac{181393 E_4^2 E_6^5 \varphi_2(\epsilon )^3 \varphi_2(\phi ) \varphi_0(\epsilon ) \varphi_0(\phi )^7}{285315214344192} \\
            &+\frac{36469 E_4^9 \varphi_2(\epsilon )^2 \varphi_2(\phi ) \varphi_0(\epsilon )^2 \varphi_0(\phi )^7}{10271347716390912}+\frac{28319 E_4^6 E_6^2 \varphi_2(\epsilon )^2 \varphi_2(\phi ) \varphi_0(\epsilon )^2 \varphi_0(\phi )^7}{380420285792256}-\frac{274339 E_4^3 E_6^4 \varphi_2(\epsilon )^2 \varphi_2(\phi ) \varphi_0(\epsilon )^2 \varphi_0(\phi )^7}{3423782572130304}+\frac{21935 E_6^6 \varphi_2(\epsilon )^2 \varphi_2(\phi ) \varphi_0(\epsilon )^2 \varphi_0(\phi )^7}{10271347716390912}+\frac{208991 E_4^7 E_6 \varphi_2(\epsilon ) \varphi_2(\phi ) \varphi_0(\epsilon )^3 \varphi_0(\phi )^7}{855945643032576} \\
            &+\frac{377953 E_4^4 E_6^3 \varphi_2(\epsilon ) \varphi_2(\phi ) \varphi_0(\epsilon )^3 \varphi_0(\phi )^7}{427972821516288}+\frac{196319 E_4 E_6^5 \varphi_2(\epsilon ) \varphi_2(\phi ) \varphi_0(\epsilon )^3 \varphi_0(\phi )^7}{855945643032576}+\frac{961 E_4^8 \varphi_2(\phi ) \varphi_0(\epsilon )^4 \varphi_0(\phi )^7}{213986410758144}+\frac{3503 E_4^5 E_6^2 \varphi_2(\phi ) \varphi_0(\epsilon )^4 \varphi_0(\phi )^7}{106993205379072}+\frac{12769 E_4^2 E_6^4 \varphi_2(\phi ) \varphi_0(\epsilon )^4 \varphi_0(\phi )^7}{213986410758144}
    \end{aligned}$}
\end{align*}
The discriminants of Picard-Fuchs operators are  $ \Delta_1 = (1-4z_2) $ and
\begin{align*}
    \scalebox{0.79}{$\begin{aligned}
            \Delta_2 &= 1-3 z_1+3 z_1^2-z_1^3-4 z_2+120 z_1 z_2-309 z_1^2 z_2+274 z_1^3 z_2-81 z_1^4 z_2-432 z_1 z_2^2 +5724 z_1^2 z_2^2-12096 z_1^3 z_2^2 \\
            &+8991 z_1^4 z_2^2-2187 z_1^5 z_2^2-19440 z_1^2 z_2^3+139968 z_1^3 z_2^3 -222588 z_1^4 z_2^3+122472 z_1^5 z_2^3-19683 z_1^6 z_2^3+5184 z_1^2 z_2^4 \\
            &-476928 z_1^3 z_2^4+1894752 z_1^4 z_2^4-1924560 z_1^5 z_2^4+551124 z_1^6 z_2^4+373248 z_1^3 z_2^5-6811776 z_1^4 z_2^5+13996800 z_1^5 z_2^5 \\
            &-6613488 z_1^6 z_2^5+9517824 z_1^4 z_2^6-53187840 z_1^5 z_2^6+44089920 z_1^6 z_2^6-2239488 z_1^4 z_2^7+103016448 z_1^5 z_2^7 \\
            &-176359680 z_1^6 z_2^7-80621568 z_1^5 z_2^8+423263232 z_1^6 z_2^8-564350976 z_1^6 z_2^9+322486272 z_1^6 z_2^{10}+27 z_3-144 z_2 z_3 \\
            &+2628 z_1 z_2 z_3+128 z_2^2 z_3-13936 z_1 z_2^2 z_3+95456 z_1^2 z_2^2 z_3+12160 z_1 z_2^3 z_3-538256 z_1^2 z_2^3 z_3+1553824 z_1^3 z_2^3 z_3 \\
            &+619776 z_1^2 z_2^4 z_3-10301760 z_1^3 z_2^4 z_3+10329984 z_1^4 z_2^4 z_3-172800 z_1^2 z_2^5 z_3+18385920 z_1^3 z_2^5 z_3-93308544 z_1^4 z_2^5 z_3 \\
            &+16003008 z_1^5 z_2^5 z_3-10257408 z_1^3 z_2^6 z_3+268918272 z_1^4 z_2^6 z_3-256048128 z_1^5 z_2^6 z_3-317011968 z_1^4 z_2^7 z_3 \\
            &+1536288768 z_1^5 z_2^7 z_3+292626432 z_1^4 z_2^8 z_3-4096770048 z_1^5 z_2^8 z_3+4096770048 z_1^5 z_2^9 z_3-1024 z_2^3 z_3^2-100352 z_1 z_2^4 z_3^2 \\
            &-3687936 z_1^2 z_2^5 z_3^2+1404928 z_1^2 z_2^6 z_3^2-61465600 z_1^3 z_2^6 z_3^2+88510464 z_1^3 z_2^7 z_3^2-1858719744 z_1^4 z_2^7 z_3^2 \\
            &+9293598720 z_1^4 z_2^8 z_3^2-7434878976 z_1^4 z_2^9 z_3^2+53971714048 z_1^4 z_2^{10} z_3^3
    \end{aligned}$}
\end{align*}
The numerator in the modular ansatz for the 3-string elliptic genus is


\newpage

\paragraph{$ d=3 $ case} The numerator in the modular ansatz for the 2-string elliptic genus is
\begin{align*}
    \scalebox{0.36}{$\begin{aligned}
            \mathcal{N}_2 &= -\frac{31 E_4^{13} \varphi_2(\epsilon )^4 \varphi_2(\phi )^6}{126806761930752}+\frac{2145883 E_4^{10} E_6^2 \varphi_2(\epsilon )^4 \varphi_2(\phi )^6}{3423782572130304}+\frac{5190883 E_4^7 E_6^4 \varphi_2(\epsilon )^4 \varphi_2(\phi )^6}{1141260857376768}+\frac{454435 E_4^4 E_6^6 \varphi_2(\epsilon )^4 \varphi_2(\phi )^6}{126806761930752}+\frac{54233 E_4 E_6^8 \varphi_2(\epsilon )^4 \varphi_2(\phi )^6}{213986410758144}-\frac{29743 E_4^{11} E_6 \varphi_2(\epsilon )^3 \varphi_0(\epsilon ) \varphi_2(\phi )^6}{31701690482688}-\frac{312677 E_4^8 E_6^3 \varphi_2(\epsilon )^3 \varphi_0(\epsilon ) \varphi_2(\phi )^6}{47552535724032}-\frac{513349 E_4^5 E_6^5 \varphi_2(\epsilon )^3 \varphi_0(\epsilon ) \varphi_2(\phi )^6}{95105071448064} \\
            &-\frac{10969 E_4^2 E_6^7 \varphi_2(\epsilon )^3 \varphi_0(\epsilon ) \varphi_2(\phi )^6}{23776267862016}+\frac{35 E_4^{12} \varphi_2(\epsilon )^2 \varphi_0(\epsilon )^2 \varphi_2(\phi )^6}{126806761930752}-\frac{1282895 E_4^9 E_6^2 \varphi_2(\epsilon )^2 \varphi_0(\epsilon )^2 \varphi_2(\phi )^6}{3423782572130304}+\frac{15857 E_4^6 E_6^4 \varphi_2(\epsilon )^2 \varphi_0(\epsilon )^2 \varphi_2(\phi )^6}{126806761930752}+\frac{272753 E_4^3 E_6^6 \varphi_2(\epsilon )^2 \varphi_0(\epsilon )^2 \varphi_2(\phi )^6}{1141260857376768}+\frac{1111 E_6^8 \varphi_2(\epsilon )^2 \varphi_0(\epsilon )^2 \varphi_2(\phi )^6}{106993205379072}+\frac{29267 E_4^{10} E_6 \varphi_2(\epsilon ) \varphi_0(\epsilon )^3 \varphi_2(\phi )^6}{95105071448064} \\
            &+\frac{152357 E_4^7 E_6^3 \varphi_2(\epsilon ) \varphi_0(\epsilon )^3 \varphi_2(\phi )^6}{71328803586048}+\frac{438125 E_4^4 E_6^5 \varphi_2(\epsilon ) \varphi_0(\epsilon )^3 \varphi_2(\phi )^6}{285315214344192}+\frac{12931 E_4 E_6^7 \varphi_2(\epsilon ) \varphi_0(\epsilon )^3 \varphi_2(\phi )^6}{142657607172096}+\frac{10609 E_4^8 E_6^2 \varphi_0(\epsilon )^4 \varphi_2(\phi )^6}{71328803586048}+\frac{4223 E_4^5 E_6^4 \varphi_0(\epsilon )^4 \varphi_2(\phi )^6}{35664401793024}+\frac{1681 E_4^2 E_6^6 \varphi_0(\epsilon )^4 \varphi_2(\phi )^6}{71328803586048}-\frac{61073 E_4^{11} E_6 \varphi_2(\epsilon )^4 \varphi_2(\phi )^5 \varphi_0(\phi )}{71328803586048} \\
            &-\frac{872515 E_4^8 E_6^3 \varphi_2(\epsilon )^4 \varphi_2(\phi )^5 \varphi_0(\phi )}{47552535724032}-\frac{19379 E_4^5 E_6^5 \varphi_2(\epsilon )^4 \varphi_2(\phi )^5 \varphi_0(\phi )}{660451885056}-\frac{788237 E_4^2 E_6^7 \varphi_2(\epsilon )^4 \varphi_2(\phi )^5 \varphi_0(\phi )}{142657607172096}+\frac{244591 E_4^{12} \varphi_2(\epsilon )^3 \varphi_0(\epsilon ) \varphi_2(\phi )^5 \varphi_0(\phi )}{190210142896128}+\frac{5114351 E_4^9 E_6^2 \varphi_2(\epsilon )^3 \varphi_0(\epsilon ) \varphi_2(\phi )^5 \varphi_0(\phi )}{190210142896128}+\frac{2743703 E_4^6 E_6^4 \varphi_2(\epsilon )^3 \varphi_0(\epsilon ) \varphi_2(\phi )^5 \varphi_0(\phi )}{63403380965376} \\
            &+\frac{1661165 E_4^3 E_6^6 \varphi_2(\epsilon )^3 \varphi_0(\epsilon ) \varphi_2(\phi )^5 \varphi_0(\phi )}{190210142896128}+\frac{655 E_6^8 \varphi_2(\epsilon )^3 \varphi_0(\epsilon ) \varphi_2(\phi )^5 \varphi_0(\phi )}{11888133931008}+\frac{133613 E_4^{10} E_6 \varphi_2(\epsilon )^2 \varphi_0(\epsilon )^2 \varphi_2(\phi )^5 \varphi_0(\phi )}{142657607172096}+\frac{1621 E_4^7 E_6^3 \varphi_2(\epsilon )^2 \varphi_0(\epsilon )^2 \varphi_2(\phi )^5 \varphi_0(\phi )}{1981355655168}-\frac{72029 E_4^4 E_6^5 \varphi_2(\epsilon )^2 \varphi_0(\epsilon )^2 \varphi_2(\phi )^5 \varphi_0(\phi )}{47552535724032}-\frac{17119 E_4 E_6^7 \varphi_2(\epsilon )^2 \varphi_0(\epsilon )^2 \varphi_2(\phi )^5 \varphi_0(\phi )}{71328803586048} \\
            &-\frac{80191 E_4^{11} \varphi_2(\epsilon ) \varphi_0(\epsilon )^3 \varphi_2(\phi )^5 \varphi_0(\phi )}{190210142896128}-\frac{1648799 E_4^8 E_6^2 \varphi_2(\epsilon ) \varphi_0(\epsilon )^3 \varphi_2(\phi )^5 \varphi_0(\phi )}{190210142896128}-\frac{2484485 E_4^5 E_6^4 \varphi_2(\epsilon ) \varphi_0(\epsilon )^3 \varphi_2(\phi )^5 \varphi_0(\phi )}{190210142896128}-\frac{431389 E_4^2 E_6^6 \varphi_2(\epsilon ) \varphi_0(\epsilon )^3 \varphi_2(\phi )^5 \varphi_0(\phi )}{190210142896128}-\frac{4429 E_4^9 E_6 \varphi_0(\epsilon )^4 \varphi_2(\phi )^5 \varphi_0(\phi )}{11888133931008}-\frac{6083 E_4^6 E_6^3 \varphi_0(\epsilon )^4 \varphi_2(\phi )^5 \varphi_0(\phi )}{5944066965504} \\
            &-\frac{4141 E_4^3 E_6^5 \varphi_0(\epsilon )^4 \varphi_2(\phi )^5 \varphi_0(\phi )}{11888133931008}+\frac{265 E_4^{12} \varphi_2(\epsilon )^4 \varphi_2(\phi )^4 \varphi_0(\phi )^2}{380420285792256}+\frac{23681963 E_4^9 E_6^2 \varphi_2(\epsilon )^4 \varphi_2(\phi )^4 \varphi_0(\phi )^2}{1141260857376768}+\frac{30255395 E_4^6 E_6^4 \varphi_2(\epsilon )^4 \varphi_2(\phi )^4 \varphi_0(\phi )^2}{380420285792256}+\frac{12933611 E_4^3 E_6^6 \varphi_2(\epsilon )^4 \varphi_2(\phi )^4 \varphi_0(\phi )^2}{380420285792256}+\frac{64129 E_6^8 \varphi_2(\epsilon )^4 \varphi_2(\phi )^4 \varphi_0(\phi )^2}{71328803586048} \\
            &-\frac{8770091 E_4^{10} E_6 \varphi_2(\epsilon )^3 \varphi_0(\epsilon ) \varphi_2(\phi )^4 \varphi_0(\phi )^2}{285315214344192}-\frac{2788789 E_4^7 E_6^3 \varphi_2(\epsilon )^3 \varphi_0(\epsilon ) \varphi_2(\phi )^4 \varphi_0(\phi )^2}{23776267862016}-\frac{4842413 E_4^4 E_6^5 \varphi_2(\epsilon )^3 \varphi_0(\epsilon ) \varphi_2(\phi )^4 \varphi_0(\phi )^2}{95105071448064}-\frac{234281 E_4 E_6^7 \varphi_2(\epsilon )^3 \varphi_0(\epsilon ) \varphi_2(\phi )^4 \varphi_0(\phi )^2}{142657607172096}-\frac{224861 E_4^{11} \varphi_2(\epsilon )^2 \varphi_0(\epsilon )^2 \varphi_2(\phi )^4 \varphi_0(\phi )^2}{380420285792256}-\frac{997549 E_4^8 E_6^2 \varphi_2(\epsilon )^2 \varphi_0(\epsilon )^2 \varphi_2(\phi )^4 \varphi_0(\phi )^2}{380420285792256} \\
            &+\frac{711281 E_4^5 E_6^4 \varphi_2(\epsilon )^2 \varphi_0(\epsilon )^2 \varphi_2(\phi )^4 \varphi_0(\phi )^2}{380420285792256}+\frac{511129 E_4^2 E_6^6 \varphi_2(\epsilon )^2 \varphi_0(\epsilon )^2 \varphi_2(\phi )^4 \varphi_0(\phi )^2}{380420285792256}+\frac{2798899 E_4^9 E_6 \varphi_2(\epsilon ) \varphi_0(\epsilon )^3 \varphi_2(\phi )^4 \varphi_0(\phi )^2}{285315214344192}+\frac{15887 E_4^6 E_6^3 \varphi_2(\epsilon ) \varphi_0(\epsilon )^3 \varphi_2(\phi )^4 \varphi_0(\phi )^2}{440301256704}+\frac{1405181 E_4^3 E_6^5 \varphi_2(\epsilon ) \varphi_0(\epsilon )^3 \varphi_2(\phi )^4 \varphi_0(\phi )^2}{95105071448064}+\frac{54511 E_6^7 \varphi_2(\epsilon ) \varphi_0(\epsilon )^3 \varphi_2(\phi )^4 \varphi_0(\phi )^2}{142657607172096} \\
            &+\frac{1849 E_4^{10} \varphi_0(\epsilon )^4 \varphi_2(\phi )^4 \varphi_0(\phi )^2}{7925422620672}+\frac{12437 E_4^7 E_6^2 \varphi_0(\epsilon )^4 \varphi_2(\phi )^4 \varphi_0(\phi )^2}{5944066965504}+\frac{46007 E_4^4 E_6^4 \varphi_0(\epsilon )^4 \varphi_2(\phi )^4 \varphi_0(\phi )^2}{23776267862016}+\frac{1189 E_4 E_6^6 \varphi_0(\epsilon )^4 \varphi_2(\phi )^4 \varphi_0(\phi )^2}{11888133931008}-\frac{604195 E_4^{10} E_6 \varphi_2(\epsilon )^4 \varphi_2(\phi )^3 \varphi_0(\phi )^3}{106993205379072}-\frac{5849939 E_4^7 E_6^3 \varphi_2(\epsilon )^4 \varphi_2(\phi )^3 \varphi_0(\phi )^3}{71328803586048}-\frac{56095 E_4^4 E_6^5 \varphi_2(\epsilon )^4 \varphi_2(\phi )^3 \varphi_0(\phi )^3}{660451885056} \\
            &-\frac{1635973 E_4 E_6^7 \varphi_2(\epsilon )^4 \varphi_2(\phi )^3 \varphi_0(\phi )^3}{213986410758144}+\frac{805445 E_4^{11} \varphi_2(\epsilon )^3 \varphi_0(\epsilon ) \varphi_2(\phi )^3 \varphi_0(\phi )^3}{95105071448064}+\frac{11584037 E_4^8 E_6^2 \varphi_2(\epsilon )^3 \varphi_0(\epsilon ) \varphi_2(\phi )^3 \varphi_0(\phi )^3}{95105071448064}+\frac{11953895 E_4^5 E_6^4 \varphi_2(\epsilon )^3 \varphi_0(\epsilon ) \varphi_2(\phi )^3 \varphi_0(\phi )^3}{95105071448064}+\frac{364261 E_4^2 E_6^6 \varphi_2(\epsilon )^3 \varphi_0(\epsilon ) \varphi_2(\phi )^3 \varphi_0(\phi )^3}{31701690482688}+\frac{366211 E_4^9 E_6 \varphi_2(\epsilon )^2 \varphi_0(\epsilon )^2 \varphi_2(\phi )^3 \varphi_0(\phi )^3}{213986410758144} \\
            &+\frac{577 E_4^6 E_6^3 \varphi_2(\epsilon )^2 \varphi_0(\epsilon )^2 \varphi_2(\phi )^3 \varphi_0(\phi )^3}{1320903770112}-\frac{146287 E_4^3 E_6^5 \varphi_2(\epsilon )^2 \varphi_0(\epsilon )^2 \varphi_2(\phi )^3 \varphi_0(\phi )^3}{71328803586048}-\frac{2603 E_6^7 \varphi_2(\epsilon )^2 \varphi_0(\epsilon )^2 \varphi_2(\phi )^3 \varphi_0(\phi )^3}{26748301344768}-\frac{27829 E_4^{10} \varphi_2(\epsilon ) \varphi_0(\epsilon )^3 \varphi_2(\phi )^3 \varphi_0(\phi )^3}{10567230160896}-\frac{10667911 E_4^7 E_6^2 \varphi_2(\epsilon ) \varphi_0(\epsilon )^3 \varphi_2(\phi )^3 \varphi_0(\phi )^3}{285315214344192}-\frac{10836397 E_4^4 E_6^4 \varphi_2(\epsilon ) \varphi_0(\epsilon )^3 \varphi_2(\phi )^3 \varphi_0(\phi )^3}{285315214344192} \\
            &-\frac{968629 E_4 E_6^6 \varphi_2(\epsilon ) \varphi_0(\epsilon )^3 \varphi_2(\phi )^3 \varphi_0(\phi )^3}{285315214344192}-\frac{23849 E_4^8 E_6 \varphi_0(\epsilon )^4 \varphi_2(\phi )^3 \varphi_0(\phi )^3}{17832200896512}-\frac{32167 E_4^5 E_6^3 \varphi_0(\epsilon )^4 \varphi_2(\phi )^3 \varphi_0(\phi )^3}{8916100448256}-\frac{15497 E_4^2 E_6^5 \varphi_0(\epsilon )^4 \varphi_2(\phi )^3 \varphi_0(\phi )^3}{17832200896512}-\frac{265 E_4^{11} \varphi_2(\epsilon )^4 \varphi_2(\phi )^2 \varphi_0(\phi )^4}{1141260857376768}+\frac{10326413 E_4^8 E_6^2 \varphi_2(\epsilon )^4 \varphi_2(\phi )^2 \varphi_0(\phi )^4}{380420285792256}+\frac{3582007 E_4^5 E_6^4 \varphi_2(\epsilon )^4 \varphi_2(\phi )^2 \varphi_0(\phi )^4}{42268920643584} \\
            &+\frac{26582677 E_4^2 E_6^6 \varphi_2(\epsilon )^4 \varphi_2(\phi )^2 \varphi_0(\phi )^4}{1141260857376768}-\frac{3877709 E_4^9 E_6 \varphi_2(\epsilon )^3 \varphi_0(\epsilon ) \varphi_2(\phi )^2 \varphi_0(\phi )^4}{95105071448064}-\frac{499229 E_4^6 E_6^3 \varphi_2(\epsilon )^3 \varphi_0(\epsilon ) \varphi_2(\phi )^2 \varphi_0(\phi )^4}{3962711310336}-\frac{1076579 E_4^3 E_6^5 \varphi_2(\epsilon )^3 \varphi_0(\epsilon ) \varphi_2(\phi )^2 \varphi_0(\phi )^4}{31701690482688}+\frac{5911 E_6^7 \varphi_2(\epsilon )^3 \varphi_0(\epsilon ) \varphi_2(\phi )^2 \varphi_0(\phi )^4}{47552535724032}+\frac{99365 E_4^{10} \varphi_2(\epsilon )^2 \varphi_0(\epsilon )^2 \varphi_2(\phi )^2 \varphi_0(\phi )^4}{1141260857376768} \\
            &-\frac{106019 E_4^7 E_6^2 \varphi_2(\epsilon )^2 \varphi_0(\epsilon )^2 \varphi_2(\phi )^2 \varphi_0(\phi )^4}{126806761930752}+\frac{237229 E_4^4 E_6^4 \varphi_2(\epsilon )^2 \varphi_0(\epsilon )^2 \varphi_2(\phi )^2 \varphi_0(\phi )^4}{380420285792256}+\frac{143119 E_4 E_6^6 \varphi_2(\epsilon )^2 \varphi_0(\epsilon )^2 \varphi_2(\phi )^2 \varphi_0(\phi )^4}{1141260857376768}+\frac{1157371 E_4^8 E_6 \varphi_2(\epsilon ) \varphi_0(\epsilon )^3 \varphi_2(\phi )^2 \varphi_0(\phi )^4}{95105071448064}+\frac{1818821 E_4^5 E_6^3 \varphi_2(\epsilon ) \varphi_0(\epsilon )^3 \varphi_2(\phi )^2 \varphi_0(\phi )^4}{47552535724032} \\
            &+\frac{1011067 E_4^2 E_6^5 \varphi_2(\epsilon ) \varphi_0(\epsilon )^3 \varphi_2(\phi )^2 \varphi_0(\phi )^4}{95105071448064}+\frac{1333 E_4^9 \varphi_0(\epsilon )^4 \varphi_2(\phi )^2 \varphi_0(\phi )^4}{11888133931008}+\frac{55655 E_4^6 E_6^2 \varphi_0(\epsilon )^4 \varphi_2(\phi )^2 \varphi_0(\phi )^4}{23776267862016}+\frac{10709 E_4^3 E_6^4 \varphi_0(\epsilon )^4 \varphi_2(\phi )^2 \varphi_0(\phi )^4}{5944066965504}+\frac{841 E_6^6 \varphi_0(\epsilon )^4 \varphi_2(\phi )^2 \varphi_0(\phi )^4}{7925422620672}-\frac{4093 E_4^9 E_6 \varphi_2(\epsilon )^4 \varphi_2(\phi ) \varphi_0(\phi )^5}{2229025112064}-\frac{1284709 E_4^6 E_6^3 \varphi_2(\epsilon )^4 \varphi_2(\phi ) \varphi_0(\phi )^5}{47552535724032} \\
            &-\frac{191657 E_4^3 E_6^5 \varphi_2(\epsilon )^4 \varphi_2(\phi ) \varphi_0(\phi )^5}{7925422620672}-\frac{147887 E_6^7 \varphi_2(\epsilon )^4 \varphi_2(\phi ) \varphi_0(\phi )^5}{142657607172096}+\frac{1571869 E_4^{10} \varphi_2(\epsilon )^3 \varphi_0(\epsilon ) \varphi_2(\phi ) \varphi_0(\phi )^5}{570630428688384}+\frac{7746575 E_4^7 E_6^2 \varphi_2(\epsilon )^3 \varphi_0(\epsilon ) \varphi_2(\phi ) \varphi_0(\phi )^5}{190210142896128}+\frac{6762949 E_4^4 E_6^4 \varphi_2(\epsilon )^3 \varphi_0(\epsilon ) \varphi_2(\phi ) \varphi_0(\phi )^5}{190210142896128}+\frac{684647 E_4 E_6^6 \varphi_2(\epsilon )^3 \varphi_0(\epsilon ) \varphi_2(\phi ) \varphi_0(\phi )^5}{570630428688384} \\
            &-\frac{18907 E_4^8 E_6 \varphi_2(\epsilon )^2 \varphi_0(\epsilon )^2 \varphi_2(\phi ) \varphi_0(\phi )^5}{47552535724032}+\frac{5659 E_4^5 E_6^3 \varphi_2(\epsilon )^2 \varphi_0(\epsilon )^2 \varphi_2(\phi ) \varphi_0(\phi )^5}{23776267862016}+\frac{7589 E_4^2 E_6^5 \varphi_2(\epsilon )^2 \varphi_0(\epsilon )^2 \varphi_2(\phi ) \varphi_0(\phi )^5}{47552535724032}-\frac{443885 E_4^9 \varphi_2(\epsilon ) \varphi_0(\epsilon )^3 \varphi_2(\phi ) \varphi_0(\phi )^5}{570630428688384}-\frac{760613 E_4^6 E_6^2 \varphi_2(\epsilon ) \varphi_0(\epsilon )^3 \varphi_2(\phi ) \varphi_0(\phi )^5}{63403380965376}-\frac{2121013 E_4^3 E_6^4 \varphi_2(\epsilon ) \varphi_0(\epsilon )^3 \varphi_2(\phi ) \varphi_0(\phi )^5}{190210142896128} \\
            &-\frac{282151 E_6^6 \varphi_2(\epsilon ) \varphi_0(\epsilon )^3 \varphi_2(\phi ) \varphi_0(\phi )^5}{570630428688384}-\frac{3565 E_4^7 E_6 \varphi_0(\epsilon )^4 \varphi_2(\phi ) \varphi_0(\phi )^5}{11888133931008}-\frac{6947 E_4^4 E_6^3 \varphi_0(\epsilon )^4 \varphi_2(\phi ) \varphi_0(\phi )^5}{5944066965504}-\frac{3277 E_4 E_6^5 \varphi_0(\epsilon )^4 \varphi_2(\phi ) \varphi_0(\phi )^5}{11888133931008}+\frac{31 E_4^{10} \varphi_2(\epsilon )^4 \varphi_0(\phi )^6}{3423782572130304}+\frac{1995541 E_4^7 E_6^2 \varphi_2(\epsilon )^4 \varphi_0(\phi )^6}{1141260857376768}+\frac{246421 E_4^4 E_6^4 \varphi_2(\epsilon )^4 \varphi_0(\phi )^6}{42268920643584}+\frac{4908413 E_4 E_6^6 \varphi_2(\epsilon )^4 \varphi_0(\phi )^6}{3423782572130304} \\
            &-\frac{250129 E_4^8 E_6 \varphi_2(\epsilon )^3 \varphi_0(\epsilon ) \varphi_0(\phi )^6}{95105071448064}-\frac{420143 E_4^5 E_6^3 \varphi_2(\epsilon )^3 \varphi_0(\epsilon ) \varphi_0(\phi )^6}{47552535724032}-\frac{181393 E_4^2 E_6^5 \varphi_2(\epsilon )^3 \varphi_0(\epsilon ) \varphi_0(\phi )^6}{95105071448064}+\frac{36469 E_4^9 \varphi_2(\epsilon )^2 \varphi_0(\epsilon )^2 \varphi_0(\phi )^6}{3423782572130304}+\frac{28319 E_4^6 E_6^2 \varphi_2(\epsilon )^2 \varphi_0(\epsilon )^2 \varphi_0(\phi )^6}{126806761930752}-\frac{274339 E_4^3 E_6^4 \varphi_2(\epsilon )^2 \varphi_0(\epsilon )^2 \varphi_0(\phi )^6}{1141260857376768}+\frac{21935 E_6^6 \varphi_2(\epsilon )^2 \varphi_0(\epsilon )^2 \varphi_0(\phi )^6}{3423782572130304} \\
            &+\frac{208991 E_4^7 E_6 \varphi_2(\epsilon ) \varphi_0(\epsilon )^3 \varphi_0(\phi )^6}{285315214344192}+\frac{377953 E_4^4 E_6^3 \varphi_2(\epsilon ) \varphi_0(\epsilon )^3 \varphi_0(\phi )^6}{142657607172096}+\frac{196319 E_4 E_6^5 \varphi_2(\epsilon ) \varphi_0(\epsilon )^3 \varphi_0(\phi )^6}{285315214344192}+\frac{961 E_4^8 \varphi_0(\epsilon )^4 \varphi_0(\phi )^6}{71328803586048}+\frac{3503 E_4^5 E_6^2 \varphi_0(\epsilon )^4 \varphi_0(\phi )^6}{35664401793024}+\frac{12769 E_4^2 E_6^4 \varphi_0(\epsilon )^4 \varphi_0(\phi )^6}{71328803586048}
    \end{aligned}$}
\end{align*}

\paragraph{$ d=4 $ case} The numerator in the modular ansatz for the 2-string elliptic genus is


\paragraph{$ d=5 $ case} The numerator in the modular ansatz for the 2-string elliptic genus is
%

\paragraph{$ d=6 $ case} The numerator in the modular ansatz for the 2-string elliptic genus is
%

\paragraph{$ d=7 $ case} The numerator in the modular ansatz for the 2-string elliptic genus is
%

\paragraph{$ d=8 $ case} The numerator in the modular ansatz for the 2-string elliptic genus is
%

\newpage

\subsection{\texorpdfstring{$ B=\mathbb{F}_1 $ with $ SU(2) $}{B=F1 with SU(2)} gauge symmetry} \label{appendix:F1-SU2}

We summarize the numerator in the modular ansatz for the elliptic genus of the string with charge $ Q=(1,0) $ considered in the subsection~\ref{subsubsec:F1SU2}.

\paragraph{$ d=1 $ case}
\begin{align*}
    \scalebox{0.6}{$\begin{aligned}
            \mathcal{N}_{(1,0)} &= \frac{E_4^6 \varphi_2(\epsilon ) \varphi_2(\phi )^5}{23328}+\frac{13 E_4^3 E_6^2 \varphi_2(\epsilon ) \varphi_2(\phi )^5}{15552}+\frac{5 E_6^4 \varphi_2(\epsilon ) \varphi_2(\phi )^5}{23328}+\frac{13 E_4^4 E_6 \varphi_2(\phi )^5 \varphi_0(\epsilon )}{373248}+\frac{11 E_4 E_6^3 \varphi_2(\phi )^5 \varphi_0(\epsilon )}{373248}-\frac{199 E_4^4 E_6 \varphi_2(\epsilon ) \varphi_2(\phi )^4 \varphi_0(\phi )}{124416} \\
            &-\frac{47 E_4 E_6^3 \varphi_2(\epsilon ) \varphi_2(\phi )^4 \varphi_0(\phi )}{41472}-\frac{13 E_4^5 \varphi_2(\phi )^4 \varphi_0(\epsilon ) \varphi_0(\phi )}{248832}-\frac{E_4^2 E_6^2 \varphi_2(\phi )^4 \varphi_0(\epsilon ) \varphi_0(\phi )}{9216}+\frac{89 E_4^5 \varphi_2(\epsilon ) \varphi_2(\phi )^3 \varphi_0(\phi )^2}{248832}+\frac{319 E_4^2 E_6^2 \varphi_2(\epsilon ) \varphi_2(\phi )^3 \varphi_0(\phi )^2}{248832} \\
            &+\frac{E_4^3 E_6 \varphi_2(\phi )^3 \varphi_0(\epsilon ) \varphi_0(\phi )^2}{10368}+\frac{83 E_4^3 E_6 \varphi_2(\epsilon ) \varphi_2(\phi )^2 \varphi_0(\phi )^3}{186624}+\frac{19 E_6^3 \varphi_2(\epsilon ) \varphi_2(\phi )^2 \varphi_0(\phi )^3}{186624}+\frac{13 E_4^4 \varphi_2(\phi )^2 \varphi_0(\epsilon ) \varphi_0(\phi )^3}{746496}+\frac{11 E_4 E_6^2 \varphi_2(\phi )^2 \varphi_0(\epsilon ) \varphi_0(\phi )^3}{746496} \\
            &-\frac{31 E_4^4 \varphi_2(\epsilon ) \varphi_2(\phi ) \varphi_0(\phi )^4}{248832}-\frac{35 E_4 E_6^2 \varphi_2(\epsilon ) \varphi_2(\phi ) \varphi_0(\phi )^4}{82944}-\frac{E_4^2 E_6 \varphi_2(\phi ) \varphi_0(\epsilon ) \varphi_0(\phi )^4}{31104}
    \end{aligned}$}
\end{align*}

\paragraph{$ d=2 $ case}
\begin{align*}
    \scalebox{0.6}{$\begin{aligned}
            \mathcal{N}_{(1,0)} &= -\frac{47 E_4^5 \varphi_2(\epsilon ) \varphi_2(\phi )^2}{6912}-\frac{89 E_4^2 E_6^2 \varphi_2(\epsilon ) \varphi_2(\phi )^2}{6912}-\frac{E_4^3 E_6 \varphi_0(\epsilon ) \varphi_2(\phi )^2}{864} +\frac{83 E_4^3 E_6 \varphi_2(\epsilon ) \varphi_2(\phi ) \varphi_0(\phi )}{2592}+\frac{19 E_6^3 \varphi_2(\epsilon ) \varphi_2(\phi ) \varphi_0(\phi )}{2592} \\
            &+\frac{13 E_4^4 \varphi_0(\epsilon ) \varphi_2(\phi ) \varphi_0(\phi )}{10368}+\frac{11 E_4 E_6^2 \varphi_0(\epsilon ) \varphi_2(\phi ) \varphi_0(\phi )}{10368}-\frac{31 E_4^4 \varphi_2(\epsilon ) \varphi_0(\phi )^2}{6912}-\frac{35 E_4 E_6^2 \varphi_2(\epsilon ) \varphi_0(\phi )^2}{2304}-\frac{E_4^2 E_6 \varphi_0(\epsilon ) \varphi_0(\phi )^2}{864} 
    \end{aligned}$}
\end{align*}

\paragraph{$ d=3 $ case}
\begin{align*}
    \scalebox{0.6}{$\begin{aligned}
            \mathcal{N}_{(1,0)} &= \frac{25 E_4^4 E_6 \varphi_2(\epsilon ) \varphi_2(\phi )^3}{20736}+\frac{E_4 E_6^3 \varphi_2(\epsilon ) \varphi_2(\phi )^3}{2304}+\frac{E_4^5 \varphi_0(\epsilon ) \varphi_2(\phi )^3}{27648}+\frac{5 E_4^2 E_6^2 \varphi_0(\epsilon ) \varphi_2(\phi )^3}{82944}-\frac{43 E_4^5 \varphi_2(\epsilon ) \varphi_2(\phi )^2 \varphi_0(\phi )}{27648}-\frac{31 E_4^2 E_6^2 \varphi_2(\epsilon ) \varphi_2(\phi )^2 \varphi_0(\phi )}{9216} \\
            &-\frac{E_4^3 E_6 \varphi_0(\epsilon ) \varphi_2(\phi )^2 \varphi_0(\phi )}{3456}+\frac{83 E_4^3 E_6 \varphi_2(\epsilon ) \varphi_2(\phi ) \varphi_0(\phi )^2}{20736}+\frac{19 E_6^3 \varphi_2(\epsilon ) \varphi_2(\phi ) \varphi_0(\phi )^2}{20736}+\frac{13 E_4^4 \varphi_0(\epsilon ) \varphi_2(\phi ) \varphi_0(\phi )^2}{82944}+\frac{11 E_4 E_6^2 \varphi_0(\epsilon ) \varphi_2(\phi ) \varphi_0(\phi )^2}{82944} \\
            &-\frac{31 E_4^4 \varphi_2(\epsilon ) \varphi_0(\phi )^3}{82944}-\frac{35 E_4 E_6^2 \varphi_2(\epsilon ) \varphi_0(\phi )^3}{27648}-\frac{E_4^2 E_6 \varphi_0(\epsilon ) \varphi_0(\phi )^3}{10368}
    \end{aligned}$}
\end{align*}

\paragraph{$ d=4 $ case}
\begin{align*}
    \scalebox{0.6}{$\begin{aligned}
            \mathcal{N}_{(1,0)} &= -\frac{37 E_4^6 \varphi_2(\epsilon ) \varphi_2(\phi )^4}{2985984}-\frac{113 E_4^3 E_6^2 \varphi_2(\epsilon ) \varphi_2(\phi )^4}{995328}-\frac{E_6^4 \varphi_2(\epsilon ) \varphi_2(\phi )^4}{93312}-\frac{E_4^4 E_6 \varphi_0(\epsilon ) \varphi_2(\phi )^4}{186624}-\frac{E_4 E_6^3 \varphi_0(\epsilon ) \varphi_2(\phi )^4}{373248}+\frac{E_4^4 E_6 \varphi_2(\epsilon ) \varphi_2(\phi )^3 \varphi_0(\phi )}{2592} \\
            &+\frac{5 E_4 E_6^3 \varphi_2(\epsilon ) \varphi_2(\phi )^3 \varphi_0(\phi )}{31104}+\frac{E_4^5 \varphi_0(\epsilon ) \varphi_2(\phi )^3 \varphi_0(\phi )}{82944}+\frac{5 E_4^2 E_6^2 \varphi_0(\epsilon ) \varphi_2(\phi )^3 \varphi_0(\phi )}{248832}-\frac{125 E_4^5 \varphi_2(\epsilon ) \varphi_2(\phi )^2 \varphi_0(\phi )^2}{497664}-\frac{283 E_4^2 E_6^2 \varphi_2(\epsilon ) \varphi_2(\phi )^2 \varphi_0(\phi )^2}{497664} \\
            &-\frac{E_4^3 E_6 \varphi_0(\epsilon ) \varphi_2(\phi )^2 \varphi_0(\phi )^2}{20736}+\frac{83 E_4^3 E_6 \varphi_2(\epsilon ) \varphi_2(\phi ) \varphi_0(\phi )^3}{186624}+\frac{19 E_6^3 \varphi_2(\epsilon ) \varphi_2(\phi ) \varphi_0(\phi )^3}{186624}+\frac{13 E_4^4 \varphi_0(\epsilon ) \varphi_2(\phi ) \varphi_0(\phi )^3}{746496}+\frac{11 E_4 E_6^2 \varphi_0(\epsilon ) \varphi_2(\phi ) \varphi_0(\phi )^3}{746496} \\
            &-\frac{31 E_4^4 \varphi_2(\epsilon ) \varphi_0(\phi )^4}{995328}-\frac{35 E_4 E_6^2 \varphi_2(\epsilon ) \varphi_0(\phi )^4}{331776}-\frac{E_4^2 E_6 \varphi_0(\epsilon ) \varphi_0(\phi )^4}{124416}
    \end{aligned}$}
\end{align*}

\paragraph{$ d=5 $ case}
\begin{align*}
    \scalebox{0.6}{$\begin{aligned}
            \mathcal{N}_{(1,0)} &= \frac{E_4^5 E_6 \varphi_2(\epsilon ) \varphi_2(\phi )^5}{497664}+\frac{7 E_4^2 E_6^3 \varphi_2(\epsilon ) \varphi_2(\phi )^5}{746496}-\frac{E_4^6 \varphi_0(\epsilon ) \varphi_2(\phi )^5}{11943936}+\frac{E_4^3 E_6^2 \varphi_0(\epsilon ) \varphi_2(\phi )^5}{1327104}-\frac{55 E_4^6 \varphi_2(\epsilon ) \varphi_2(\phi )^4 \varphi_0(\phi )}{11943936}-\frac{1675 E_4^3 E_6^2 \varphi_2(\epsilon ) \varphi_2(\phi )^4 \varphi_0(\phi )}{35831808} \\
            &-\frac{25 E_6^4 \varphi_2(\epsilon ) \varphi_2(\phi )^4 \varphi_0(\phi )}{4478976}-\frac{5 E_4^4 E_6 \varphi_0(\epsilon ) \varphi_2(\phi )^4 \varphi_0(\phi )}{2239488}-\frac{5 E_4 E_6^3 \varphi_0(\epsilon ) \varphi_2(\phi )^4 \varphi_0(\phi )}{4478976}+\frac{235 E_4^4 E_6 \varphi_2(\epsilon ) \varphi_2(\phi )^3 \varphi_0(\phi )^2}{2985984}+\frac{35 E_4 E_6^3 \varphi_2(\epsilon ) \varphi_2(\phi )^3 \varphi_0(\phi )^2}{995328} \\
            &+\frac{5 E_4^5 \varphi_0(\epsilon ) \varphi_2(\phi )^3 \varphi_0(\phi )^2}{1990656}+\frac{25 E_4^2 E_6^2 \varphi_0(\epsilon ) \varphi_2(\phi )^3 \varphi_0(\phi )^2}{5971968}-\frac{205 E_4^5 \varphi_2(\epsilon ) \varphi_2(\phi )^2 \varphi_0(\phi )^3}{5971968}-\frac{475 E_4^2 E_6^2 \varphi_2(\epsilon ) \varphi_2(\phi )^2 \varphi_0(\phi )^3}{5971968}-\frac{5 E_4^3 E_6 \varphi_0(\epsilon ) \varphi_2(\phi )^2 \varphi_0(\phi )^3}{746496} \\
            &+\frac{415 E_4^3 E_6 \varphi_2(\epsilon ) \varphi_2(\phi ) \varphi_0(\phi )^4}{8957952}+\frac{95 E_6^3 \varphi_2(\epsilon ) \varphi_2(\phi ) \varphi_0(\phi )^4}{8957952}+\frac{65 E_4^4 \varphi_0(\epsilon ) \varphi_2(\phi ) \varphi_0(\phi )^4}{35831808}+\frac{55 E_4 E_6^2 \varphi_0(\epsilon ) \varphi_2(\phi ) \varphi_0(\phi )^4}{35831808}-\frac{31 E_4^4 \varphi_2(\epsilon ) \varphi_0(\phi )^5}{11943936} \\
            &-\frac{35 E_4 E_6^2 \varphi_2(\epsilon ) \varphi_0(\phi )^5}{3981312}-\frac{E_4^2 E_6 \varphi_0(\epsilon ) \varphi_0(\phi )^5}{1492992}
    \end{aligned}$}
\end{align*}

\paragraph{$ d=6 $ case}
\begin{align*}
    \scalebox{0.6}{$\begin{aligned}
    \mathcal{N}_{(1,0)} &= \frac{E_4^7 \varphi_2(\epsilon ) \varphi_2(\phi )^6}{15925248}-\frac{49 E_4^4 E_6^2 \varphi_2(\epsilon ) \varphi_2(\phi )^6}{143327232}-\frac{E_4 E_6^4 \varphi_2(\epsilon ) \varphi_2(\phi )^6}{1492992}-\frac{E_4^2 E_6^3 \varphi_0(\epsilon ) \varphi_2(\phi )^6}{17915904}+\frac{5 E_4^5 E_6 \varphi_2(\epsilon ) \varphi_2(\phi )^5 \varphi_0(\phi )}{5971968}+\frac{29 E_4^2 E_6^3 \varphi_2(\epsilon ) \varphi_2(\phi )^5 \varphi_0(\phi )}{5971968} \\
    &-\frac{E_4^6 \varphi_0(\epsilon ) \varphi_2(\phi )^5 \varphi_0(\phi )}{23887872}+\frac{E_4^3 E_6^2 \varphi_0(\epsilon ) \varphi_2(\phi )^5 \varphi_0(\phi )}{2654208}-\frac{17 E_4^6 \varphi_2(\epsilon ) \varphi_2(\phi )^4 \varphi_0(\phi )^2}{15925248}-\frac{1663 E_4^3 E_6^2 \varphi_2(\epsilon ) \varphi_2(\phi )^4 \varphi_0(\phi )^2}{143327232}-\frac{7 E_6^4 \varphi_2(\epsilon ) \varphi_2(\phi )^4 \varphi_0(\phi )^2}{4478976} \\
    &-\frac{5 E_4^4 E_6 \varphi_0(\epsilon ) \varphi_2(\phi )^4 \varphi_0(\phi )^2}{8957952}-\frac{5 E_4 E_6^3 \varphi_0(\epsilon ) \varphi_2(\phi )^4 \varphi_0(\phi )^2}{17915904}+\frac{29 E_4^4 E_6 \varphi_2(\epsilon ) \varphi_2(\phi )^3 \varphi_0(\phi )^3}{2239488}+\frac{E_4 E_6^3 \varphi_2(\epsilon ) \varphi_2(\phi )^3 \varphi_0(\phi )^3}{165888}+\frac{5 E_4^5 \varphi_0(\epsilon ) \varphi_2(\phi )^3 \varphi_0(\phi )^3}{11943936} \\
    &+\frac{25 E_4^2 E_6^2 \varphi_0(\epsilon ) \varphi_2(\phi )^3 \varphi_0(\phi )^3}{35831808}-\frac{203 E_4^5 \varphi_2(\epsilon ) \varphi_2(\phi )^2 \varphi_0(\phi )^4}{47775744}-\frac{53 E_4^2 E_6^2 \varphi_2(\epsilon ) \varphi_2(\phi )^2 \varphi_0(\phi )^4}{5308416}-\frac{5 E_4^3 E_6 \varphi_0(\epsilon ) \varphi_2(\phi )^2 \varphi_0(\phi )^4}{5971968} \\
    &+\frac{83 E_4^3 E_6 \varphi_2(\epsilon ) \varphi_2(\phi ) \varphi_0(\phi )^5}{17915904}+\frac{19 E_6^3 \varphi_2(\epsilon ) \varphi_2(\phi ) \varphi_0(\phi )^5}{17915904}+\frac{13 E_4^4 \varphi_0(\epsilon ) \varphi_2(\phi ) \varphi_0(\phi )^5}{71663616}+\frac{11 E_4 E_6^2 \varphi_0(\epsilon ) \varphi_2(\phi ) \varphi_0(\phi )^5}{71663616}-\frac{31 E_4^4 \varphi_2(\epsilon ) \varphi_0(\phi )^6}{143327232} \\
    &-\frac{35 E_4 E_6^2 \varphi_2(\epsilon ) \varphi_0(\phi )^6}{47775744}-\frac{E_4^2 E_6 \varphi_0(\epsilon ) \varphi_0(\phi )^6}{17915904}
\end{aligned}$}
\end{align*}

\newpage

\subsection{\texorpdfstring{$ B=\mathbb{F}_1 $ with $ SU(2)\times SU(2) $}{B=F1 with SU(2)×SU(2)} gauge symmetry} \label{appendix:F1-SU2SU2}

We present the numerator in the modular ansatz for the elliptic genus of the string with charge $ Q=(1,0) $ and $ Q=(0,2) $ considered in the subsection~\ref{subsubsec:F1-SU2SU2}.


%
\bibliographystyle{JHEP}
\bibliography{refs}
\end{document}